\documentclass[a4paper,12pt]{article}
\usepackage{graphicx,rotating,hyperref,slashed,verbatim,amsmath,xcolor,amssymb,amsfonts,expdlist,colortbl,cite,youngtab}
\usepackage{amsfonts}
\usepackage{halloweenmath}
\usepackage{physics}
\usepackage{amsmath}
\usepackage{mathtools}
\usepackage{hhline}
\usepackage{comment}
\usepackage{colortbl}
\usepackage{multicol}
\usepackage{pdflscape}
\usepackage{multirow}
\usepackage{booktabs}    
\usepackage{calc}        
\usepackage{tikz}
\usepackage[compat=1.1.0]{tikz-feynman}
\usepackage{subcaption}
\usepackage{fancyhdr}

\hypersetup{colorlinks,bookmarksopen,bookmarksnumbered,
	linkcolor=blus,pdfstartview=FitH,urlcolor=rossos,citecolor=verde}

\newcommand{\BR}{\hbox{BR}}
\usepackage{tabularx}

\definecolor{gold}{RGB}{187,161,79}
\definecolor{silver}{RGB}{192,192,192}
\definecolor{darkgreen}{RGB}{0, 130, 0}

\def\greenmedal{\textcolor{darkgreen}{\huge $\bullet$}}

\newcommand{\sfootnote}[1]{}
\definecolor{bluc}{cmyk}{1,1,0,0.1}
\definecolor{rossoCP3}{cmyk}{0,.88,.77,.40}
\definecolor{rosso}{cmyk}{0,1,1,0.4}
\definecolor{giallo}{cmyk}{0,.33,1,0}
\definecolor{rossos}{cmyk}{0,1,1,0.55}
\definecolor{rossoc}{cmyk}{0,1,1,0.2}
\definecolor{verdes}{cmyk}{0.92,0,0.59,0.4}

\hypersetup{colorlinks, bookmarksopen, bookmarksnumbered,
	citecolor=verdes, linkcolor=bluc, pdfstartview=FitH, urlcolor=rossos}

\renewcommand{\Re}{{\rm Re}\,}
\newcommand{\mio}[1]{}

\newcommand{\fig}[1]{~\ref{fig:#1}}

\definecolor{Gray}{gray}{0.95}

\usepackage{multicol}
\usepackage{color}
\definecolor{rosso}{cmyk}{0,1,1,0.4}
\definecolor{rossos}{cmyk}{0,1,1,0.55}
\definecolor{rossoc}{cmyk}{0,1,1,0.2}
\definecolor{blu}{cmyk}{1,1,0,0.3}
\definecolor{blus}{cmyk}{1,1,0,0.6}
\definecolor{bluc}{cmyk}{1,1,0,0.1}
\definecolor{verde}{cmyk}{0.92,0,0.59,0.25}
\definecolor{verdec}{cmyk}{0.92,0,0.59,0.15}
\definecolor{verdes}{cmyk}{0.92,0,0.59,0.4}

\renewcommand\&{&}

\newcommand{\MeV}{\,{\rm MeV}}
\newcommand{\GeV}{\,{\rm GeV}}
\newcommand{\TeV}{\,{\rm TeV}}

\def\circa#1{\,\raise.3ex\hbox{$#1$\kern-.75em\lower1ex\hbox{$\sim$}}\,}

\newcommand{\beq}{\begin{equation}}
	\newcommand{\eeq}{\end{equation}}

\newcommand{\bea}{\begin{eqnarray}}
	\newcommand{\eea}{\end{eqnarray}}
\newcommand{\be}{\begin{equation}}
	\newcommand{\ee}{\end{equation}}
\font\tenrsfs=rsfs10 at 12pt
\font\sevenrsfs=rsfs7 at 10 pt
\font\fiversfs=rsfs5
\newfam\rsfsfam
\textfont\rsfsfam=\tenrsfs
\scriptfont\rsfsfam=\sevenrsfs
\scriptscriptfont\rsfsfam=\fiversfs
\def\mathscr#1{{\fam\rsfsfam\relax#1}}

\makeatletter

\def\hhref#1{\href{http://arxiv.org/abs/#1}{arXiv:#1}} 

\newcommand{\doi}[1]{\href{http://dx.doi.org/#1}{[doi]}}

\def\hhref#1{\href{http://arxiv.org/abs/#1}{arXiv:#1}}

\def\art{\@ifnextchar[{\eart}{\oart}}
\def\eart[#1]#2#3#4#5#6{{\rm #2}, {\em #3 \bf #4} {\rm (#6) #5} ({\em #1})}

\def\article{\@ifnextchar[{\earticle}{\oarticle}}
\def\oarticle#1#2#3#4#5#6{{\rm #1}, {\em ``#6''}, {\rm #2 #3 (#5) #4}}
\def\earticle[#1]#2#3#4#5#6#7{{\rm #2}, {\em ``#7''}, {\rm #3 #4 (#6) #5}  [\hhref{#1}]}
\def\hepart[#1]#2{{\rm #2, \em#1}}
\def\heparticle[#1]#2#3{#2, {\em ``#3''} [\hhref{#1}]}

\newcounter{alphaequation}[equation]
\def\thealphaequation{\theequation\hbox to
	0.6em{\hfil\alph{alphaequation}\hfil}}
\def\eqnsystem#1{
	\def\@eqnnum{{\rm (\thealphaequation)}}
	\def\@@eqncr{\let\@tempa\relax \ifcase\@eqcnt \def\@tempa{& & &} \or
		\def\@tempa{& &}\or \def\@tempa{&}\fi\@tempa
		\if@eqnsw\@eqnnum\refstepcounter{alphaequation}\fi
		\global\@eqnswtrue\global\@eqcnt=0\cr}
	\refstepcounter{equation} \let\@currentlabel\theequation \def\@tempb{#1}
	\ifx\@tempb\empty\else\label{#1}\fi
	\refstepcounter{alphaequation}
	\let\@currentlabel\thealphaequation
	\global\@eqnswtrue\global\@eqcnt=0 \tabskip\@centering\let\\=\@eqncr
	$$\halign to \displaywidth\bgroup \@eqnsel\hskip\@centering
	$\displaystyle\tabskip\z@{##}$&\global\@eqcnt\@ne
	\hskip2\arraycolsep\hfil${##}$\hfil& \global\@eqcnt\tw@\hskip2\arraycolsep
	$\displaystyle\tabskip\z@{##}$\hfil
	\tabskip\@centering&\llap{##}\tabskip\z@\cr}
\def\endeqnsystem{\@@eqncr\egroup$$\global\@ignoretrue} \makeatother

\newcommand{\SU}{\,{\rm SU}}

\definecolor{fiorentina}{rgb}{.5,0,.5}

\usepackage{tikz-feynman}

\let\oldheadrule\headrule
\renewcommand{\headrule}{\color{white}\oldheadrule}
\begin{document}

	\begin{center}
		\boldmath
		
		{\textbf{\LARGE\color{rossoCP3}
				Standard model anomalies:} \\~\\ \textbf{\Large\color{rossoCP3}Lepton  flavour non-universality, $g-2$ and W-mass
			}}
		\unboldmath
		
		\bigskip\bigskip

	 	{ Alessandra D'Alise$^1$, Guglielmo~De~Nardo$^1$, Maria Grazia~Di Luca$^{1,2}$, Giuseppe~Fabiano$^1$, Domenico~Frattulillo$^1$,  Giovanni Gaudino$^{1,2}$, Davide~Iacobacci$^{1}$,  Mario Merola$^1$, Francesco~Sannino$^{1,2,3,4}$, Pietro~Santorelli$^1$, Natascia~Vignaroli$^1$\\[5mm]}

\small{$^1$ Dipartimento di Fisica ``E. Pancini", Università di Napoli Federico II - INFN sezione di Napoli, Complesso Universitario di Monte S. Angelo Edificio 6, via Cintia, 80126 Napoli, Italy}\\
\small{$^2$ Scuola Superiore Meridionale, Largo S. Marcellino, 10, 80138 Napoli NA, Italy}\\
\small{$^3$ CP$^3$-Origins and D-IAS, Univ. of Southern Denmark, Campusvej 55, 5230 Odense M, Denmark}\\
\small{$^4$ CERN, Theoretical Physics Department, 1211 Geneva 23, Switzerland}\\

		\bigskip\bigskip
		
		\thispagestyle{empty}\large
		{\bf\color{blus} Abstract}
		\begin{quote} 
		\normalsize
		We critically analyze the body of results that hints to the existence of New Physics from possible violations of lepton universality observed by the LHCb experiment in the $\mu/e$ ratios $R_{K}$ and  $R_{K^*}$  to the $g-2$ lepton anomalies.   The analysis begins with a theoretical, in depth, study of the $\mu/e$ ratios $R_{K}$ and  $R_{K^*}$  as well as the process $B_s \rightarrow \mu^+ \mu^-$. Here we consider the impact of complex Wilson coefficients and  derive constraints on their imaginary and real parts. We then move to a comprehensive comparison with experimental results. We show that, by fitting a single Wilson coefficient, the deviations from the Standard Model are at the $4.7\sigma$ level when including only the hadronic insensitive observables while it increases to $6.1\sigma$ when including also the hadronic sensitive ones. When switching on all relevant Wilson coefficients and combining both hadronic sensitive and insensitive data into the fit, the deviation from the Standard Model peaks at $7.2\sigma$ and decreases at the $4.9\sigma$ level if we assume that the central values of $R_K$ and $R_{K^{\ast}}$ are taken to be unity. We further estimate other unaccounted for SM contributions and show that their inclusion still requires New Physics to fit the data.   We then introduce the $g-2$ lepton anomalies  as well as the most recent $W$-mass results.  Different theoretical models are considered that can explain the discrepancies from the Standard Model. In the final part of our work we  estimate the impact of the forthcoming data from LHCb (coming from LHC Run3) and Belle II, when it will have accumulated about $5~ab^{-1}$.   
		\end{quote}
		\thispagestyle{empty}
	\end{center}
	
	\setcounter{page}{1}
	\setcounter{footnote}{0}
	
	
\thispagestyle{fancy}
\rhead{CERN-TH-2022-063}

	\newpage
	\tableofcontents

	\newpage
 
	\section{Introduction}
	The phenomenological success of the Standard Model (SM) of particle interactions is being challenged by a number of experimental results. These include significant deviations with respect to the SM predictions in rare $B$ meson decays~\cite{1403.8044,BELLE:2019xld,LHCb:2017avl,LHCb:2021trn,LHCb:2021awg,2105.14007,LHCb:2020gog,Belle:2019oag,ATLAS:2018cur,CMS:2019qnb,1606.04731,1312.5364,hep-ex/0503044,CMS:2017ivg,ATLAS:2017dlm} as well as lepton anomalous magnetic moments \cite{Muong-2:2006rrc}.    
	
These anomalies have attracted a great deal of interest.  We aim at providing an in depth study of the anomalies including possible SM interpretations and then use this knowledge to investigate the impact for New Physics (NP) by way of a number of benchmark models. Last but not least we will consider predictions for future experiments. Following  and expanding on Ref.~\cite{DAmico:2017mtc} we summarise observables involved in rare $B$ meson decays directly related to  the observed anomalies:

	\begin{enumerate}
		\item The $R_{K}$ and $R_{K^*}$ ratios~\cite{BELLE:2019xld,LHCb:2021trn,LHCb:2017avl,Belle:2019oag}, respectively
		\beq \label{eq:RKRK*}
		R_K =\frac{{\rm BR} \left ( B^+ \to K^+ \mu ^+ \mu ^-\right )}{{\rm BR} \left ( B^+ \to K^+ e^+ e^-\right )}, \quad 	R_{K^*} =\frac{{\rm BR} \left ( B \to K^* \mu ^+ \mu ^-\right )}{{\rm BR} \left ( B \to K^* e^+ e^-\right )} .
		\eeq
		\item The branching ratio $\text{BR}(B_s \to \mu^+ \mu^-)$ \cite{LHCb:2021awg,ATLAS:2018cur,CMS:2019qnb} .
		\item The differential branching ratios of the semi-leptonic decays $B \to K^{(*)} \mu^+ \mu^-$~\cite{1403.8044,1606.04731} and $B_s \to \phi \mu^+ \mu^-$~\cite{2105.14007}; in particular, the latter shows a deviation of more than $3.5$ $\sigma$ from the Standard Model prediction in several bins.
		\item The angular distributions of the decay rate of $B \to K^* \mu^+ \mu^-$ ~\cite{LHCb:2020gog,CMS:2017ivg,ATLAS:2017dlm}.
	
	\end{enumerate}
The $R_K$ and $R_{K^*}$ measures are a direct test of lepton flavour universality.
The ratio of branching ratios is introduced to strongly suppress SM theoretical uncertainties associated with hadronic effects, as suggested first in ref.~\cite{Hiller:2003js} and along with the   $B_s \to \mu^+ \mu^-$ branching ratio constitute theoretically hadronic insensitive \cite{DAmico:2017mtc} observables to test lepton flavour universality. Following \cite{DAmico:2017mtc} here by `Hadronic Insensitive' (HI) we refer to those observables featuring few percent SM induced theoretical errors. Differently, the observables in $3.$ and $4.$ are strongly affected by theoretical uncertainties stemming from the proper evaluation of the associated form factors and estimates of the non-factorizable hadronic contributions. It is for this reason that these are referred to as Hadronic Sensitive ~(HS) observables.  It is therefore  important to gain control of these corrections ~\cite{1006.4945,1101.5118,1207.2753,1211.0234,1212.2263,1406.0566,1407.8526,1412.3183,1512.07157,1701.08672,1702.02234,Bobeth:2017vxj,Gubernari:2020eft}. 
Overall, as we shall see, we will further confirm  that these anomalies constitute consistent growing evidence for deviations from the SM related to lepton non-universal flavor processes. In fact, the consistent deviations from the SM  was pointed out in a number or research papers \cite{Ghosh:2014awa,Altmannshofer:2014rta,Descotes-Genon:2015uva,Altmannshofer:2013foa,Hurth:2013ssa,Hiller:2014yaa,Alonso:2014csa,Hurth:2014vma,Ciuchini:2017mik,Capdevila:2017bsm,Altmannshofer:2017fio,Alguero:2019ptt,Aebischer:2019mlg,Ciuchini:2020gvn}, while the look elsewhere effect has been recently investigated in \cite{Isidori:2021vtc}.

At same time tantalizing hints of NP are emerging in another relevant sector of the SM. In fact, the Fermilab collaboration recently presented their first  measurement of the muon anomalous magnetic moment \cite{Abi:2021gix}, confirming a discrepancy  from the SM value (by $3.3 \sigma$).  Combining this result with the BNL E821 one \cite{Bennett:2006fi} leads to the experimental average of 
\begin{equation}
a_{\mu} ({\mathrm{Exp}}) = 116 \, 592 \, 061 (41) \times 10^{-11} \quad (0.35\,{\mathrm{ppm}}) \ ,
\end{equation} 
 with an overall deviation from the SM central value \cite{Aoyama:2020ynm} of
  \begin{equation}
\Delta a_{\mu} = a_{\mu} ({\mathrm{Exp}})  - a_{\mu} ({\mathrm{SM}}) = (251 \pm 59) \times 10^{-11} \ ,
\end{equation} 
corresponding to a significance of $4.2\sigma$ \cite{Abi:2021gix}. However,  if the latest lattice results from the BMW collaboration \cite{Borsanyi:2020mff}, not included in the world average, are considered as the true contribution to the hadronic vacuum polarisation of the photon \cite{Davier:2010nc,Davier:2017zfy,Davier:2019can}, the discrepancy from the SM reduces. 
 
 For the electron case, a recent measurement of the fine-structure constant  based on Rubidium atoms \cite{Morel:2020dww} leads to a deviation of 1.7$\sigma$ from the SM in the associated magnetic moment:
  \begin{equation}
\Delta a_{e} = (0.48 \pm 0.30) \times 10^{-12} \ .
\end{equation}
A previous measurement of $\alpha$ based on Cs atoms \cite{Parker:2018vye} has given a deviation on $a_e$ from the SM with opposite sign, corresponding to about -2.5$\sigma$. \\

Last but not least, recently the CDF collaboration provided the 
  best  measure of the  $W$ boson mass  
\cite{CDF:2022hxs}
\begin{equation} \label{eq:CDF}
    \left. M_W \right|_{\rm CDF} = 80,433.5 \pm 6.4_{\rm stat} \pm 6.9_{\rm syst}= 80,433.5 \pm 9.4\, {\rm MeV}  \ . 
\end{equation}
Shockingly the central value deviates by more than $7\sigma$  from the expected SM value \cite{ParticleDataGroup:2020ssz}:
\begin{equation} \label{eq:SM}
    \left. M_W \right|_{\rm SM} = 80,357 \pm 4_{\rm inputs} \pm 4_{\rm theory}  \, {\rm MeV}  \ . 
\end{equation}

\bigskip
	The paper is structured as follows.
	In section~\ref{sec:TheorPred} we analyse the HI observables from a theoretical viewpoint and we express them in terms of effective operators. We also provide an, in depth, analysis of the impact of complex Wilson coefficients. The analytic study forms the bedrock on which we build our comparison to the experimental results.
	
	The data analysis is performed in section~\ref{fit}. Here we first analyse the HI data and show that for a single muon Wilson coefficient the deviation from the SM is at the $4.7 \sigma$ level while this value increases to $6.1 \sigma$ when HS observables are added. 
	Different non-zero combinations of Wilson coefficients are also investigated without dramatically changing the overall picture. 
	
	Because the HS data may be affected by  unaccounted for hadronic effects \cite{Ciuchini:2020gvn,Ciuchini:2019usw,Ciuchini:2021smi,Bobeth:2017vxj,Gubernari:2020eft} we propose a new way of estimating them leading to a result which is still consistent with the emergence of NP. In our analysis all four Wilson coefficients are simultaneously turned on and a simple estimate of the overall deviation from the SM is at the $7.1\sigma$ level. The deviation from the Standard Model remains at the $4.9\sigma$ level even assuming that the central values of $R_K$ and $R_{K^{\ast}}$ are taken to be unity.
	 
We then introduce the experimental results for the $g-2$ lepton anomalies in section~\ref{g-2} and the CDF $W$-mass result in section~\ref{CDFWmass}.

Prompted by the deviations from the SM above and taking into account our results, we consider in section~\ref{th} several theoretical  models apt to explain these anomalies such as:  $Z'$,  leptoquarks, composite Higgs dynamics as well as radiative models. We will show that models with scalar leptoquarks and technicolor-like theories are the most favourable to explain simultaneously all of the anomalies, or at least to alleviate the tension with the SM. 

 We finally discuss in section~\ref{Belle} the impact of the future data collected by LHCb at run3 and Belle II after having collected $5~ab^{-1}$.   We offer out conclusions in section~\ref{conclusions}. In Appendix~\ref{appendixA} we report the observables used in the data analysis.

	\section{Hadronic insensitive   observables}\label{sec:TheorPred}

To explore NP emerging around the electroweak energy scale affecting the bottom to strange transitions involving leptons one is led to introduce the following class of effective operators	\beq {\cal O}_{b_X \ell_Y} = (\bar s \gamma_\mu P_X b)(\bar \ell \gamma_\mu P_Y \ell) \ ,\eeq
 which can be written as $\SU(2)_L$-invariant operators. A more general discussion can be found in ~\cite{Aebischer:2017gaw,Aebischer:2018bkb,1407.7044}.
 These operators are  incorporated  in the following effective Hamiltonian 
	\beq \mathscr{H}_{\rm eff}
	=  - V_{tb} V_{ts}^*  \frac{\alpha_{\rm em}}{4\pi v^2}
	\sum_{\ell, X, Y} C_{b_X \ell_Y}  {\cal O}_{b_X \ell_Y}\, + \mathrm{h.c.} \, , \eeq
	where the sum runs over leptons $\ell =   \{e,\mu , \tau\}$ and over their chiralities $X,Y = \{L,R\}$. Additionally, it is convenient to define dimensionless coefficients $C_I$ related to the dimensionful $c_I$ coefficients appearing in the equivalent Lagrangian formulation 
	\begin{equation}
	\mathscr{L}_{\rm eff}=
	\sum_{\ell, X, Y} c_{b_X \ell_Y}  {\cal O}_{b_X \ell_Y} \ ,\eeq
and
	\beq c_I = V_{tb} V_{ts}^*  \frac{\alpha_{\rm em}}{4\pi v^2} C_I   = - \frac{C_I}{(36\TeV)^2} \, ,
	\eeq
	where
	$V_{ts}= - 0.0412 \pm 0.0006$  and 
	$v= 1/(2\sqrt{2}G_{\rm F}) = (174\GeV)^2$ is the square of the Higgs vacuum expectation value and $G_{\rm F}$ the Fermi constant.

 Motivated by the SM prediction that $|C^{\rm SM}_{b_L \ell_R}| \ll |C^{\rm SM}_{b_L \ell_L}|$, since  $C^{\rm SM}_{b_L \ell_L} = 8.64$ and $C^{\rm SM}_{b_L \ell_R} = -0.18$, as in \cite{DAmico:2017mtc} we use the chiral basis related to the standard one~\cite{1411.3161}  via
	\begin{eqnarray}
	 2C_9 = {C_{b_L \mu_{L+R}}}, \quad  2C_{10}=-{C_{b_L \mu_{L-R}}}, \quad 2C'_9 = {C_{b_R \mu_{L+R}}}, \quad  2C'_{10}=-{C_{b_R \mu_{L-R}}} \ .
	 \end{eqnarray}
 In the SM we have $C_9^{\rm SM} \approx - C_{10}^{\rm SM}$.  We also use the following  notation \cite{DAmico:2017mtc} 
	 \begin{eqnarray}
	C_{b_{L\pm R} \ell_{Y}} &\equiv & C_{b_L \ell_Y} \pm C_{b_R\ell_Y}, \quad
	C_{b_{L+R} \ell_{L\pm R}} \equiv C_{b_L \ell_L}+ C_{b_R \ell_L} \pm C_{b_L \ell_R} \pm C_{b_R \ell_R}, \nonumber \\C_{b_X (\mu-e)_Y} &\equiv & C_{b_{X} \mu_{Y}}- C_{b_{X} e_{Y}}.
    	\end{eqnarray}

	Experimental deviations from the SM predictions are encoded by splitting  the $C_{b_X\ell_Y}$  Wilson coefficients as follows
	\begin{equation}
	    C_I=C_I^{SM}+\Delta C_I \ .
	    \label{WilsonCoefficient}
	\end{equation}
	
{ We define with $C_I^{SM}$ the coefficients arising from the RG analysis performed by {\sc Flavio}~\cite{Straub:2018kue,Aebischer:2017gaw,Aebischer:2018bkb} within the SM.  }

We further divide $\Delta C_I$ into a part that contains potentially unknown  hadronic   effects \cite{Ciuchini:2020gvn}  {(not contained in  {\sc Flavio}~\cite{Straub:2018kue})} plus a genuine beyond standard model (BSM) contribution:
	\begin{equation}
	   \Delta C_I=C_I^H+C_I^{BSM} \ .
	    \label{WilsonCoefficient-modified}
	\end{equation}
Although, in principle, one needs a non-perturbative computation of the hadronic  contributions \cite{Colangelo:1995jv,Gubernari:2020eft}, it is a fact that any SM-induced hadronic corrections  preserve lepton flavour universality (LFU) and therefore it cannot explain $R_K$ and $R_{K^*}$. We use this point, in one of our analyses, to phenomenologically estimate both the hadronic contributions and the ones stemming from BSM able to explain the observed lepton flavour universality violations (LFV)s. 
	
 We move now to an anatomic study of the various relevant observables starting with the HI ones.

\begin{figure}[ht!] 
	    \centering
	    \begin{subfigure}{0.5\textwidth}
	    \centering
	    \begin{tikzpicture}
    \begin{feynman}
    \vertex (a1) {\(\quad \overline b\)};
    \vertex[right=2cm of a1] (a2);
    \vertex[right=2cm of a2] (a3);
    \vertex[right=2cm of a3] (a4) {\(\overline s\quad\)};
    \vertex[below=2em of a1] (b1) {\(d/u\)};
    \vertex[below=2em of a4] (b2) {\(d/u\)};
    \vertex at ($(a2)!0.5!(a3)!1cm!90:(a3)$) (d);
    \vertex[above=1.5cm of a4] (c1) {\(\ell^-\)};
    \vertex[above=4em of c1] (c3) {\(\ell^+\)};
    \vertex at ($(c1)!0.5!(c3) - (1cm, 0)$) (c2);
    \diagram* {
    (a4) -- [fermion] (a3) -- [red, boson, edge label=\(W^+\)] (a2) -- [fermion] (a1),
    (b1) -- [fermion] (b2),
    (c3) -- [fermion, out=180, in=45] (c2) -- [fermion, out=-45, in=180] (c1),
    (a2) -- [anti fermion, quarter left,edge label=\(\overline{u}\text{,}\overline{c}\text{,}\overline{t}\)] (d) -- [anti fermion, quarter left] (a3),
    (d) -- [boson, red, bend left, edge label=\(\gamma/Z^0\), momentum'={[arrow style=red, label distance=0.5mm]\(q^2\)}] (c2),
    };
    \draw [decoration={brace}, decorate] (b1.south west) -- (a1.north west)
    node [pos=0.5, left] {\(B^0/B^+\)};
    \draw [decoration={brace}, decorate] (a4.north east) -- (b2.south east)
    node [pos=0.5, right] {\(K^*/K\)};
    \end{feynman}
    \end{tikzpicture}
    \subcaption{Penguin diagram of $B\rightarrow K^{(*)}\ell^+\ell^-$.}
    \label{fig:RKs1a}
    \end{subfigure}
    \\
    	\begin{subfigure}{0.5\textwidth}
	    \centering
	    \begin{tikzpicture}
\begin{feynman}
\vertex (a1) {\( \quad \, \overline b \)};
\vertex[right=2cm of a1] (a2);
\vertex[right=2cm of a2] (a3);
\vertex[right=2cm of a3] (a4) {\(\overline s \quad \,\)};
\vertex[below=3em of a1] (b1) {\(d/u\)};
\vertex[right=2cm of b1] (b2);
\vertex[right=2cm of b2] (b3);
\vertex[right=2cm of b3] (b4) {\(d/u\)};
 \vertex at ($(a4) + (0cm, 3cm)$)  (c1){\(\ell^+\)};
 \vertex at ($(a3) + (1cm, 3cm)$) (c2);
 \vertex at ($(a3) + (1cm, 1cm)$) (c3);
 \vertex at ($(a4) + (0cm, 1cm)$)  (c4){\(\ell^-\)};
\diagram* {
{[edges=fermion]
(c1) -- (c2) -- [fermion, pos=0.5, edge label= \(\nu_\ell \)] (c3) -- (c4),
(a4) -- (a3) -- [fermion, pos=0.5, edge label= \(\overline u\text{,}\overline c\text{,}\overline t\)] (a2) -- (a1),
(b1) -- (b4),
},
(a2) -- [red, boson, edge label=\(W\)] (c2),
(a3) -- [red, boson, edge label=\(W\)] (c3),
};
\draw [decoration={brace}, decorate] (b1.south west) -- (a1.north west)
node [pos=0.5, left] {\(B^{0}/B^{+}\)};
\draw [decoration={brace,mirror}, decorate] (b4.south east) -- (a4.north east)
node [pos=0.5, right] {\(K^{*}/K\)};

\end{feynman}
\end{tikzpicture}
	    \subcaption{Box diagram for $B\rightarrow K^{(*)}\ell^+\ell^-$. }
	    \label{fig:RKs1b}
	\end{subfigure}
	\caption{Leading Feynman diagrams for the decays $B^0/B^+\to K^\ast /K\,\, \ell^+ \ell^-$.}
	\label{rkrkstardiagrams}
	\end{figure}
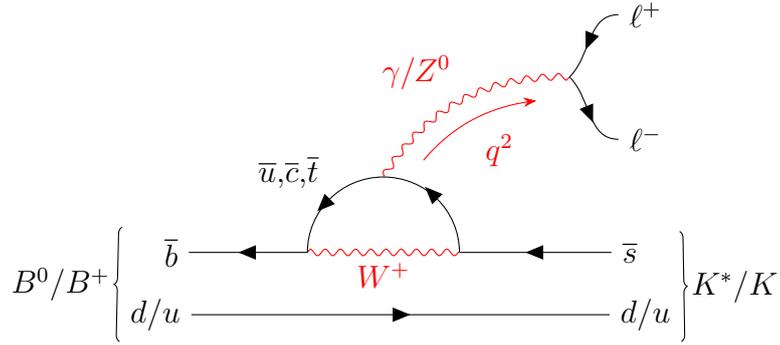
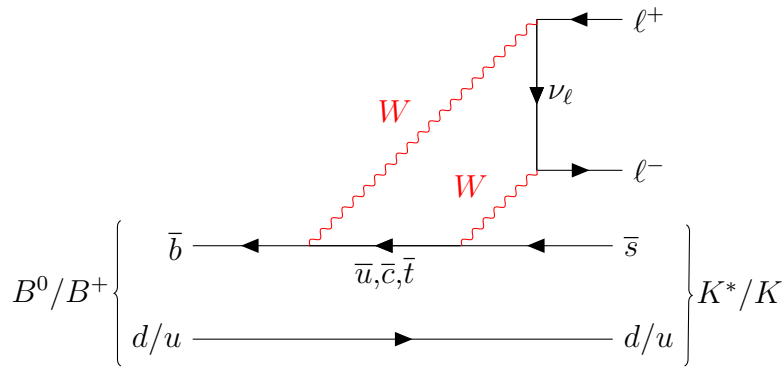
	
	\subsection{\texorpdfstring{$R_K$} - anatomy}
	
	The LHCb collaboration presented a recent measurement of
	the $R_K$ ratio (see Figure \ref{fig:RKs1a} and Figure \ref{fig:RKs1b} representing the leading Feynman diagrams for the relevant $B$ decay) with full Run 1 and Run 2 data. 
    The result~\cite{LHCb:2021trn} is

	\beq \label{eq:RKexpnew}
		R_K =\frac{{\rm BR} \left ( B^+ \to K^+ \mu ^+ \mu ^-\right )}{{\rm BR} \left ( B^+ \to K^+ e^+ e^-\right )}= 0.846^{+0.042}_{-0.039}\text{(stat)}^{+0.013}_{-0.012}\text{(syst)}
		\qquad [1.1,6]\,\text{GeV}^2\ .
		\eeq
	
		The central value in~(\ref{eq:RKexpnew}) is $3.5$ standard deviations below the SM unity prediction.
	Furthermore, the {\sc Belle} collaboration presented a preliminary 
	results for this quantity, that is~\cite{BELLE:2019xld}  $$R_K=1.03^{+0.28}_{-0.24}\pm 0.01 \quad [1.1,6]\,\text{GeV}^2.$$
	
Because	the experimental analysis provides results in bins of the squared invariant mass of the lepton system $q^2\equiv (P_{\ell^-} + P_{\ell^+})^2$  we consider the related $q^2$-dependence  via
	\begin{equation}\label{eq:RKq2}
		R_K[q_{\rm min}^2,q_{\rm max}^2]\equiv \frac{\int_{q_{\rm min}^2}^{q_{\rm max}^2}dq^2 d\Gamma(B^+ \to K^+ \mu^+ \mu^-)/dq^2}{\int_{q_{\rm min}^2}^{q_{\rm max}^2}dq^2 d\Gamma(B^+ \to K^+ e^+ e^-)/dq^2} \, .
	\end{equation}
	The energy bin reported by the experiments used in eq.~(\ref{eq:RKexpnew}) refers to 
	$R_K\! \equiv\! R_K[1.1\,{\rm GeV}^2,6\,{\rm GeV}^2]$. More precisely, the decay rate leading to eq.~(\ref{eq:RKq2}), in the limit of vanishing lepton masses, reads~\cite{1411.3161}
	
	\begin{equation}\label{eq:RKfull}
			\frac{ d\Gamma(B^+ \to K^+ \mu^+ \mu^-)}{dq^2} = \frac{G_{\rm F}^2\alpha_{\rm em}^2|V_{tb}V_{ts}^*|^2}{2^{10}\pi^5 M_B^3}\lambda^{3/2}(M_B^2, M_K^2, q^2)
			\left(
			|F_V|^2 + |F_A|^2
			\right)~,
		\end{equation}
		where  $\lambda(a,b,c) \equiv a^2 + b^2 + c^2 -2(ab+bc+ac)$, $M_B \approx 5.279$ GeV, $M_K \approx  0.494$ GeV,
		$|V_{tb}V_{ts}^*| \approx 40.58\times 10^{-3}$.    The form factors $F_{V/A}$ collect the contribution of the standard form factors $f_{+/T}$ as well as the relevant Wilson coefficients and possible non-factorazible hadronic terms
		\begin{eqnarray}
			F_A(q^2) &=& \left(C_{10} + C_{10}^{\prime}\right)f_+(q^2)~,\\
			F_V(q^2) &=& (C_9 + C_9^{\prime})f_+(q^2) + \underbrace{\frac{2m_b}{M_B + M_K}\left(C_7 +
				C_7^{\prime}\right)f_T(q^2)}_{\mathclap{\rm SM\,electromagnetic\,dipole\,contribution}}\qq{+} \underbrace{h_K(q^2)}_{\mathclap{\rm non-factorizable\,term}}~.
		\label{nonfact}
		\end{eqnarray}
	
It is possible to neglect the SM contributions from the electromagnetic dipole operator (encoded in the coefficients $C_7^{(\prime)}$) because  of the experimental cut $q_{\rm min}^2 = 1.1$ GeV$^2$.  

{{ It is important to establish which part of the  non-factorizable contribution, stemming from $h_K(q^2)$, affects the scaling dependence of  $C_9^{SM}$  and which enters  the hadronic contribution in $C_9^H$. In fact, one can even define an effective   $C_9^{\rm eff}$  that eliminates the overall scale-dependence due to its contribution as shown in \cite{Gubernari:2020eft}. This happens because there is a part of $h_K(q^2)$ which is proportional to $f_+(q^2)$ \cite{Gubernari:2020eft}. There is, however, still an undetermined hadronic long-distance contribution estimated, for example in \cite{Colangelo:1995jv} and discussed also in  \cite{Gubernari:2020eft}. Here, we use a pragmatic approach according to which $C_9^{SM}$ is taken to be the result of the RG analysis of   {\sc Flavio}~\cite{Straub:2018kue}. We relegate the undetermined hadronic contribution in $C_9^H$. Once precise non-perturbative techniques will be able to determine the hadronic contributions they can be compared with our fitted results.   }  }

We will return to the non-factorizable hadronic contributions in the later sections but we neglect them here since they do not affect the LFV processes.  One finally arrives at the following expression for  $R_K$ in terms of the relevant Wilson coefficients:
	\beq \label{eq:RKTheory}
	R_K 
	= \frac{|C_{b_{L+R} \mu_{L-R}}|^2  + |C_{b_{L+R} \mu_{L+R}}|^2}{|C_{b_{L+R} e_{L-R}}|^2  + |C_{b_{L+R} e_{L+R}}|^2} \ .
	\eeq
	In the SM it is expected to be $R_K=1$ with QED corrections in the relevant $q^2$ bin of the order of few percents~\cite{1605.07633}.

To learn more about the $\Delta C_I$ structure we  switch on muon and electron coefficients' deviations separately. We also assume $C_I^{H}=0$ since here we are interested in understanding the contribution to LFVs. 
We expand (\ref{eq:RKTheory}) by splitting the Wilson coefficient  in  SM and BSM contributions. Focusing, at first, on assuming NP in the muon sector we have: 
	 	 \beq \label{eq:RKanalytical}
	  \begin{aligned}
R_K=1+\frac{2C_{b_L\mu_L}^{SM}}{\textcolor{blue}{(C_{{b_Le_L}}^{SM})^2+(C_{{b_Le_R}}^{SM})^2}}\textcolor{red}{[Re(C_{{b_L\mu_L}}^{BSM})+Re(C_{{b_R\mu_L}}^{BSM})]}+&\frac{\textcolor{red}{[Re(C_{b_L\mu_L}^{BSM})+Re(C_{b_R\mu_L}^{BSM})]}^2}{\textcolor{blue}{(C_{b_Le_L}^{SM})^2+(C_{b_Le_R}^{SM})^2}}+ \\
\frac{2C_{b_L\mu_R}^{SM}}{\textcolor{blue}{(C_{b_Le_L}^{SM})^2+(C_{b_Le_R}^{SM})^2}}\textcolor{olive}{[Re(C_{b_L\mu_R}^{BSM})+Re(C_{b_R\mu_R}^{BSM})]}+&\frac{\textcolor{olive}{[Re(C_{b_L\mu_R}^{BSM})+Re(C_{b_R\mu_R}^{BSM})]}^2}{\textcolor{blue}{(C_{b_Le_L}^{SM})^2+(C_{b_Le_R}^{SM})^2}}+\\
\frac{{[Im(C_{b_L\mu_L}^{BSM})+Im(C_{b_R\mu_L}^{BSM})]}^2}{\textcolor{blue}{(C_{b_Le_L}^{SM})^2+(C_{b_Le_R}^{SM})^2}}+
&\frac{{[Im(C_{b_L\mu_R}^{BSM})+Im(C_{b_R\mu_R}^{BSM})]}^2}{\textcolor{blue}{(C_{b_Le_L}^{SM})^2+(C_{b_Le_R}^{SM})^2}}  ,
\end{aligned}
	 \eeq
	 where we have separated real and imaginary parts for BSM coefficients in order to gain further insight on which NP operators might be responsible for the observed discrepancies.

\vskip .2cm
    We can immediately make the following preliminary observations:
\begin{itemize}
    \item There is the $b_L$ into $b_R$ exchange symmetry in the BSM corrections. 
    
     \item Terms linear in the BSM coefficients  involve only their real parts and are the only ones that can reduce the SM value of $R_K$.   
    \item Contributions from the imaginary parts of the coefficients push  $R_K$ towards higher values than the SM one unless they cancel each other as follows 
    \begin{equation}Im(C_{b_L\mu_R}^{BSM})\simeq -Im(C_{b_R\mu_R }^{BSM}) \ , \quad 
        Im(C_{b_L\mu_L}^{BSM})\simeq -Im(C_{b_R\mu_L }^{BSM}) 
        \label{imagine}\ .\end{equation}
    
\end{itemize}

	 By inserting the numerical values for the SM contributions, equation (\ref{eq:RKanalytical}) specializes to
	 
\beq
    \begin{aligned}
         R_K=1+0.2314\textcolor{red}{[Re(C_{b_L\mu_L}^{BSM})+Re(C_{b_R\mu_L}^{BSM})]}-0.0049\textcolor{olive}{[Re(C_{b_L\mu_R}^{BSM})+Re(C_{b_R\mu_R}^{BSM})]}+\\
        0.0134\{\textcolor{red}{[Re(C_{b_L\mu_L}^{BSM})+Re(C_{b_R\mu_L}^{BSM})]}^2+{[Im(C_{b_L\mu_L}^{BSM})+Im(C_{b_R\mu_L}^{BSM})]}^2+ \\
        \textcolor{olive}{[Re(C_{b_L\mu_R}^{BSM})+Re(C_{b_R\mu_R}^{BSM})]}^2+{[Im(C_{b_L\mu_R}^{BSM})+Im(C_{b_R\mu_R}^{BSM})]}^2\}  \ .
    \end{aligned}
\eeq
We can now obtain some interesting constraints on the NP coefficients by first considering one coefficient at a time while setting the remaining to zero.  Requiring, for example, the final value of $R_K$ to be within one standard deviation from the experimental value in (\ref{eq:RKexpnew}) we deduce the following analytic constraint on the real parts of the left-handed muon coefficients
    \beq 
    \begin{aligned}
    \label{recbxmul}
    0.807\leq1+&0.2314\textcolor{black}{[Re(C_{b_X\mu_L}^{BSM})]}+0.0134\textcolor{black}{[Re(C_{b_X\mu_L}^{BSM})]}^2\leq 0.888\\
    \Longrightarrow \quad  &\textcolor{red}{-0.8786\leq {Re(C_{b_X\mu_L}^{BSM})}\leq -0.4983}
    \end{aligned}
    \eeq 
   where $X$ is either $L$ or $R$.  Because of the quadratic nature of the equation we typically find two solutions. The physically acceptable one is the one for which $|C_{b_X \mu_L}|$ is less or of the order unity for the effective approach to be sensible. Still insisting on the single non-vanishing coefficient approximation no other solution is possible either real or complex.  
   
   If we assume NP in the electron sector only, we obtain an expression for $1/R_K$ analogous to (\ref{eq:RKanalytical}). Within one standard deviation from the experimental value we deduce
    \beq 
    \begin{aligned}
    1.126\leq1+&0.2314\textcolor{black}{[Re(C_{b_Xe_L}^{BSM})]}+0.0134\textcolor{black}{[Re(C_{b_Xe_L}^{BSM})]}^2\leq 1.238\\
    \Longrightarrow \quad  &\textcolor{red}{0.5283\leq {Re(C_{b_Xe_L}^{BSM})}\leq 0.9736} \ .
    \end{aligned}
    \eeq
   We clearly experience a mirror like behaviour for the electron coefficient with respect to the muon one in trying to explain the SM deviation.  In these two extreme limits the values of the coefficients are not very small. In fact, it is the large value of $C^{\rm SM}_{b_L \ell_L} = 8.64$  that justifies the chiral-linear approximation
	\beq\label{eq:RKChiral} 
	{R_K \simeq 1 +2  \frac{ \Re C^{\rm BSM}_{b_{L+R} (\mu-e)_{L}}}{C^{\rm SM}_{b_L \mu_L}}} \ .
	\eeq
	The overall picture is consistent with earlier findings according to which the dominant effect stems from couplings to left-handed leptons.

	\subsection{\texorpdfstring{$R_{K^*}$} -  anatomy}\label{sec:AnatomyOfRKStar}\label{RK*an}
	The LHCb collaboration reported the following experimental result~\cite{LHCb:2017avl} for $R_{K^*}$ (see Figure \ref{fig:RKs1a} and Figure \ref{fig:RKs1b}) in two bins of di-lepton invariant mass
	\beq\label{eq:RK*LHCB}
	R_{K^*}  = \left\{ \begin{array}{ll}
		0.660^{+0.110}_{-0.070} \pm 0.024 & (2m_\mu)^2 < q^2 < 1.1 \GeV^2  \\[2mm]
		0.685^{+0.113}_{-0.069} \pm 0.047& 1.1  \GeV^2 < q^2 < 6   \GeV^2 \, .
	\end{array}
	\right.
	\eeq
	Moreover, the {\sc Belle} collaboration presented preliminary 
	results about $R_{K^*}$.  Averaging over $B^0$ and $B^+$, 
	the {\sc Belle} result is~\cite{Belle:2019oag}
	\beq R_{K^*}[0.045,1.1]=0.52^{+0.36}_{-0.26}\pm0.05,\qquad
	R_{K^*}[1.1,6]=0.96^{+0.45}_{-0.29}\pm0.11.\eeq
	These values have to be compared with the SM predictions \cite{1605.07633}
	\beq
	R^{SM}_{K^*}  = \left\{\begin{array}{ll}
		0.906 \pm 0.028 & (2m_\mu)^2 < q^2 < 1.1 \GeV^2  \\[2mm]
		1.00 \pm 0.01 & 1.1  \GeV^2 < q^2 < 6   \GeV^2 \, .
	\end{array}
	\right.
	\eeq
	Given that the $K^*$ hadron has spin 1 and mass $M_{K^*} = 892\MeV$, the theoretical prediction for the $R_{K^*}$ ratio given in eq.(\ref{eq:RKRK*}) is
	\beq\label{eq:RK*}
	R_{K^*}
	= \frac{(1-p) (|C_{b_{L+R} \mu_{L-R}}|^2  + |C_{b_{L+R} \mu_{L+R}}|^2 ) + p \left( |C_{b_{L-R} \mu_{L-R}}|^2  + |C_{b_{L-R} \mu_{L+R}}|^2 \right) }
	{(1-p)( |C_{b_{L+R} e_{L-R}}|^2  + |C_{b_{L+R} e_{L+R}}|^2 ) + p \left( |C_{b_{L-R}e_{L-R}}|^2  + |C_{b_{L-R}e_{L+R}}|^2 \right)}
	\eeq
	where $p\approx 0.86$ is the ``polarization fraction"~\cite{0805.2525,1308.4379,1411.4773}, that is defined as
	\begin{equation}
		p =\frac{g_0 +g_{\parallel}}{g_0 +g_{\parallel}+g_{\perp}} \, .
	\end{equation}
	The $g_i$ are the contributions to the decay rate (integrated over the intermediate bin) of the different helicities of the $K^*$. The index $i$ distinguishes the various helicities: longitudinal ($i=0$), parallel ($i=\parallel$) and perpendicular ($i=\perp$).
	
	Trying to address simultaneously $R_K$ and $R_{K^*}$ here we consider $R_{K^*}-R_K$ for which the analytic expression under the assumption that NP hides in the muon sector reads:
    \beq
    \begin{aligned}
 \label{eq:RK*-RK}
   \frac{ R_{K^*}-R_K}{4p}=-\frac{C_{b_L\mu_L}^{SM}}{\textcolor{blue}{(C_{b_Le_L}^{SM})^2+(C_{b_Le_R}^{SM})^2}}\textcolor{red}{Re(C_{b_R\mu_L}^{BSM})}-&\frac{\textcolor{red}{Re(C_{b_L\mu_L}^{BSM})Re(C_{b_R\mu_L}^{BSM})}}{\textcolor{blue}{(C_{b_Le_L}^{SM})^2+(C_{b_Le_R}^{SM})^2}}+\\ ~~\\
    -\frac{C_{b_L\mu_R}^{SM}}{\textcolor{blue}{(C_{b_Le_L}^{SM})^2+(C_{b_Le_R}^{SM})^2}}\textcolor{olive}{Re(C_{b_R\mu_R}^{BSM})}-&\frac{\textcolor{olive}{Re(C_{b_L\mu_R}^{BSM})Re(C_{b_R\mu  _R}^{BSM})}}{\textcolor{blue}{(C_{b_Le_L}^{SM})^2+(C_{b_Le_R}^{SM})^2}}+\\~~ \\
    -\frac{Im(C_{b_L\mu_L}^{BSM})Im(C_{b_R\mu_L}^{BSM})}{\textcolor{blue}{(C_{b_Le_L}^{SM})^2+(C_{b_Le_R}^{SM})^2}}-&\frac{Im(C_{b_L\mu_R}^{BSM})Im(C_{b_R\mu_R}^{BSM})}{\textcolor{blue}{(C_{b_Le_L}^{SM})^2+(C_{b_Le_R}^{SM})^2}} \ .
    \end{aligned}
    \eeq
     Given that the left-hand side is negative we have that:
    \begin{itemize}
        \item For the terms linear  in  $Re(C^{BSM}_{b_R\mu_X})$ the product $C_{b_L\mu_X}^{SM}Re(C^{BSM}_{b_R\mu_X})$ should be positive if all the other coefficients are set to zero. This is, however, in conflict with considerations made in the previous section for $R_K$. 
        
        \item To resolve the conflict above, one more coefficient must be turned on. The minimal set of non-vanishing coefficients are $Re(C_{b_R\mu_L}^{BSM})$ and $Re(C_{b_L\mu_L}^{BSM})$. One can also imagine to turn on $Re(C_{b_R\mu_R}^{BSM})$, but as we shall soon see this contribution is highly suppressed via the SM coefficients. Additionally, note also that now we can determine the chirality of the $b$ quark while it is not possible by the solely analysis on $R_K$, as seen in the previous section.

        \item Interestingly we also observe that the imaginary parts must be suppressed in absolute value since the  $R_K$ solution in \eqref{imagine} would provide positive corrections here.
    \end{itemize}
Substituting the SM values we have 
\beq \label{KK*difference}
\begin{aligned}
    R_{K^*}-R_K=&-0.3981\textcolor{red}{Re(C_{b_R\mu_L}^{BSM})}+0.0084\textcolor{olive}{Re(C_{b_R\mu_R}^{BSM})}\\
    &-0.0461[\textcolor{red}{Re(C_{b_L\mu_L}^{BSM})Re(C_{b_R\mu_L}^{BSM})}+Im(C_{b_L\mu_L}^{BSM})Im(C_{b_R\mu_L}^{BSM})+\\
    &+\textcolor{olive}{Re(C_{b_L\mu_R}^{BSM})Re(C_{b_R\mu_R}^{BSM})}+Im(C_{b_L\mu_R}^{BSM})Im(C_{b_R\mu_R}^{BSM})] \ .
\end{aligned}
\eeq
A clear SM induced hierarchy emerges, according to which we can neglect the NP contribution associated to $Re(C_{b_R\mu_R}^{BSM})$ as it will also be clear from the final numerical analysis. 

Deviations coming from the electron sector can be studied in a fashion similar to the muon case using $1/R_{K^*}-1/R_K$, which will be of the same form as (\ref{KK*difference}), but with electronic BSM coefficients instead of muonic ones. 

All in all, only real coefficients involving left-handed leptons can be used to agree with data within a one sigma range with the current analytic considerations.

It is instructive, following \cite{DAmico:2017mtc}, to
visualize the constraints discussed above via the parametric
plots in  Figure \ref{pIMRKRK*} and Figure \ref{fig:RKRKStar}. The left panel of Figure \ref{pIMRKRK*} shows the parametric 
curve in the \linebreak 
$[R_K(\textcolor{violet}{Im(C_{b_X\mu_Y}^{BSM})})$,  
$R_{K^*}(\textcolor{violet}{Im(C_{b_X\mu_Y}^{BSM})})]$ plane 
obtained by turning on the imaginary part of one muon 
coefficient. From equations 
\eqref{eq:RKanalytical} and \eqref{eq:RK*-RK} it is clear that such 
a curve is the same for any $b$ and muon chirality. For 
increasing values of 
$\abs{\textcolor{violet}{Im(C_{b_X\mu_Y}^{BSM})}}$, the dots 
move along the diagonal in the $R_{K^*}$ vs $R_K$ plane, away 
from the experimental point~\cite{LHCb:2021trn,LHCb:2017avl
}, reported with its error bars. In the right panel, 
increasing values of 
$\abs{\textcolor{violet}{Im(C_{b_Xe_Y}^{BSM})}}$ point in the 
desired direction, towards the experimental values. However, 
in order to explain the data, we would require an imaginary 
coefficient of order $\sim 5$, which is  too big in the 
context of the effective approximation, where we expect 
Wilson coefficients' deviations to be, in absolute value, 
of order unity.  
	
	In Figure \ref{fig:RKRKStar} we see the analogous plots corresponding to the real parts of each coefficient associated to NP in the muon (left panel) and electron (right panel) sector. The  left-handed real part of the operator $C^{\rm BSM}_{b_L\mu_L}$ (purple lines) can reduce both $R_K$ and $R_{K^{\ast}}$.  The arrows correspond to $Re(C^{\rm BSM}_{b_L\mu_L}) = \pm 1$ (see caption for details).
	Conversely, as previously mentioned, a deviation of $R_{K^*}$ from $R_K$ signals the presence of $Re(C^{\rm BSM}_{b_R\mu_L})$ (red line).
	Finally, as observed above,  the reduced value of $R_K$ measured in equation \eqref{eq:RKexpnew} cannot be explained by $Re(C^{\rm BSM}_{b_R\mu_R})$ and $Re(C^{\rm BSM}_{b_L\mu_R})$.

	It is also instructive to summarise the case in which NP directly affects the electron sector, as can be appreciated in the right panel of Figure \ref{fig:RKRKStar}, which is a mirror-like image of the muon case.

	As discussed above, the  $Re(C^{\rm BSM}_{b_L e_L})$ can explain the reduced value of $R_K$ but alone cannot explain also $R_{K^{\ast}}$.  However, differences between $R_K$ and  $R_{K^*}$ can
	be achieved by a non-zero value of $Re(C^{\rm BSM}_{b_R e_L})$. Notice that also $Re(C^{\rm BSM}_{b_{L,R} e_R})$  points towards the observed experimental data but
	they require larger numerical values.

\begin{figure}
\centering
		\minipage{0.45\textwidth}
		\includegraphics[width=0.95\textwidth]{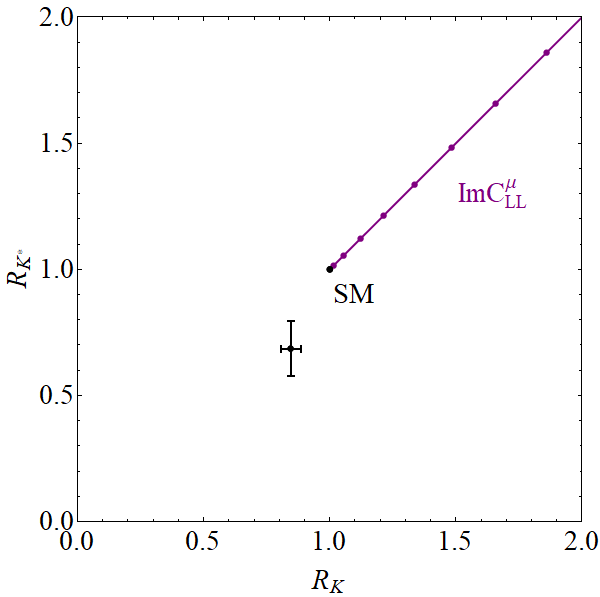}
		\endminipage 
		\minipage{0.45\textwidth}
		\includegraphics[width=0.95\textwidth]{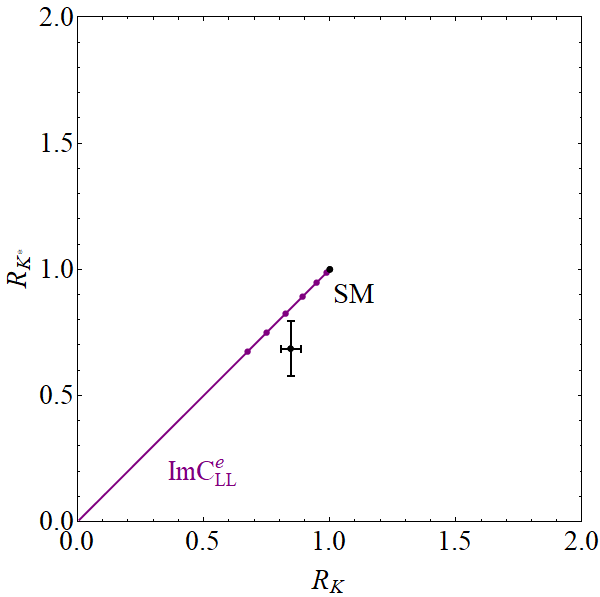}
		\endminipage
    \caption{\em We show the experimental point in the $R_{K^*}$ vs $R_K$ plane, along with purple parametric curves, obtained by switching on $Im(C_{b_L\mu_L})$ and $Im(C_{b_Le_L})$ (switching on $Im(C_{b_Xl_Y})$ produces the same result). For the muon case, increasing values of the parameter, point in the upper-right direction, moving away from the experimental measure. In the electronic case, increasing values of the parameter, point in the desired direction, but would imply a coefficient of order $\sim 5$, unphysical in the context of our approximation.}
    \label{pIMRKRK*}
\end{figure}

	\begin{figure}[t]
	\centering
		\minipage{0.45\textwidth}
		\includegraphics[width=0.95\textwidth]{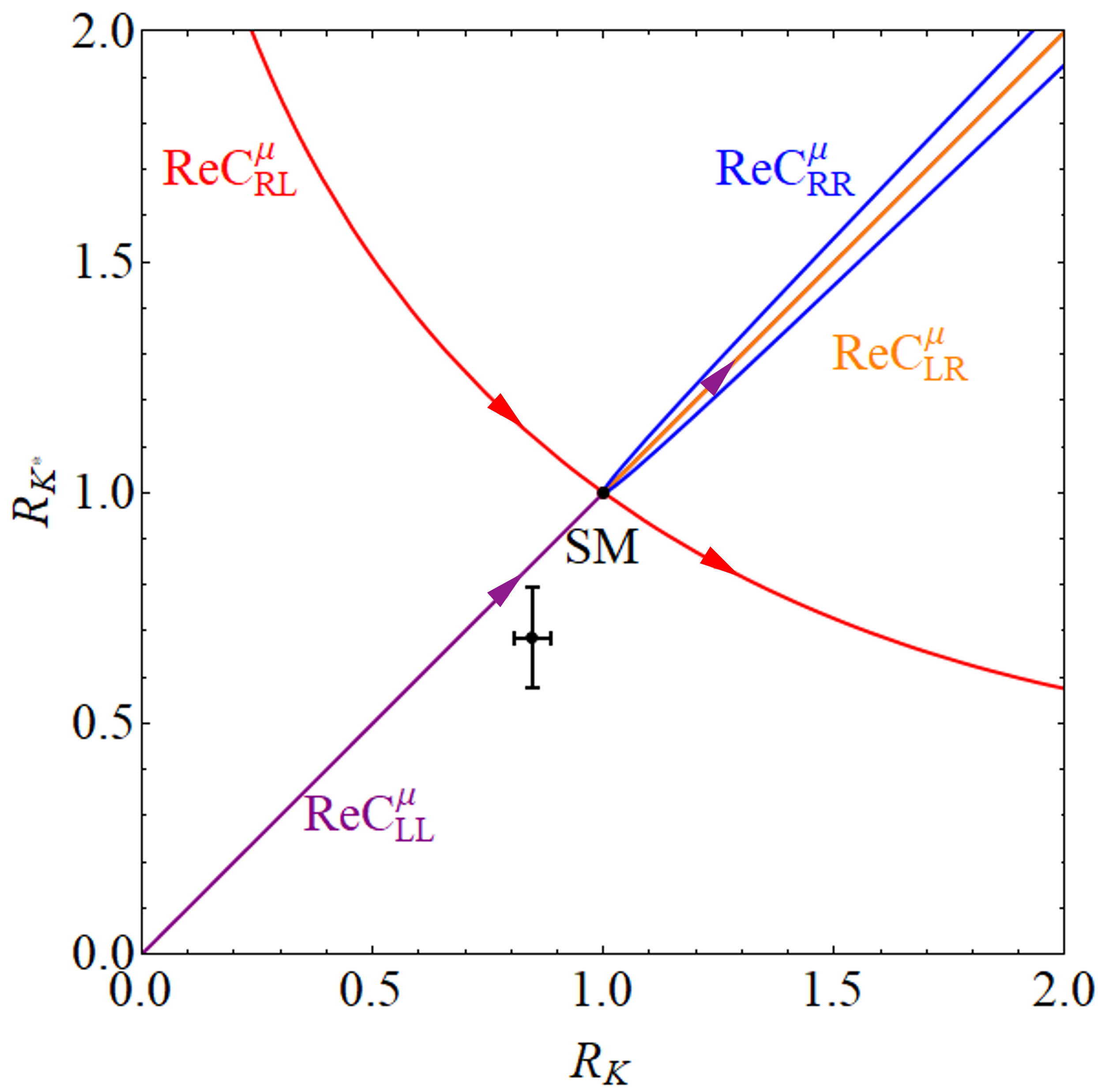}
		\endminipage 
		\minipage{0.45\textwidth}
		\includegraphics[width=0.95\textwidth]{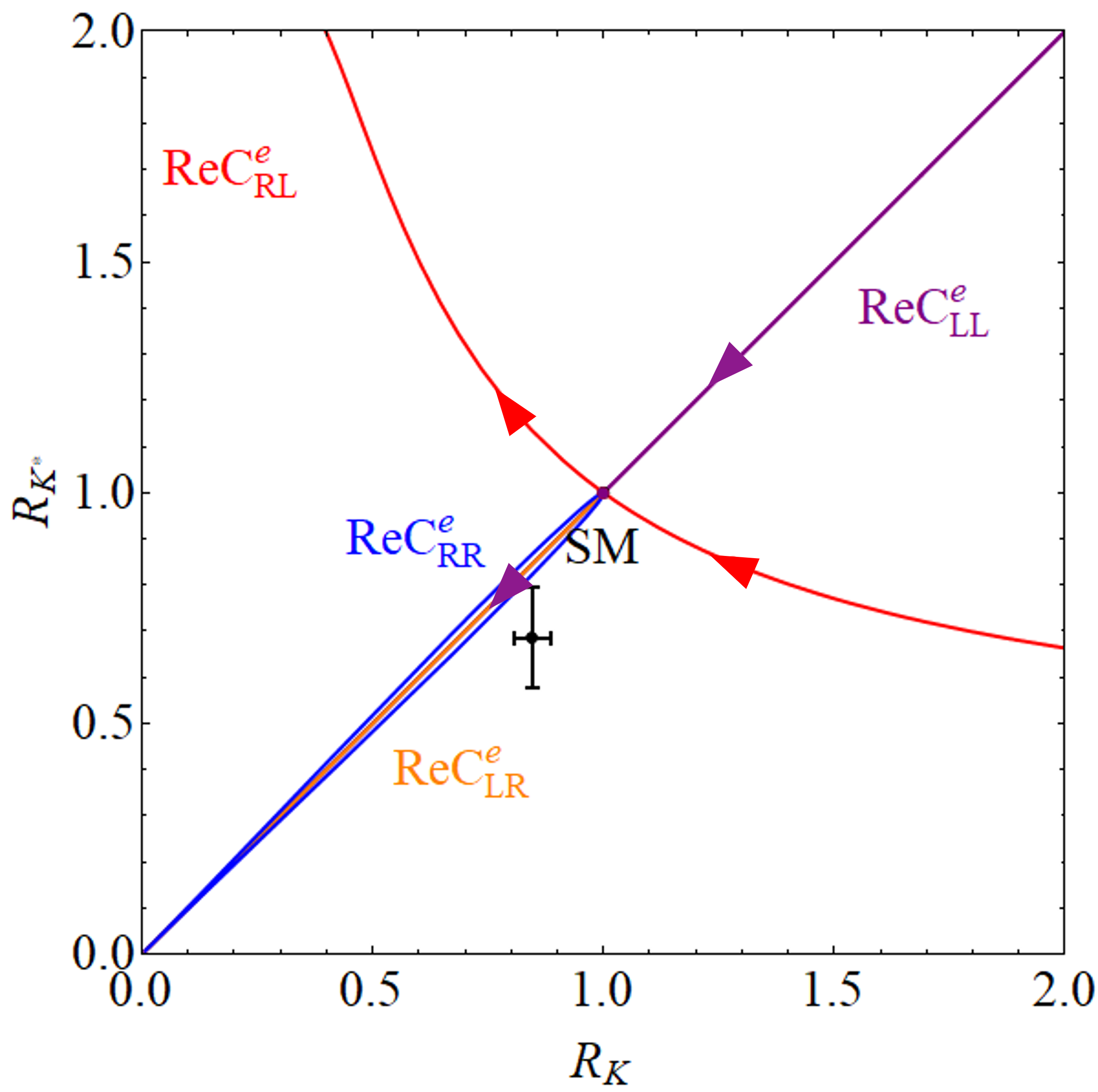}
		\endminipage
		\caption{\em Deviations from the SM value $R_K = R_{K^*} =1$ due to the various chiral operators possibly generated by new physics in the muon (left panel) and electron (right panel) sector. In this plot, we focus on the real parts of the coefficients. The arrows indicate the direction in which the coefficients grow.
			Both ratios refer to the $[1.1,6]$ bin.
			\label{fig:RKRKStar}}
	\end{figure}
	
	Since $R_{K^*}$ is provided for two different bins of di-lepton invariant mass we consider these two measures as independent in the numerical analysis (see \cite{DAmico:2017mtc} for further details).

	\subsection{\texorpdfstring{$B_s \to \mu^+ \mu^-$} - anatomy}
	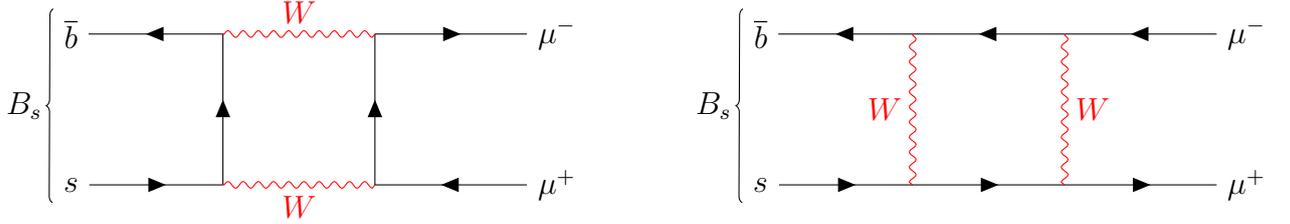
\begin{figure}[tbp] 
	    \centering
	    \begin{subfigure}{0.4\textwidth}
	    \centering
	    \begin{tikzpicture}
\begin{feynman}
\vertex (a1) {\(\overline b\)};
\vertex[right=2cm of a1] (a2);
\vertex[right=2cm of a2] (a3);
\vertex[right=2cm of a3] (a4) {\(\mu^-\)};
\vertex[below=2cm of a1] (b1) {\(s\)};
\vertex[right=2cm of b1] (b2);
\vertex[right=2cm of b2] (b3);
\vertex[right=2cm of b3] (b4) {\(\mu^+\)};
\diagram* {
{[edges=fermion]
(b1) -- (b2) -- (a2) -- (a1),
(b4) -- (b3) -- (a3) -- (a4),
},
(a2) -- [red, boson, edge label=\(W\)] (a3),
(b2) -- [red, boson, edge label'=\(W\)] (b3),
};
\draw [decoration={brace}, decorate] (b1.south west) -- (a1.north west)
node [pos=0.5, left] {\(B_s\)};
\end{feynman}
\end{tikzpicture}
\end{subfigure}
\hspace{2cm}
\begin{subfigure}{0.4\textwidth}
	    \centering
	    \begin{tikzpicture}
\begin{feynman}
\vertex (a1) {\(\overline b\)};
\vertex[right=2cm of a1] (a2);
\vertex[right=2cm of a2] (a3);
\vertex[right=2cm of a3] (a4) {\(\mu^-\)};
\vertex[below=2cm of a1] (b1) {\(s\)};
\vertex[right=2cm of b1] (b2);
\vertex[right=2cm of b2] (b3);
\vertex[right=2cm of b3] (b4) {\(\mu^+\)};
\diagram* {
{[edges=fermion]
(a4) -- (a3) -- (a2) -- (a1),
(b1) -- (b2) -- (b3) -- (b4),
},
(a2) -- [red, boson, edge label'=\(W\)] (b2),
(a3) -- [red, boson, edge label=\(W\)] (b3),
};
\draw [decoration={brace}, decorate] (b1.south west) -- (a1.north west)
node [pos=0.5, left] {\(B_s\)};
\end{feynman}
\end{tikzpicture}
	    \end{subfigure}

	    \caption{Feynman diagrams of $B_s\rightarrow\mu^+\mu^-$.}
	    \label{fig:bsmumu}
	\end{figure}
	To enrich our analysis, we now investigate the further interesting hadronic insensitive observable involving $C_{b_X \mu_Y}$: $\textrm{BR}(B_s \to \mu^+ \mu^-)$ (see Figure \ref{fig:bsmumu}). The rate is theoretically predicted to be
	\beq \label{eq:BSmumu} {\frac{\textrm{BR}(B_s \to \mu^+ \mu^-) }{\textrm{BR}(B_s \to \mu^+ \mu^-) _{\rm SM}} =
		\bigg|\frac{C_{b_{L-R}\mu_{L-R}}}{C^{\rm SM}_{b_{L-R}\mu_{L-R}}}\bigg|^2} \ , \eeq
	where in the SM we have 
	$
	\textrm{BR}(B_s \to \mu^+ \mu^-)_{\textrm{SM}} = (3.65 \pm 0.23) \times 10^{-9} $ while the experimental value is found to be
	$\textrm{BR}(B_s \to \mu^+ \mu^-)_{\textrm{exp}} = (3.09^{+0.46+0.15}_{-0.43-0.11}) \times 10^{-9}$. The latter has been measured by the LHCb collaboration~\cite{LHCb:2021awg}.
Following the previous analyses for $R_K$ and $R_{K^*}$, we  expand (\ref{eq:BSmumu}) in terms of NP coefficients in the muon sector:
	\begin{equation}
	\label{eq:analyticalBs}
    	\begin{split} 
    	\frac{\textrm{BR}(B_s \to \mu^+ \mu^-) }{\textrm{BR}(B_s \to \mu^+ \mu^-) _{\rm SM}} &=1+\frac{2\textcolor{red}{[Re(C^{BSM}_{b_L\mu_L})+Re(C^{BSM}_{b_R\mu_R})-Re(C^{BSM}_{b_L\mu_R})- 
	Re(C^{BSM}_{b_R\mu_L})]}}{\textcolor{blue}{[Re(C^{SM}_{b_L\mu_L})-Re(C^{SM}_{b_L\mu_R})]}}+ \\
        	&+\frac{\textcolor{red}{[Re(C^{BSM}_{b_L\mu_L})+Re(C^{BSM}_{b_R\mu_R})-  Re(C^{BSM}_{b_L\mu_R})-Re(C^{BSM}_{b_R\mu_L})]}^2}{\textcolor{blue}{[Re(C^{SM}_{b_L\mu_L})-Re(C^{SM}_{b_L\mu_R})]}^2}+ \\
        	&+\frac{\textcolor{purple}{[Im(C^{BSM}_{b_L\mu_L})+Im(C^{BSM}_{b_R\mu_R})- Im(C^{BSM}_{b_L\mu_R})-Im(C^{BSM}_{b_R\mu_L})]}^2}{\textcolor{blue}{[Re(C^{SM}_{b_L\mu_L})-Re(C^{SM}_{b_L\mu_R})]}^2} \ . 
        		\end{split}
	\end{equation}
  Inserting the SM values for the coefficients we have:       
        	\beq
        	\label{eq:SManalyticalBs}
        		\begin{split} 
        \frac{\textrm{BR}(B_s \to \mu^+ \mu^-) }{\textrm{BR}(B_s \to \mu^+ \mu^-) _{\rm SM}} 	&=1+0.227\textcolor{red}{[Re(C^{BSM}_{b_L2mu_L})+Re(C^{BSM}_{b_R\mu_R})-Re(C^{BSM}_{b_L\mu_R})- Re(C^{BSM}_{b_R\mu_L})]}+\\
        &+0.013\textcolor{red}{[Re(C^{BSM}_{b_L\mu_L})+Re(C^{BSM}_{b_R\mu_R})-Re(C^{BSM}_{b_L\mu_R})- Re(C^{BSM}_{b_R\mu_L})]}^2+ \\
        &+0.013\textcolor{purple}{[Im(C^{BSM}_{b_L\mu_L})+Im(C^{BSM}_{b_R\mu_R})- Im(C^{BSM}_{b_L\mu_R})-Im(C^{BSM}_{b_R\mu_L})]}^2 \ .
    	\end{split}
	\eeq
	We notice that the imaginary parts of the Wilson coefficients tend to shift this quantity away from the experimental value; once again the data privileges BSM Wilson coefficients to be real. Furthermore, equation (\ref{eq:analyticalBs}) is symmetric under the exchange $L\longleftrightarrow R$. We conclude this subsection by reporting the  relevant physical range for $Re(C^{BSM}_{b_X\mu_Y})$ at the one standard deviation level:
	\beq \begin{aligned}
	    \textcolor{red}{-1\leq Re(C^{BSM}_{b_L\mu_L}),Re(C^{BSM}_{b_R\mu_R})\leq 0.166}  \ ,\qquad \textcolor{red}{-0.166\leq Re(C^{BSM}_{b_L\mu_R}),Re(C^{BSM}_{b_R\mu_L})\leq 1} \ .
	\end{aligned}
	\eeq
	 In particular the constraints for $Re(C^{BSM}_{b_L\mu_L})$ and $Re(C^{BSM}_{b_R\mu_L})$ are compatible with the ones given by $R_K$ and $R_{K^*}$ respectively.

We are now ready to summarise the analytical investigation of the Hadronic Insensitive observables. 
	  
	\subsection{Summary of theoretical findings}
	
From the pedagogical theoretical investigations above we have extracted a number of relevant information that we summarise below. 	 
	
	\begin{enumerate}
			
			\item NP is needed to account for the deficit in $R_K$ and $R_{K^{\ast}}$ with respect to the SM value. 
			
		\item  If NP appears in either the muon or electron sector it privileges real Wilson coefficients. 
		
		\item In agreement with the results of \cite{DAmico:2017mtc} one can explain the experimental result for $R_K$ with NP in the muon sector by turning on the real part of $C^{\rm BSM}_{b_L \mu_L} \sim - 1$. However, because of the simultaneous deviation of $R_{K^*}$ from the SM and from $R_K$ itself, another coefficient is needed which is the real part of $C^{\rm BSM}_{b_R \mu_L}$.
		
		\item To the extent that one focuses only on the HI observables, one can explain the deviations also with NP in the electron sector with order unity values for the real parts of $C^{\rm BSM}_{b_L e_L}$ and $C^{\rm BSM}_{b_R e_L}$. 
		
	\item A more precise measure of $B_s \to \mu^+ \mu^-$  will select either the muon or the electron solution within the HI observables. As we shall see in the following sections, HS (angular observables) results suggest NP in the muon sector. 
			
	\end{enumerate}

The time is ripe to provide a numerical analysis that takes into account both the HS and HI observables.

	\section{Comparison with experimental data}\label{fit}

In the previous section we have assessed the main properties of the `Hadronic Insensitive' observables in order to guide our expectations when comparing with the bulk of experimental results.  

Here we use the  {\sc Flavio}~\cite{Straub:2018kue} toolkit to compare theory and experiments. This as an open source code aimed for flavour physics analyses. In the basic {\sc Flavio} package NP contributions are encoded in deviations from the SM Wilson coefficients introduced in section \ref{sec:TheorPred}.

The statistical method employed is that of the `Fast Likelihood' ~\cite{Straub:2018kue}, which is based on the following form of the likelihood  function
\begin{equation}
  -2\ln L(\mathbf{C})=\sum_i \mathbf{x}_i^T(\mathbf{C})[C_{\text{exp}}+C_{\text{th}}(\mathbf{\theta}_0)]^{-1}\mathbf{x}_i(\mathbf{C}) \qquad \mathbf{x}_i(\mathbf{C})=\mathbf{O}_i^{\text{exp}}-\mathbf{O}_i^{\text{th}}(\mathbf{C},\mathbf{\theta}_0)
\end{equation}
where

    \begin{enumerate}
    \item$L(\mathbf{C})$ is the likelihood function.
       \item $\mathbf{C}$ is the set of Wilson Coefficients.
        \item $\mathbf{\theta}_0$ is the set of the SM parameters' central values.
        \item $\mathbf{O}^{\text{th}}$ are the theory predictions for the observables.
        \item $\mathbf{O}^{\text{exp}}$ are the experimental measurements for the observables. 
        
        \item $C_{\text{exp}}$ is the covariance matrix of the experimental measurements.
        \item $C_{\text{th}}$ is the covariance matrix of theory predictions in the SM for the central values $\mathbf{\theta_0}$ of the theory parameters. 
    \end{enumerate}

In particular, $C_{\text{th}}$ is obtained from randomly sampling the observables in the theory parameter space. Finally, $C_{\text{exp}}$ is obtained via a Gaussian approximation of the experimental Probability Distribution Functions (PDFs).

Following the overall strategy first outlined in \cite{DAmico:2017mtc},  we divide the flavour sensitive experimental data in two, hadronic insensitive and hadronic sensitive. The HI are $R_K$, $R_{K^*}$ and $\BR(B_s\to\mu^+\mu^-)$.  The HS list of variables counts 257 observables (see Appendix~\ref{appendixA}). 	We report in Table \ref{tab:br_obs} the differential branching ratios of semi-leptonic $B$-meson decays  and in Table \ref{tab:ang_obs} the physical quantities extracted by the angular analysis of $B$-mesons decay products. Specifically, observables deviating more than $2.5 \sigma$ from the SM prediction are highlighted in blue, while those deviating more than $3.5 \sigma$ are highlighted in red.  Some of the observables, such as $S_5$ and $P_5^\prime$, provide similar information as discussed in \cite{Matias:2012xw, Descotes-Genon:2013vna} and therefore are expected to be correlated. To compensate for the lack of knowledge about the experimental correlation we have redone the fit switching off $S_5$ and find identical results. Furthermore, if we consider only one set of observables, $S_i$ rather than $P_i$ or vice-versa, we find results compatible with each other within one sigma.
		
	In the first step of our analysis we use only  the HI set of observables while in the second and third step we estimate the effects of the HS observables and combine them all in a global fit. 
	Finally, we propose an approach to obtain some insight on possible long distance hadronic effects based on the available data.

	\subsection{Step one: The hadronic insensitive  story}
	\label{fit:clean}
      HI observables are affected by small uncertainties in the SM, allowing to extract solid information about possible existence of NP. Due to the limited amount of data we set to zero all the Wilson coefficients but one. The chosen operator violates LFU since it is either for NP in the muon or the electron sector. Following the analytic analysis we can focus on real coefficients. 
	In HI columns of Table \ref{tab:Fit2021}  we present the central values of the Wilson coefficient, their $1$-$\sigma$ intervals as well as a measure of the deviation from the SM in terms of  $\sqrt{\Delta\chi^2} \equiv \sqrt{\chi^2_{\rm SM} - \chi^{2}_{\rm best}}$. We further highlight, in green, the rows corresponding to a significance of at least $4 \sigma$. The top part of the table corresponds to the muon fit while the bottom part to the electron fit.  The results are in agreement with the analytic investigation provided in   section \ref{sec:AnatomyOfRKStar}; the left-handed coefficient $C^{\rm BSM}_{b_L \mu_L}$  is favoured by the measured anomalies in $R_K$ and $R_{K^*}$, with a significance of $4.7 \sigma$.
The electron sector fit equally favours $C^{\rm BSM}_{b_L e_L}$, $C^{\rm BSM}_{b_L e_R}$ and $C^{\rm BSM}_{b_R e_{R}}$.
	However, except for two quite large coefficients  (with similar significance) the most reasonable is ${C}_{b_L e_L}$ which deviates less from the SM.   As for the muon case, the statistical preference with respect to the SM case is in the $4$-$5$ $\sigma$ range. For completeness, we show in Table \ref{tab:Fit2021axial} the results of $1$-parameter fits in the vector-axial basis.

	In conclusion, (in agreement with previous results \cite{DAmico:2017mtc,Ghosh:2014awa,Altmannshofer:2014rta,Descotes-Genon:2015uva,Altmannshofer:2013foa,Hurth:2013ssa,Hiller:2014yaa,Alonso:2014csa,Hurth:2014vma,Ciuchini:2017mik,Capdevila:2017bsm,Altmannshofer:2017fio,Alguero:2019ptt,Aebischer:2019mlg,Ciuchini:2020gvn})  by restricting the analysis to the selected subset of HI observables $R_K$, $R_{K^*}$ and ${\rm BR}(B_s \to \mu^+\mu^-)$, we find a  preference for a new neutral current that couples left-handed $b$, $s$ quarks and left-handed muons/electrons.

	\begin{table}[pt]
		\begin{center}
			\begin{tabular}{crrrrrrrrrc}
                \toprule
				& \multicolumn{9}{c}{\textbf{New physics in the muon sector (Chiral basis)}} &    \\
				\cmidrule{2-10}
				& \multicolumn{3}{c}{Best-fit} & \multicolumn{3}{c}{1-$\sigma$ range} &
				\multicolumn{3}{c}{$\sqrt{\chi^2_{\rm SM} - \chi^2_{\rm best}}$}   &\\ \cmidrule{2-10}
				& \emph{HI}  & \emph{HS} & \emph{all}  & \emph{HI}  & \emph{HS} & \emph{all} & \emph{HI} & \emph{HS} & \emph{all} &   \\ \cmidrule{2-10}
				\multirow{2}{*}{$C_{b_L \mu_L}^{\rm BSM}$} &  \multirow{2}{*}{$-0.75$} & \multirow{2}{*}{$-1.31$} &  \multirow{2}{*}{$-0.89$}  & $-0.64$ & $-1.05$  & $-0.78$ & \multirow{2}{*}{$4.7$} &  \multirow{2}{*}{$4.1$} & \multirow{2}{*}{$6.1$}  & \multirow{2}{*}{\greenmedal}\\
				& & & & $-0.86$ & $-1.56$ & $-0.99$ & & & &\\  \cmidrule{2-10}
				\multirow{2}{*}{$C_{b_L \mu_R}^{\rm BSM}$} &  \multirow{2}{*}{$0.78$} & \multirow{2}{*}{$-0.66$} &  \multirow{2}{*}{$-0.08$}  & $1.02$ & $-0.47$  & $0.09$ & \multirow{2}{*}{$2.3$} &  \multirow{2}{*}{$2.6$} & \multirow{2}{*}{$0.3$} & \\
				& & & & $0.54$ & $-0.85$ & $-0.24$ & & && \\  \cmidrule{2-10}
				\multirow{2}{*}{$C_{b_R \mu_L}^{\rm BSM}$} &  \multirow{2}{*}{$-0.16$} & \multirow{2}{*}{$0.08$} &  \multirow{2}{*}{$-0.09$}  & $-0.06$ & $0.19$  & $0.01$ & \multirow{2}{*}{$1.1$} &  \multirow{2}{*}{$0.5$} & \multirow{2}{*}{$0.8$}  & \\
				& & & & $-0.26$ & $-0.03$ & $-0.16$ & && & \\  \cmidrule{2-10}
				\multirow{2}{*}{$C_{b_R \mu_R}^{\rm BSM}$} &  \multirow{2}{*}{$-0.45$} & \multirow{2}{*}{$0.30$} &  \multirow{2}{*}{$-0.01$}  & $-0.21$ & $0.52$  & $0.14$ & \multirow{2}{*}{$1.3$} &  \multirow{2}{*}{$1.6$} & \multirow{2}{*}{$0.1$} &  \\
				& & & & $-0.97$ & $0.18$ & $0.04$ & && & \\ 
				\toprule
				& \multicolumn{9}{c}{\textbf{New physics in the electron sector (Chiral basis)}} &    \\
				\cmidrule{2-10}
				& \multicolumn{3}{c}{Best-fit} & \multicolumn{3}{c}{1-$\sigma$ range} &
				\multicolumn{3}{c}{$\sqrt{\chi^2_{\rm SM} - \chi^2_{\rm best}}$}   &\\ \cmidrule{2-10}
				& \emph{HI}  & \emph{HS} & \emph{all}  & \emph{HI}  & \emph{HS} & \emph{all} & \emph{HI} & \emph{HS} & \emph{all} &   \\ \cmidrule{2-10}
				\multirow{2}{*}{$C_{b_L e_L}^{\rm BSM}$} &  \multirow{2}{*}{$0.82$} & \multirow{2}{*}{$0.94$} &  \multirow{2}{*}{$0.84$}  & $0.97$ & $1.45$  & $0.99$ & \multirow{2}{*}{$4.4$} &  \multirow{2}{*}{$1.2$} & \multirow{2}{*}{$4.6$}  & \multirow{2}{*}{\greenmedal}\\
				& & & & $0.67$ & $0.43$ & $0.79$ & & & &\\  \cmidrule{2-10}
				\multirow{2}{*}{$C_{b_L e_R}^{\rm BSM}$} &  \multirow{2}{*}{$-3.22$} & \multirow{2}{*}{$-2.71$} &  \multirow{2}{*}{$-3.18$}  & $0.11$ & $-1.03$  & $-0.59$ & \multirow{2}{*}{$4.7$} &  \multirow{2}{*}{$1.3$} & \multirow{2}{*}{$4.9$} & \multirow{2}{*}{\greenmedal}\\
				& & & & $-2.88$ & $-1.73$ & $-2.87$ & & && \\  \cmidrule{2-10}
				\multirow{2}{*}{$C_{b_R e_L}^{\rm BSM}$} &  \multirow{2}{*}{$0.29$} & \multirow{2}{*}{$-3.87$} &  \multirow{2}{*}{$0.31$}  & $0.41$ & $-2.86$  & $0.41$ & \multirow{2}{*}{$1.8$} &  \multirow{2}{*}{$1.4$} & \multirow{2}{*}{$2.0$}  & \\
				& & & & $0.18$ & $-4.88$ & $0.21$ & && & \\  \cmidrule{2-10}
				\multirow{2}{*}{$C_{b_R e_R}^{\rm BSM}$} &  \multirow{2}{*}{$-3.56$} & \multirow{2}{*}{$-3.94$} &  \multirow{2}{*}{$-3.63$}  & $-3.20$ & $-2.92$  & $-3.28$ & \multirow{2}{*}{$4.2$} &  \multirow{2}{*}{$1.4$} & \multirow{2}{*}{$4.5$} &   \multirow{2}{*}{\greenmedal}\\
				& & & & $-3.92$ & $-4.96$ & $-3.98$ & && & \\  \bottomrule
			\end{tabular}
		\end{center}
		\caption{\em Best fits turning on a single operator at a time in the chiral basis, using the `hadronic insensitive' observables `HI',  the `hadronic sensitive' observables `HS', or all the observables `all'. Green dots correspond to a significance $>4 \sigma$ in the `hadronic insensitive' fit.
		The full list of observables can be found in Appendix~\ref{appendixA}.
			\label{tab:Fit2021}}
	\end{table}%
		\newpage
		
			\begin{table}[pt]
		\begin{center}
			\begin{tabular}{crrrrrrrrrc}
                \toprule
				& \multicolumn{9}{c}{\textbf{New physics in the muon sector (Vector Axial basis)}} &    \\
				\cmidrule{2-10}
				& \multicolumn{3}{c}{Best-fit} & \multicolumn{3}{c}{1-$\sigma$ range} &
				\multicolumn{3}{c}{$\sqrt{\chi^2_{\rm SM} - \chi^2_{\rm best}}$}   &\\ \cmidrule{2-10}
				& \emph{HI}  & \emph{HS} & \emph{all}  & \emph{HI}  & \emph{HS} & \emph{all} & \emph{HI} & \emph{HS} & \emph{all} &   \\ \cmidrule{2-10}
				\multirow{2}{*}{$C_{9,\mu}^{\rm BSM}$} &  \multirow{2}{*}{$-0.79$} & \multirow{2}{*}{$-0.94$} &  \multirow{2}{*}{$-0.91$}  & $-0.64$ & $-0.82$  & $-0.81$ & \multirow{2}{*}{$4.1$} &  \multirow{2}{*}{$4.9$} & \multirow{2}{*}{$6.5$}  & \multirow{2}{*}{\greenmedal}\\
				& & & & $-0.94$ & $-1.06$ & $-1.01$ & & & &\\  \cmidrule{2-10}
				\multirow{2}{*}{$C_{10,\mu}^{\rm BSM}$} &  \multirow{2}{*}{$0.63$} & \multirow{2}{*}{$-0.23$} &  \multirow{2}{*}{$0.51$}  & $0.73$ & $0.36$  & $0.59$ & \multirow{2}{*}{$4.7$} &  \multirow{2}{*}{$1.2$} & \multirow{2}{*}{$4.5$} & \multirow{2}{*}{\greenmedal}\\
				& & & & $0.53$ & $0.09$ & $0.43$ & & && \\  \cmidrule{2-10}
				\multirow{2}{*}{$C_{9,\mu}^{'\rm BSM}$} &  \multirow{2}{*}{$-0.30$} & \multirow{2}{*}{$0.20$} &  \multirow{2}{*}{$-0.11$}  & $-0.19$ & $0.33$  & $-0.02$ & \multirow{2}{*}{$1.8$} &  \multirow{2}{*}{$1.1$} & \multirow{2}{*}{$0.9$}  & \\
				& & & & $-0.41$ & $0.07$ & $-0.20$ & && & \\  \cmidrule{2-10}
				\multirow{2}{*}{$C_{10,\mu}^{'\rm BSM}$} &  \multirow{2}{*}{$0.05$} & \multirow{2}{*}{$0$} &  \multirow{2}{*}{$0.05$}  & $0.13$ & $0.08$  & $0.11$ & \multirow{2}{*}{$0.4$} &  \multirow{2}{*}{$0$} & \multirow{2}{*}{$0.6$} &  \\
				& & & & $-0.03$ & $-0.08$ & $-0.01$ & && & \\ 
				\toprule
				& \multicolumn{9}{c}{\textbf{New physics in the electron sector (Vector Axial basis)}} &    \\
				\cmidrule{2-10}
				& \multicolumn{3}{c}{Best-fit} & \multicolumn{3}{c}{1-$\sigma$ range} &
				\multicolumn{3}{c}{$\sqrt{\chi^2_{\rm SM} - \chi^2_{\rm best}}$}   &\\ \cmidrule{2-10}
				& \emph{HI}  & \emph{HS} & \emph{all}  & \emph{HI}  & \emph{HS} & \emph{all} & \emph{HI} & \emph{HS} & \emph{all} &   \\ \cmidrule{2-10}
				\multirow{2}{*}{$C_{9,e}^{\rm BSM}$} &  \multirow{2}{*}{$0.81$} & \multirow{2}{*}{$1.01$} &  \multirow{2}{*}{$0.84$}  & $0.96$ & $1.55$  & $0.88$ & \multirow{2}{*}{$4.3$} &  \multirow{2}{*}{$1.4$} & \multirow{2}{*}{$4.4$}  & \multirow{2}{*}{\greenmedal}\\
				& & & & $0.67$ & $0.47$ & $0.70$ & & & &\\  \cmidrule{2-10}
				\multirow{2}{*}{$C_{10,e}^{\rm BSM}$} &  \multirow{2}{*}{$-0.76$} & \multirow{2}{*}{$-0.79$} &  \multirow{2}{*}{$-0.77$}  & $-0.62$ & $-0.36$  & $-0.64$ & \multirow{2}{*}{$4.5$} &  \multirow{2}{*}{$1.2$} & \multirow{2}{*}{$4.7$} & \multirow{2}{*}{\greenmedal}\\
				& & & & $-0.89$ & $-1.21$ & $-0.90$ & & && \\  \cmidrule{2-10}
				\multirow{2}{*}{$C_{9,e}^{'\rm BSM}$} &  \multirow{2}{*}{$0.34$} & \multirow{2}{*}{$0.20$} &  \multirow{2}{*}{$0.36$}  & $0.46$ & $0.33$  & $0.49$ & \multirow{2}{*}{$2.0$} &  \multirow{2}{*}{$1.4$} & \multirow{2}{*}{$2.1$}  & \\
				& & & & $0.21$ & $0.07$ & $0.24$ & && & \\  \cmidrule{2-10}
				\multirow{2}{*}{$C_{10,e}^{'\rm BSM}$} &  \multirow{2}{*}{$-0.29$} & \multirow{2}{*}{} &  \multirow{2}{*}{$-0.31$}  & $-0.17$ &   & $-0.20$ & \multirow{2}{*}{$1.9$} &  \multirow{2}{*}{} & \multirow{2}{*}{$2.0$} &  \\
				& & & & $-0.40$ &  & $-0.42$ & && & \\  \bottomrule
			\end{tabular}
		\end{center}
		\caption{\em Best fits turning on a single operator at a time in the vector-axial basis, using the `hadronic insensitive' observables `HI',  the `hadronic sensitive' observables `HS', or all the observables `all'. Green Dots correspond to a significance $>4 \sigma$ in the `hadronic insensitive' fit.
		\label{tab:Fit2021axial}}
	\end{table}%

	\subsection{Step two: The {hadronic sensitive}   story}\label{sec:Dirty}

    Using the observables in the tables of Appendix~\ref{appendixA}, we can now perform fits featuring  multiple coefficients turned on, although we started with one coefficient at the time to check consistency with the earlier results.   

    Focusing on the green rows of Table~\ref{tab:Fit2021}, in the muon sector we still find a preference for $C_{b_L\mu_L}^{BSM}\sim-1$ with a slightly larger error. On the other hand, in the electron sector we witness strongly reduced values of the significance. In fact, most HS observables involve muons (detailed measurements are much more difficult with electrons), so that they are less sensitive to effects in the electron sector. 
	Since HS data still highlight significant deviations with respect to SM predictions in the muon sector, we perform a multiple coefficient fit involving all 4 muonic operators.
	Given the reduced sensitivity of our observables to electronic coefficients, the result remains essentially unchanged assuming non-zero deviations of the 4 electronic coefficients, as we have verified with the {\sc Flavio} code. 
	For this reason, even if we are focusing on the muon sector, the electronic coefficients cannot be constrained. Therefore we fit directly the sum $\Delta C_I=C_I^H+C_I^{\text{BSM}}$ for the muon coefficients with the results for their mean values, the associated chi-square $\tilde{\chi}^2$  compared with the SM  one $\chi^{2}_{SM}$ and associated covariance matrix $\rho$:
	\beq
	\begin{array}{rcl}
		\Delta C_ {b_L\mu _L} &=& -1.16 \pm 0.20,\\ 
		\Delta C_ {b_L\mu _R} &=& -0.79 \pm 0.18,\\ 
		\Delta C_ {b_R\mu _L} &=& 0.27 \pm 0.14,\\ 
		\Delta C_ {b_R\mu _R} &=& 0.06 \pm 0.26.\\ 
	\end{array} \quad \begin{aligned}
     &\chi^{2}_{SM}=258.05,\\
     &\tilde{\chi}^{2}=232.99.
     \end{aligned} \quad\rho = 
	\left(
\begin{array}{cccc}
 1 & -0.01 & -0.002 & 0.003 \\
 -0.01 & 1 & 0.02 & 0.07 \\
 -0.002 & 0.02 & 1 & 0.11 \\
 0.003 & 0.07 & 0.11 & 1 \\
\end{array}
\right)
	\label{HSmuon}
	.\eeq
We further report the number of degrees of freedom and reduced chi-square for both the SM and the four-coefficient fit:
\beq
\frac{\chi^{2}_{SM}}{\text{\# d.o.f.}}=\frac{258.05}{257}=1.004, \qquad \frac{\tilde{\chi}^{2}}{\text{\# d.o.f.}}=\frac{232.99}{253}=0.921
\eeq
The comparison between these two values shows a preference for non-zero deviations $\Delta C_I$.
In particular, the only coefficients that are significantly different from zero are those involving the left-handed $b$-quark. At this stage one could still imagine that these deviations could be attributed to   hardonic effects  rather than NP. In fact, one expects these effects to be strongly suppressed for the right handed $b$-quark. Both the HI and HS observables are therefore needed to disentangle NP contributions present in $\Delta C_I$.

	\subsection{Step three: NP only HI-HS observables interplay} \label{Global}
	We are now ready to combine HS and HI observables in a global fit. As a first step we neglect hadronic   contributions that will be, however, discussed in the next subsection.  This approach is consistent with having neglected the electronic contribution.  Here, we can immediately say that the results favour a deviation in the SM in $C^{\rm BSM}_{b_L\mu_L}$ as well as $C^{\rm BSM}_{b_L\mu_R}$. We can also further set bounds on the remaining muon NP coefficients. The result of the global fit is summarized as
		\beq 
	\begin{array}{rcl}
		C_ {b_L\mu _L}^{\text{BSM}} &=& -1.19 \pm 0.12,\\ 
		C_ {b_L\mu _R}^{\text{BSM}} &=& -0.78 \pm 0.16,\\ 
		C_ {b_R\mu _L}^{\text{BSM}} &=& 0.40 \pm 0.11,\\ 
		C_ {b_R\mu _R}^{\text{BSM}} &=& 0.15 \pm 0.24.\\ 
		\label{Globalmuon}
	\end{array}\quad \begin{aligned}
     &\chi^{2}_{SM}=289.10,\\
     &\tilde{\chi}^{2}=237.46.
     \end{aligned} \quad
	\rho = 
	\left(
\begin{array}{cccc}
 1 & 0.02 & -0.05 & -0.03 \\
 0.02 & 1 & 0.02 & 0.08 \\
 -0.05 & 0.02 & 1 & 0.11 \\
 -0.03 & 0.08 & 0.11 & 1 \\
\end{array}
\right)
	.\eeq
As in the previous case, we report the reduced chi-square:
\beq \label{eq:redglobal}
\frac{\chi^{2}_{SM}}{\text{\# d.o.f.}}=\frac{289.10}{269}=1.075, \qquad \frac{\tilde{\chi}^{2}}{\text{\# d.o.f.}}=\frac{237.46}{265}=0.896
\eeq

Here we restricted the analysis to a fit for the muon sector since there are not enough electron-sensitive data able to pin down all their   coefficients. Under this particular hypothesis, when comparing the $\chi^2$ values, we find a deviation from the Standard Model at the $7.2\sigma$ level. Here again the reduced chi-square values support the BSM scenario.
	 
	\subsection{Estimating hadronic  effects}
 It is  relevant to estimate the hadronic contributions to the Wilson coefficients in order to disentangle contributions of NP as advocated in \cite{Ciuchini:2019usw,Gubernari:2020eft,Ciuchini:2020gvn,Ciuchini:2021smi}.\\We parametrize the hadronic  effects with a shift of the Wilson coefficients $C_I^H$. Following \cite{Gubernari:2020eft,Ciuchini:2020gvn}  the hadronic contributions appear only in $C^{H}_{9}$ and therefore in the following analysis we assume $C^{H}_{10}=C^{'H}_{10}=C^{'H}_{9}=0$. We additionally assume that NP is present only in the muon sector. This working hypothesis can be tested once new observables sensitive to NP in the electronic sector will be available. {{Recalling the discussion below \eqref{nonfact}, given that our $C_9^{SM}$ is not defined to include hadronic contributions that eliminate its scale-dependence \cite{Gubernari:2020eft} our results will contain both these and long distance hadronic effects. These can be interpreted as LFU contributions from the SM not present in  {\sc Flavio}~\cite{Straub:2018kue}. }}
 
We set the muon Wilson coefficients $\Delta C_\mu$ to the values of Eq. (\ref{HSmuon}), obtained by fitting only the HS observables, which do not involve the electron Wilson coefficients. Keeping the muonic coefficients fixed, we fit $\Delta C_{9,e}=C_{9,e}^H$ using HI observables. As mentioned in Sec. \ref{sec:TheorPred},  hadronic   effects are expected to be LFU, so that $C_{9,e}^H=C_{9,\mu}^H=C_{9}^H$.

The NP contribution to the Wilson coefficients is:
\begin{equation}
    C_\mu^{\text{BSM}}=\Delta C_\mu-C^{H}_\mu.
    \label{CmuBSM}
\end{equation}

The hadronic insensitive observables give
$$
C^{H}_{9}=-0.16 \pm 0.15=C^{H}_{b_L\ell _L}=C^{H}_{b_L\ell _R} \ ,
$$
which is compatible with a small hadronic contribution. Although the overall impact is a slight decrease of the central value of the NP contributions, this is  within a one sigma effect. 

	\subsection{The \texorpdfstring{$R_K=R_{K^{*}}=1$} - case vs HS observables}
We find instructive to consider the impact of a potential disappearance of the LFU violations  on the overall fit to the HS and HI observables. We will still allow for uncertainties on $R_K=R_{K^{\ast}}$ of the order of the ones measured now, while changing the central value to unity. The results are reported below 
	\beq
	\begin{array}{rcl}
		C_ {b_L\mu _L}^{\text{BSM}} &=& -0.72 \pm 0.14,\\ 
		C_ {b_L\mu _R}^{\text{BSM}} &=& -0.87 \pm 0.07,\\ 
		C_ {b_R\mu _L}^{\text{BSM}} &=& 0.45 \pm 0.07,\\ 
		C_ {b_R\mu _R}^{\text{BSM}} &=& 0.06 \pm 0.14\\ 
		\label{GlobalmuonRKRKs1}
	\end{array}\quad \begin{aligned}
     &\chi^{2}_{SM}=279.03,\\
     &\tilde{\chi}^{2}=254.96.
     \end{aligned} \quad
	\rho = 
	\left(
\begin{array}{cccc}
1 	 &
-0.04 	 &
-0.12 	 &
-0.12 	 
 \\  
-0.04 	 &
1 	 &
-0.02 	 &
0.03 	 
 \\  
-0.12 	 &
-0.02 	 &
1 	 &
0.04 	 
 \\  
-0.12 	 &
0.03 	 &
0.04 	 &
1 	 
 \\ 
\end{array}
\right)
	.\eeq

The reduced chi-squared values are
\beq
\frac{\chi^{2}_{SM}}{\text{\# d.o.f.}}=\frac{279.03}{275}=1.015, \qquad \frac{\tilde{\chi}^{2}}{\text{\# d.o.f.}}=\frac{254.96}{271}=0.941
\eeq
It is interesting to compare this with the result obtained in step three. Even though the values for $R_K,R_{K^{\ast}}$ have been set to unity, by $\chi^2$ value comparison, we still find a deviation of $4.9\sigma$ from the SM. As expected we find lower values for both the full and reduced SM chi-square compared to the previous case (see \eqref{Globalmuon} and \eqref{eq:redglobal}) because  $R_K$ and $R_{K^\ast}$ equal to unity fit better with the SM. Nevertheless the HS observables still deviate from the SM predictions and the BSM hypothesis is still preferred.  


	\section{g-2 anomalies}
\label{g-2}
Besides the deviations observed in the flavor observables, tantalizing hints of NP are emerging, at the same time, in another relevant sector of the SM, the anomalous magnetic momenta of the muon $a_{\mu}$ and of the electron $a_{e}$.
The measurements of the anomalous magnetic moment of the charged leptons, $a_\ell$, represent long-standing precision tests of the SM. New physics at a scale $\Lambda$ is expected to affect $a_\ell$ as $\delta a_\ell \propto m^2_\ell/\Lambda^2$. Therefore, although the  largest BSM contributions should appear in the $\tau$ anomalous magnetic moment, its short lifetime renders its measurement difficult. This is the reason why one concentrates on measuring $a_\mu$ and $a_e$ as probes of NP. In particular, $a_\mu$ is measured very precisely, in magnetic storage ring experiments, by examining the precession of muons that are subjected to a magnetic field. In the electron case the $a_e$ experimental value is known at the 0.24 ppb level in one-electron quantum cyclotron experiment \cite{Hanneke:2008tm}.

 The value of the muon $(g-2)$  receives contributions from different sectors of the SM that can be summarized as: 
\begin{equation}
    a^{SM}_\mu= a^{QED}_\mu + a^{EW}_\mu + a^{HVP}_\mu+ a^{HLbL}_\mu \, .
\end{equation}
In particular, $a^{HVP}_\mu$  is the hadronic vacuum polarization contribution and $a^{HLbL}_\mu$ include contributions from hadronic light-by-light scattering.
The overall uncertainty on  $a^{SM}_\mu$ is dominated by $a^{HVP}_\mu$.
$a^{HVP}_\mu$ can be determined from lattice calculations but also by a data-driven approach, considering $e^+ e^- \to$ hadrons events. A data-driven alternative opportunity to examine the HVP contributions would be offered by the MUonE experiment \cite{CarloniCalame:2015obs}, which has been proposed to determine the leading hadronic corrections to $a_\mu$
 purely from experiments, by examining the scattering $\mu e \to \mu e$.\\
Recently, the Fermilab collaboration reported  their newer   measurement of the muon anomalous magnetic moment \cite{Abi:2021gix}, lending further support to the early discrepancy with respect to the SM value. Combining this result, with a deviation at the $3.3\sigma$,  with the earlier BNL E821 one \cite{Bennett:2006fi}, one arrives at  the experimental average of 
\begin{equation}
a_{\mu} ({\mathrm{Exp}}) = 116 \, 592 \, 061 (41) \times 10^{-11} \quad (0.35\,{\mathrm{ppm}}) \ ,
\end{equation} 
 with an overall deviation from the SM central value \cite{Aoyama:2020ynm} of
  \begin{equation}
\Delta a_{\mu} = a_{\mu} ({\mathrm{Exp}})  - a_{\mu} ({\mathrm{SM}}) = (251 \pm 59) \times 10^{-11} \ ,
\end{equation} 
corresponding to a significance of $4.2\sigma$ \cite{Abi:2021gix}.  Note however, that  the  lattice results from the BMW collaboration \cite{Borsanyi:2020mff} are not included in the world average. If these are considered as the true contribution to the HVP of the photon \cite{Davier:2010nc,Davier:2017zfy,Davier:2019can} the discrepancy from the SM reduces. {{Very recently an independent lattice simulation has appeared  \cite{Ce:2022kxy} agreeing with the BMW result.}}

The theoretical SM prediction for $a_e$ is sensitive to the
experimentally measured value of the fine-structure constant. A recent measurement of $\alpha$ \cite{Morel:2020dww} leads to a deviation of 1.7$\sigma$ from the SM in $a_e$:

  \begin{equation}
\Delta a_{e} = (0.48 \pm 0.30) \times 10^{-12} \ .
\end{equation}

	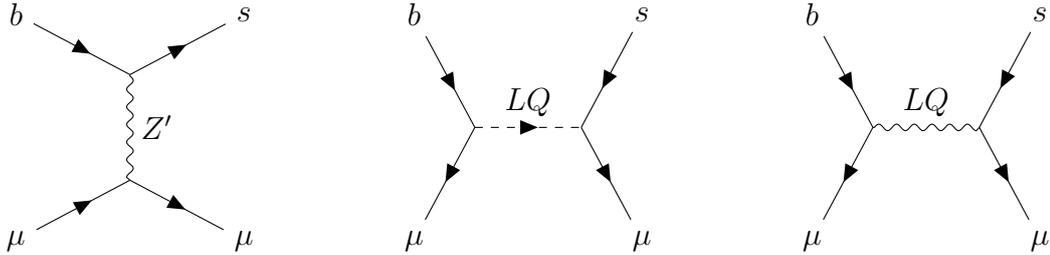
\begin{figure}[tbp]
	\centering
	\begin{subfigure}{0.3\textwidth}
	\centering
	\begin{tikzpicture}
\begin{feynman}
\vertex (a1) {\(b\)};
\vertex[right=3cm of a1] (a2) {\(s\)};
\vertex[below=3cm of a1] (b1) {\(\mu\)};
\vertex[below=3cm of a2] (b2) {\(\mu\)};
\vertex at ($(a1)!0.5!(a2) - (0, 0.8cm)$) (c1);
\vertex at ($(b1)!0.5!(b2) + (0, 0.8cm)$) (c2);
\diagram* {
(a1) -- [fermion,black] (c1) -- [fermion,black] (a2),
(b1) -- [fermion,black!50!black] (c2) -- [fermion,black!50!black] (b2),
(c1) -- [boson,black, edge label=\(Z'\)] (c2)
};
\end{feynman}
\end{tikzpicture}
	\end{subfigure}
	\begin{subfigure}{0.3\textwidth}
	\centering
	\begin{tikzpicture}
\begin{feynman}
\vertex (a1) {\(b\)};
\vertex[right=3cm of a1] (a2) {\(s\)};
\vertex[below=3cm of a1] (b1) {\(\mu\)};
\vertex[below=3cm of a2] (b2) {\(\mu\)};
\vertex at ($(a1)!0.5!(b1) + (0.8cm, 0)$) (c1);
\vertex at ($(a2)!0.5!(b2) - (0.8cm, 0)$) (c2);
\diagram* {
(a1) -- [fermion,black] (c1) -- [fermion,black!50!black] (b1),
(a2) -- [fermion,black] (c2) -- [fermion,black!50!black] (b2),
(c1) -- [charged scalar,black, edge label=\(LQ\)] (c2)
};
\end{feynman}
\end{tikzpicture}
	\end{subfigure}
	\begin{subfigure}{0.3\textwidth}
	\centering
	\begin{tikzpicture}
\begin{feynman}
\vertex (a1) {\(b\)};
\vertex[right=3cm of a1] (a2) {\(s\)};
\vertex[below=3cm of a1] (b1) {\(\mu\)};
\vertex[below=3cm of a2] (b2) {\(\mu\)};
\vertex at ($(a1)!0.5!(b1) + (0.8cm, 0)$) (c1);
\vertex at ($(a2)!0.5!(b2) - (0.8cm, 0)$) (c2);
\diagram* {
(a1) -- [fermion,black] (c1) -- [fermion,black!50!black] (b1),
(a2) -- [fermion,black] (c2) -- [fermion,black!50!black] (b2),
(c1) -- [boson,black, edge label=\(LQ\)] (c2)
};
\end{feynman}
\end{tikzpicture}
	\end{subfigure}
	\caption{\em Particles that can mediate $R_K$ at tree level: a $Z'$ or a lepto-quark, scalar or vector.
				\label{fig:FeynZLQ}}
	\end{figure}
	\bigskip
	
{ { In Ref.~\cite{Aebischer:2021uvt}	the authors performed an effective analysis of the magnetic and electric dipole moments of the muon and electron. There they gave the expressions for the dipole moments in terms of operator coefficients of the low-energy effective field theory and the SM effective field theory. Using one-loop renormalization group improved perturbation theory with one-loop matching conditions they showed that semileptonic four-fermion operators involving light quarks give sizable non-perturbative contributions to the dipole moments. They further showed  that  a very limited set of the SM effective operators can generate the observed deviation of the magnetic moment of the muon from its SM expectation, as we shall also appreciate from direct model constructions.}}

	\section{W-mass}
\label{CDFWmass}
Another anomaly appeared recently due to the CDF best precise measure, so far, of the  $W$ boson mass $M_W$ obtained using data relative to 8.8 inverse femtobarns (fb$^{-1}$) of integrated luminosity. These data were collected in proton-antiproton collisions at an energy in the center-of-mass of $1.96$~TeV, via the CDF II detector at the Fermilab Tevatron collider. Via a sample of about 4-million $W$ bosons, the collaboration reported the value 
\cite{CDF:2022hxs}
\begin{equation} 
    \left. M_W \right|_{\rm CDF} = 80,433.5 \pm 6.4_{\rm stat} \pm 6.9_{\rm syst}= 80,433.5 \pm 9.4\, {\rm MeV}  \ . 
\end{equation}
Surprisingly the central value is much larger than   the SM expectation \cite{ParticleDataGroup:2020ssz}:
\begin{equation} 
    \left. M_W \right|_{\rm SM} = 80,357 \pm 4_{\rm inputs} \pm 4_{\rm theory}  \, {\rm MeV}  \ . 
\end{equation}
 The custodial symmetry in the Higgs sector together with high-precision measurements including the Higgs and Z boson masses, the top-quark mass, the electromagnetic coupling, the muon lifetime and collider asymmetries are responsible for the SM value.  The latter is further affected by uncertainties in the input data and by limited knowledge of the higher-order perturbative computations from the theory side.  In the end the deviation of the CDF measurement from the SM is 7 standard deviations (7$\sigma$) 
 \cite{CDF:2022hxs}.

Shockingly the accuracy of the new CDF measurement exceeds that of all previous measurements combined, coming from the Large Electron Positron collider (LEP)  and previous Tevatron analyses \cite{ALEPH:2010aa}:
\begin{equation}
    \left. M_W \right|_{\rm LEP} = 80,385 \pm 15 \, {\rm MeV}  \ .
    \end{equation}
At the Large Hadron Collider (LHC), the LHCb  \cite{LHCb:2021bjt} and ATLAS collaborations \cite{ATLAS:2017rzl} reported: \begin{eqnarray}
   \left. M_W \right|_{\rm LHCb} &=& 80,354 \pm 32 \, {\rm MeV}\, , \nonumber \\
   \left. M_W \right|_{\rm ATLAS} &=& 80,370 \pm 19 \, {\rm MeV}  \ . 
    \end{eqnarray}
Following \cite{Cacciapaglia:2022xih} a weighted average of these 4 results yields: 
\begin{equation} \label{eq:AVG}
    \left. M_W \right|_{\rm AVG} = 80,409\pm  \, 17\,{\rm MeV} \ .
\end{equation}
\begin{figure}[ht]
	\centering
	\includegraphics[keepaspectratio,height=7 cm]{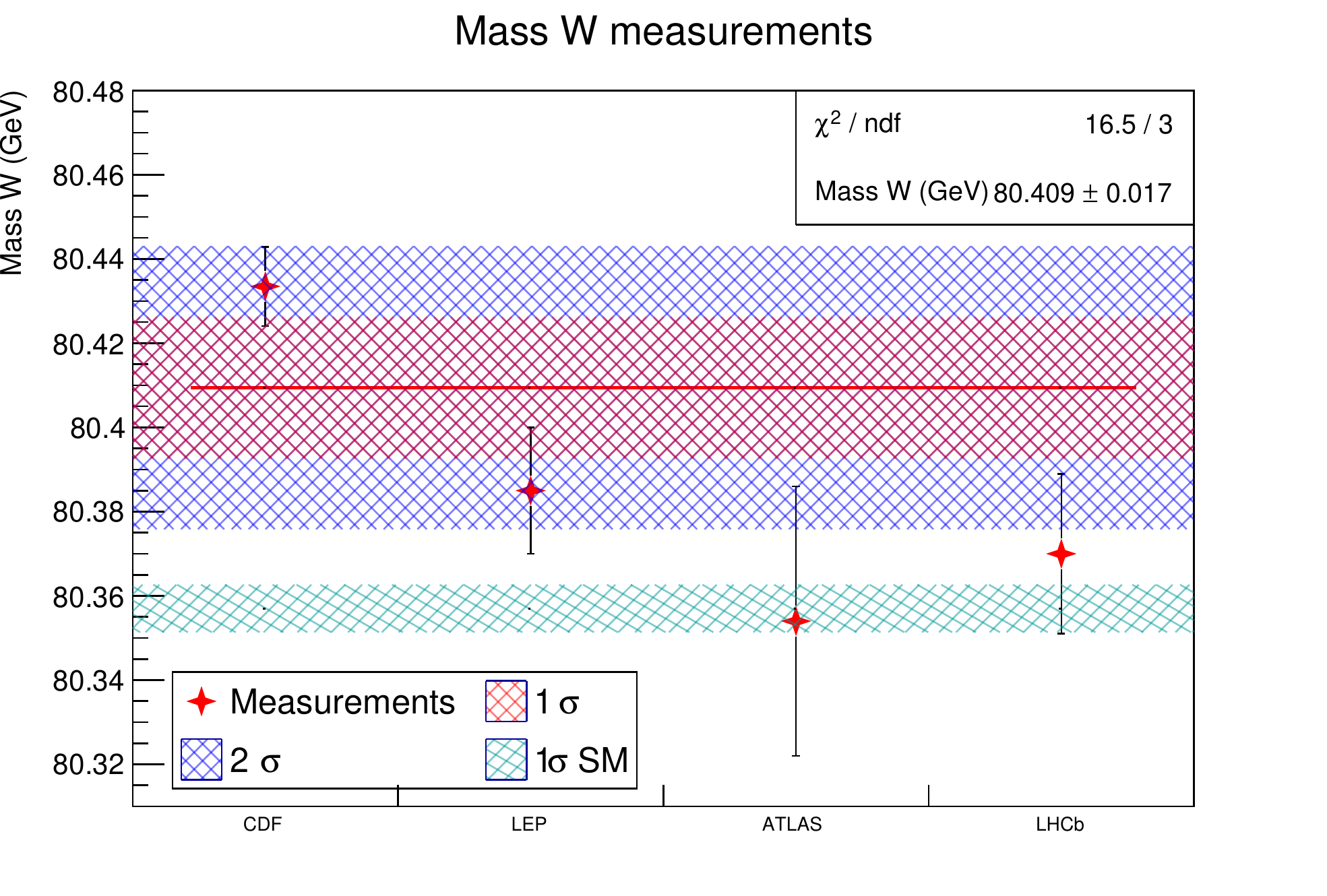}
	\caption{$W$ mass measurements and their weighted result with the final uncertainty corrected due to the large discrepancies across the experimental measures. }
	\label{fig:Wmass}
\end{figure}
Comparing our naive average with the SM in Eq.~\eqref{eq:SM}, the deviation from the SM reduces to around 3 standard deviations. { Following \cite{ParticleDataGroup:2020ssz} we determined the uncertainty on the averaged $W$ mass by increasing the weighted value by  the $\sqrt{\chi^2/\text{ndf}}$ factor. This procedure takes into account the large discrepancies among the four measures. In Fig.~\ref{fig:Wmass} we summarize the fit and the overall result.}
 It was pointed in \cite{Isaacson:2022rts} that the CDF experiment used an older version of the ResBos code \cite{Balazs:1997xd} that is only accurate at NNLL+NLO, while the state-of-the-art ResBos2 code is able to make predictions at N$^3$LL+NNLO accuracy. The authors of \cite{Isaacson:2022rts}  demonstrate that the data-driven techniques used by CDF capture most of the higher order corrections. This means that including higher order corrections can decrease the value reported by CDF by at most 10 MeV which may reduce the disagreement with the SM to $6\sigma$.

	\section{Anomalies driven theoretical models }\label{th}


	New physics is needed to explain the violation of LFU and the measured deviation from the SM of the lepton anomalous magnetic moments. It is therefore natural to imagine this physics to appear at some energy scales $
\Lambda$ that are higher than the electroweak (EW) scale taken to be around  $250$~GeV. Because the observed deviations occur at a much lower energy scale, i.e. either $B$-meson scale or the muon mass scale one has to take into account SM and BSM radiative corrections appearing over several order of magnitudes.  This investigation can be performed, model by model, and requires matching the NP theory first with effective operators around the EW scale and then running them down to the desired scale relevant for the experimental measures taking into account SM corrections. Although this procedure is well defined we prefer to concentrate on the leading BSM contributions to the desired Wilson coefficients responsible for the observed anomalies.

	We discuss now the most compelling theoretical scenarios which can account for the over-mentioned anomalies.
	The two first models include a new heavy particle, a $Z^\prime$ or a leptoquark, which mediate at tree level the relevant interaction generating the effective operator $(\bar s \gamma_\mu P_X b)(\bar \ell \gamma_\mu P_Y \ell)$. The latter can also be generated radiatively via loops of fermions and scalars. In general, related interactions also induce the operators 
	\beq c_{b_L b_L} (\bar s \gamma_\mu P_L b)^2+
	c_{\mu_L\nu_\mu}
	( \bar{\mu}\gamma^{\mu} P_L \mu )(\bar{\nu}_\mu \gamma_\mu P_L\nu_\mu) \ ,
	\eeq
	which can affect other observables. In particular, the first operator contributes to $B_s$ mass mixing and the relative coefficient is consequently bounded as 	$c^{\rm BSM}_{b_L b_L} = (-0.09 \pm 0.08)/(110\TeV)^2$ ~\cite{1408.4097,1608.07832}. The second operator affects the neutrino trident cross section. CCFR data yields the bound 	$|c^{\rm BSM}_{\mu_L\nu_\mu}|< 1/(490\GeV)^2$ at 95\% C.L. by ~\cite{1406.2332}. The scenarios we will examine, considering NP contributions in muon interactions, will also affect the $(g-2)_\mu$ observable. 
	\subsection{Can the \texorpdfstring{$Z^\prime$} - help?}
	Perhaps one of the simplest phenomenological ways  to address an anomaly has been to resort to the introduction of $Z^\prime$ bosons. 	The  latter, as illustrated in figure\fig{FeynZLQ}, couples to the neutral current 
		\beq [g_{bs} (\bar s \gamma_\mu P_L b) +\hbox{h.c.}] + g_{\mu_L} (\bar\mu \gamma_\mu P_L \mu) \ ,\eeq
 generating the $b \to s \ell^+ \ell^-$ anomalies. At tree level, a $Z^\prime$ with the above interactions and mass $M_{Z^\prime}$ yields	$c_{b_L \mu_L} =- g_{bs} g_{\mu_L}/M_{Z'}^2$. A long list of  models featuring such a kind of extra vector boson  appeared  in the literature \cite{1310.1082,1311.6729,1403.1269,1501.00993,1503.03477,1503.03865,1505.03079,1506.01705,1508.07009,1509.01249,1510.07658,1512.08500,1601.07328,1602.00881,1604.03088,1608.01349,1608.02362,1611.03507,1702.08666,1703.06019}. Attention must, however, be paid to complementary measures constraining the $Z^\prime$ induced interactions. In fact, the $Z^\prime$ contributes to the $B_s$ mass mixing $\Delta M_{B_s}$  via the coefficient 
	$c_{b_L b_L} =- g_{bs}^2/2 M_{Z'}^2$.  Therefore,  in order to satisfy the bound from  $\Delta M_{B_s}$,  requiring a small $g_{bs}$ coupling, one must consider large enough $g_{\mu L}$ couplings to address the flavor anomalies. By gauge invariance, also the neutrino operator is generated, with $c_{\mu_L \nu_\mu} =- g_{\mu_L}^2/M_{Z'}^2$ without, however,  yielding   a strong constraint on $g_{\mu_L}$. The coupling $g_{bs}$ can be generated via a flavor rotation in the left-handed down quark sector as $g_{bs} = (g_t-g_q) (U_{Q_d})_{ts}$, where $g_t$ ($g_q$) denotes the $Z^\prime$ coupling to the 3-rd generation (lighter) left-handed quarks and $U_{Q_d}$ the flavor rotation matrix. If the CKM matrix $V = U_{Q_u} U_{Q_d}^\dagger$ is dominated by the rotation among left-handed down quarks,
	rather than by the rotation $U_{Q_u}$ among left-handed up quarks, the  matrix element $(U_{Q_d})_{ts}$ is  not much larger than $V_{ts}$. Because of the coupling to lighter quarks, unless we restrict to the limiting case $g_q=0$, the $Z^\prime$ is largely produced in $pp$ collisions at the LHC and receive severe bounds from $pp\to Z'\to \mu\bar\mu$. The LHC constraints, however, can be significantly relaxed if the $Z^\prime$ can decay into other BSM particles, as for example into invisible DM particles ~\cite{1511.07447}.\\

		Besides the flavor anomalies, the introduction of a $Z^\prime$ could   simultaneously explain both the electron and muon $g-2$ anomalies \cite{Bodas:2021fsy, Hammad:2021mpl}. In this case, however, the $Z^\prime$ is generally required to be light, with a mass of the order of MeVs, and feebly coupled to the SM. Alternatively, a massive $Z^\prime$ of the order of 100 GeVs could explain simultaneously the flavor anomalies and $(g-2)_\mu$ if the $Z^\prime$ couplings to the muons are loop-induced \cite{Belanger:2015nma} or generated by mixing with extra vector-like leptons \cite{Navarro:2021sfb}. Improved limits on $B \to K^{(*)} \nu\nu$ and forthcoming new results from Belle II have the potential to  exclude the $Z^\prime$ explanation of the flavor anomalies  for $m_{Z^\prime} \lesssim 4$ GeV \cite{Crivellin:2022obd}.  \\

		Summarizing, a straightforward introduction of a new $Z^\prime$ is  insufficient to explain all the anomalies simultaneously. Additionally even addressing just the LFU anomaly the $B_s$ mass mixing and LHC bounds hamper the simplest explanation. However, one can imagine more involved models featuring several new particles/sectors that would rehabilitate the $Z^\prime$ solution. 
		
	\subsection{The promising leptoquarks landscape}\label{LQ}
	
	Leptoquarks appeared very early in particle physics in the attempt to achieve a more unified picture of fundamental interactions \cite{Pati:1974yy}. We will now briefly summarize their impact in trying to explain  both flavour anomalies and deviations in the muon and electron anomalous magnetic moments.

	These are bosonic particles carrying both quark and lepton numbers naturally coupling to a lepton and a quark. Leptoquark (LQ) induced explanations of the flavor anomalies have been extensively discussed in the literature ~\cite{ 0910.1789,1412.1791, 1503.01084, 1703.09226, 1408.1627,1501.05193, 1505.05164,1512.01560, 1608.08501, 1704.05849,1609.04367,1611.04930,  1706.07808, 1803.10972, 1806.05689, 1808.10309, 1912.00899,2004.09464, 2203.10111 }.

More recently, it has  been further shown \cite{Bigaran:2020jil, Dorsner:2020aaz,Perez:2021ddi, Crivellin:2022mff, Saad:2020ihm} that  leptoquarks can also accommodate the deviations in $(g-2)_\ell$. 	
	In the following, we discuss how LQs can explain the anomalies in  $b \to s \ell^+ \ell^-$ transitions and in $(g-2)_\ell$. We start by defining the LQs quantum numbers under the SM gauge group $(SU(3)_c, SU(2)_L, U(1)_Y)$, with the electric charge given by the sum of the hypercharge and of the third component of the isospin, $Q=Y+T_3$. 
	
\subsubsection{Scalar Leptoquarks}

Scalar LQs are the most promising kinds of LQ for explaining all of the anomalies. Interestingly, they can naturally appear as pseudo-Goldstone bosons associated to the breaking of a global symmetry of a new strongly coupled sector \cite{0910.1789, 1412.1791}. We identify three types of scalar LQs which can account for the anomalies.

\begin{itemize}
\item{$S_3=(\mathbf{\Bar{3}},\mathbf{3},1/3)$}\\
The Yukawa Lagrangian of this weak triplet is given by:
\begin{equation}
    y^{L}_{ij}\Bar{Q}^{C\ i,a}_L\ (i\tau^2)^{ab}\left(\tau^k S^k_3\right)^{bc}\ L^{j,c}_L+\ z^{L}_{ij}\Bar{Q}^{C\ i,a}_L\ (i\tau^2)^{ab}\left(\left(\tau^k S^k_3\right)^\dagger\right)^{bc}\ Q^{j,c}_L+ \qq{h.c.}
\end{equation}c'
Note that the couplings to diquarks must be suppressed, $z^{LL}_3 \approx 0$, in order to guarantee the proton stability. 
$S_3$ can mediate neutral current transitions $b \to s \mu \mu$ at tree-level, which can accommodate the anomalies in $R_{K^{(*)}}$. In particular, after integrating out the LQ, one finds 
\begin{equation}
     C_9^{\mathrm{BSM}} =  -  C_{10}^{\mathrm{BSM}} = \frac{2 \pi v^2}{V_{tb} V^*_{ts} \alpha_{em}}\frac{y_{b\mu}^L (y_{s\mu}^L)^*}{m^2_{S_3}}\, .
\end{equation}


\item{$R_2=(\mathbf{3},\mathbf{2},7/6)$}\\
$R_2$ is a weak doublet, its interactions, which have the property to conserve the baryon number, read:
\begin{equation}
    -y^{RL}_{ij}\Bar{u}^i_R R^a_2(i\tau^2)^{ab}L^{j,b}_L+\ y^{LR}_{ ij}\Bar{e}^i_R R^{a\ \ast}_2\  Q^{j,a}_L+\qq{h.c.}\ .
\end{equation}
We can include three more LQs in the model if we imagine to reproduce the right-handed neutrinos with some electrically neutral fields $\nu_R$.\\
$R_2$ can generate at tree level a contribution $ C_9^{\mathrm{BSM}} =  C_{10}^{\mathrm{BSM}}$ which is in conflict with data. However, setting $y_{LR}=0$, $R_2$ can also generate at loop level $ C_9^{\mathrm{BSM}} = -  C_{10}^{\mathrm{BSM}}$, which can accommodate the anomalies in $R_{K^{(*)}}$ \cite{Becirevic:2017jtw}. 

As we will discuss in the paragraph below, this LQ can also explain deviations from the SM in $(g-2)_{\ell}$.

\item{$S_1=(\mathbf{\Bar{3}},\mathbf{1},1/3)$}\\
The most general interactions of the weak singlet scalar LQ read:
\begin{equation}
\begin{split}
    &y^{LL}_{ij}\Bar{Q}^{C\ i,a}_L S_1 (i\tau^2)^{ab}L^{j,b}_L+\ y^{RR}_{ij}\Bar{u}^{C\ i}_R S_1 e^j_R+\ y^{\overline{RR}}_{ ij}\Bar{d}^{C\ i}_R S_1 \nu^j_R+\\&+z^{LL}_{ij}\Bar{Q}^{C\ i,a}_L S^\ast_1 (i\tau^2)^{ab}Q^{j,b}_L+\ z^{RR}_{ij}\Bar{u}^{C\ i}_R S^\ast_1 d^j_R+\qq{h.c.}\ .
    \end{split}
\end{equation}
Similarly to the $S_3$ case, one has to impose baryon number conservation, $z \approx 0$, to preserve proton stability. The $S_1$ interactions can mediate at tree-level charged transitions $b \to c \ell \bar \nu_{\ell^\prime}$.
Neutral current transitions $b \to s \ell \ell$ can be also generated, but only at loop level. 
In this way, it has been shown \cite{Bauer:2015knc} that $S_1$ can explain the anomalies in $R_{K^{(*)}}$. \\ 
As we will discuss below, $S_1$ can also accommodate the anomalies in $(g-2)_{\ell}$.

\end{itemize}
\paragraph{$g-2$ anomalies in the scalar LQ sector\\}
 Here we provide scenarios where LQ models can accomodate both $(g-2)_\ell$ anomalies as well as LFU violation. These models are constrained by experimental bounds on the $\mu\to e\gamma$ process, the stringest one coming from the MEG experiment \cite{Crivellin:2018qmi}. One finds that independently  $R_2$ and $S_1$ offer the best explanation of the anomalies \cite{Bigaran:2020jil, Perez:2021ddi}. The study in \cite{Dorsner:2020aaz} further shows that mixing through the Higgs any two types of LQs cannot explain the anomalies. In particular, the combination $S_1\And S_3$, which is not in conflict with $\mu \to e \gamma$, is excluded by the interplay between $K_L^0 \to e\mu$, $Z\to e^+e^-$ and $Z \to \mu^+\mu^-$ data.   This recent study \cite{Crivellin:2022mff} shows, however, that with three generations of a weak doublet scalar LQ $S_2$ it is possible to accommodate simultaneously all of the anomalies while being compatible with the other experimental data.  
 The deviation in $(g-2)$ for the muon is addressed by loops, as shown in Fig. \ref{fig:LQanomalyg-2}, containing the LQ and the top quark. While, for $(g-2)$ of the electron the loops can in principle involve either the top or the charm. However, in the case of top-containing loops for $(g-2)_e$, a large contribution to $\mu \to e \gamma$ decay would be also induced, which is not compatible with the MEG bound \cite{Bigaran:2020jil}. We summarize these points in Table \ref{tableg-2scalar}.
 
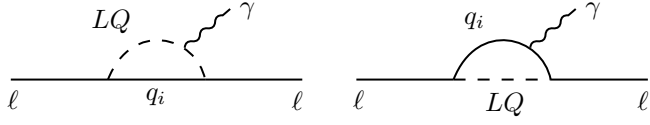
\begin{figure}[htbp]
    \centering
\tikzset{every picture/.style={line width=0.75pt}} 

\begin{tikzpicture}[x=0.75pt,y=0.75pt,yscale=-1,xscale=1]

\draw    (101.67,130.5) -- (150.17,130.5) ;
\draw    (150.17,130.5) -- (198.67,130.5) ;
\draw    (198.67,130.5) -- (247.17,130.5) ;
\draw  [dash pattern={on 4.5pt off 4.5pt}]  (150.17,130.5) .. controls (162.17,100) and (194.17,108.5) .. (198.67,130.5) ;
\draw    (188.5,114.5) .. controls (188.67,112.15) and (189.94,111.06) .. (192.29,111.24) .. controls (194.64,111.42) and (195.91,110.33) .. (196.08,107.98) .. controls (196.25,105.63) and (197.52,104.54) .. (199.87,104.72) .. controls (202.22,104.9) and (203.49,103.81) .. (203.66,101.46) .. controls (203.83,99.11) and (205.1,98.02) .. (207.45,98.2) -- (211.17,95) -- (211.17,95) ;
\draw    (274.17,130.5) -- (322.67,130.5) ;
\draw  [dash pattern={on 4.5pt off 4.5pt}]  (322.67,130.5) -- (371.17,130.5) ;
\draw    (371.17,130.5) -- (419.67,130.5) ;
\draw    (322.67,130.5) .. controls (334.67,100) and (366.67,108.5) .. (371.17,130.5) ;
\draw    (361,114.5) .. controls (361.17,112.15) and (362.44,111.06) .. (364.79,111.24) .. controls (367.14,111.42) and (368.41,110.33) .. (368.58,107.98) .. controls (368.75,105.63) and (370.02,104.54) .. (372.37,104.72) .. controls (374.72,104.9) and (375.99,103.81) .. (376.16,101.46) .. controls (376.33,99.11) and (377.6,98.02) .. (379.95,98.2) -- (383.67,95) -- (383.67,95) ;

\draw (98.5,134.4) node [anchor=north west][inner sep=0.75pt]  [font=\footnotesize]  {$\ell $};
\draw (240.5,134.9) node [anchor=north west][inner sep=0.75pt]  [font=\footnotesize]  {$\ell $};
\draw (167.5,133.4) node [anchor=north west][inner sep=0.75pt]  [font=\footnotesize]  {$q_{i}$};
\draw (140.5,93.9) node [anchor=north west][inner sep=0.75pt]  [font=\footnotesize]  {$LQ$};
\draw (214.5,90.4) node [anchor=north west][inner sep=0.75pt]  [font=\footnotesize]  {$\gamma $};
\draw (271,134.4) node [anchor=north west][inner sep=0.75pt]  [font=\footnotesize]  {$\ell $};
\draw (413,134.9) node [anchor=north west][inner sep=0.75pt]  [font=\footnotesize]  {$\ell $};
\draw (326.5,95.4) node [anchor=north west][inner sep=0.75pt]  [font=\footnotesize]  {$q_{i}$};
\draw (337,135.9) node [anchor=north west][inner sep=0.75pt]  [font=\footnotesize]  {$LQ$};
\draw (387,90.4) node [anchor=north west][inner sep=0.75pt]  [font=\footnotesize]  {$\gamma $};

\end{tikzpicture}
    \caption{Radiative LQs contributions to the $(g-2)_\ell$. Here $q_i$ is either a top or a charm quark.}
    \label{fig:LQanomalyg-2}
\end{figure}

\begin{table}[htbp!]
\centering
\begin{tabular}{@{}llll@{}}
\toprule
LQ                  & $(g-2)_e$ & $(g-2)_\mu$ & MEG bound  \\ \midrule
$S_1$               & top       & top         & disagreement    \\
$S_1$               & charm     & top         & agreement  \\
$R_2^{\frac{5}{3}}$ & top       & top         & disagreement     \\
$R_2^{\frac{5}{3}}$ & charm     & top         & agreement  \\ \bottomrule
\end{tabular}
\caption{Schematic results of the LQs scalar models simultaneously explaining $(g-2)_e$ and $(g-2)_\mu$: the first column indicates the specific LQ considered, the second and third columns define the type of quark involved in the loop. The last column shows the dis/agreement with the MEG bound \cite{Bigaran:2020jil}.  
}
\label{tableg-2scalar}
\end{table}

    



\subsubsection{Vector Leptoquarks}

Vector LQs have been also considered to explain the flavor anomalies \cite{1505.05164,1512.01560,1609.04367,1611.04930}, and typically appear as new composite states. A theory with a vector LQ in isolation is non-renormalizable. LQ-containing loops lead, indeed, to UV divergences~\cite{1607.07621}. This means that an UV embedding of the LQs in a more complete theory must be envisaged. Therefore, model-dependent aspects, such as the study of loop effects in physics observables, for example in $\Delta M_{B_s}$, must be considered in the specific UV-completed scenarios. In the following, we list the two types of vector LQs which can address the LFU violation.

\begin{itemize}
\item{$U_1=(\mathbf{3},\mathbf{1},2/3)$}\\
$U_1$ does not couple to diquarks and its genuine lepto-quark interactions read:
\begin{equation}
    x^{LL}_{1\ ij}\bar{Q}^{i,a}_L\gamma^\mu U_{1,\mu}L^{j,a}_L+\ x^{RR}_{1\ ij}\bar{d}^i_R\gamma^\mu U_{1,\mu}e^j_R+\ x^{\overline{RR}}_{1\ ij}\bar{u}^i_R\gamma^\mu U_{1,\mu}\nu^j_R+\qq{h.c.}\ .
\end{equation}
The absence of diquark interactions, and thus of proton stability issues, render this type of LQ models appealing. In particular, in the simplified assumption $x^{RR}=0$, and neglecting the presence of right-handed neutrinos, one predicts the following Wilson's coefficients  
\begin{equation}
 C_9^{\mathrm{BSM}}=-C_{10}^{\mathrm{BSM}}=-\frac{2 \pi v^2}{V_{tb}V^\ast_{ts}\alpha}\frac{x^{LL}_{s\mu} x^{LL\ \ast}_{b\mu}}{m^2_{U_1}} \, ,
\end{equation}
which can explain the flavor anomalies.

\item{$U_3=(\mathbf{3},\mathbf{3},2/3)$}\\
The weak triplet $U_3$  couples only to left-handed particles via the interaction terms
\begin{equation}
    x^{LL}_{3\ ij}\Bar{Q}^{i,a}_L\gamma^\mu\big(\tau^kU^k_{3, \mu}\big)^{ab}L^{j,b}_L+\qq{h.c.}\ .
\end{equation}
Like in the previous case diquark couplings are absent. These interactions  lead to the following contribution to the $b\to s\mu\mu$ process 
\begin{equation}
  C_9^{\mathrm{BSM}}=-C_{10}^{\mathrm{BSM}}=-\frac{2 \pi v^2}{V_{tb}V^\ast_{ts}\alpha}\frac{x^{LL}_{s\mu} x^{LL\ \ast}_{b\mu}}{m^2_{U_3}} \, ,
\end{equation}
 which is able to account for the $R_K$ and $R_{K^\ast}$ anomalies. 


	\end{itemize}

	\medskip


The  Minimal Supersymmetric Standard Model is not preferred by the LFU anomalies because the only sparticle (left-handed squark) that can contribute to these processes is unable to explain simultaneously $R_K$ and $R_{K^*}$.

	\begin{figure}[tbp]
	\centering
	\begin{subfigure}{0.3\textwidth}
	\centering
	\begin{tikzpicture}
\begin{feynman}
\vertex (a1);
\vertex at ($(a1) + (1.5cm,-0.5cm)$) (a2);
\vertex[right=1.5cm of a2] (a3);
\vertex[right=4.5cm of a1] (a4);
\vertex[below=2.5cm of a1] (b1);
\vertex at ($(b1) + (1.5cm,0.5cm)$) (b2);
\vertex[right=1.5cm of b2] (b3);
\vertex[right=4.5cm of b1] (b4);
\diagram* {
{
(a4) --[black,edge label'=\(Q\)] (a3) --[scalar, black, edge label'=\(\mathcal{S}_{D^c}\)] (a2) --[black,edge label'=\(Q\)] (a1),
(b1) --[black!50!black,edge label'=\(L\)] (b2) --[scalar, black, edge label'=\(\mathcal{S}_{E^c}\)] (b3) --[black!50!black,edge label'=\(L\)] (b4),
},
(a2) -- [very thick,black, edge label'=\(\mathcal{F}_{L}\)] (b2),
(a3) -- [very thick,black, edge label=\(\mathcal{F}_{L}\)] (b3),
};
\end{feynman}
\end{tikzpicture}
	\end{subfigure}
\hspace{0.3cm}
	\begin{subfigure}{0.3\textwidth}
	\centering
	\begin{tikzpicture}
\begin{feynman}
\vertex (a1);
\vertex at ($(a1) + (1cm,-0.5cm)$) (a2);
\vertex[right=1.5cm of a2] (a3);
\vertex[right=3.5cm of a1] (a4);
\vertex[below=2.5cm of a1] (b1);
\vertex at ($(b1) + (1cm,0.5cm)$) (b2);
\vertex[right=1.5cm of b2] (b3);
\vertex[right=3.5cm of b1] (b4);
\diagram* {
{
(a4) --[black,edge label'=\(Q\)] (a3) --[scalar, black, edge label'=\(\mathcal{S}_{D^c}\)] (a2) --[black,edge label'=\(Q\)] (a1),
(b1) --[black,edge label'=\(Q\)] (b2) --[scalar, black, edge label'=\(\mathcal{S}_{D^c}\)] (b3) --[black,edge label'=\(Q\)] (b4),
},
(a2) -- [very thick,black, edge label'=\(\mathcal{F}_{L}\)] (b2),
(a3) -- [very thick,black, edge label=\(\mathcal{F}_{L}\)] (b3),
};
\end{feynman}
\end{tikzpicture}
	\end{subfigure}
\hspace{0.3cm}
	\begin{subfigure}{0.3\textwidth}
	\centering
         \begin{tikzpicture}
\begin{feynman}
\vertex (a1);
\vertex[below=2.5cm of a1] (b1);
\vertex at ($(a1) + (1.5cm,0)$) (a2);
\vertex at ($(a1) + (3.2cm,-1.25cm)$) (c1);
\vertex at ($(b1) + (1.5cm,0)$) (b2);
\vertex[right=1.2cm of c1] (c2);
\diagram* {
{
(a1) --[black!50!black,edge label=\(L\)] (a2) --[scalar, black, edge label=\(\mathcal{S}_{E^c}\)] (c1) --[boson,edge label=\(\gamma\)] (c2),
(b1) --[black!50!black,edge label'=\(L\)] (b2) --[scalar, black, edge label'=\(\mathcal{S}_{E^c}\)] (c1),
},
(a2) -- [very thick,black, edge label'=\(\mathcal{F}_{L}\)] (b2),
};
\end{feynman}
\end{tikzpicture}

	\end{subfigure}
	\caption{\em Feynman diagrams contributing to $R_K$, $\Delta M_{B_s}$ and the muon $g-2$ in models with extra fermions ${\cal F}$
				and extra scalars ${\cal S}$.
				In Fundamental Composite Higgs models these diagrams will be dressed by further new composite dynamic contributions.
				\label{fig:FeynRK}}
	\end{figure}
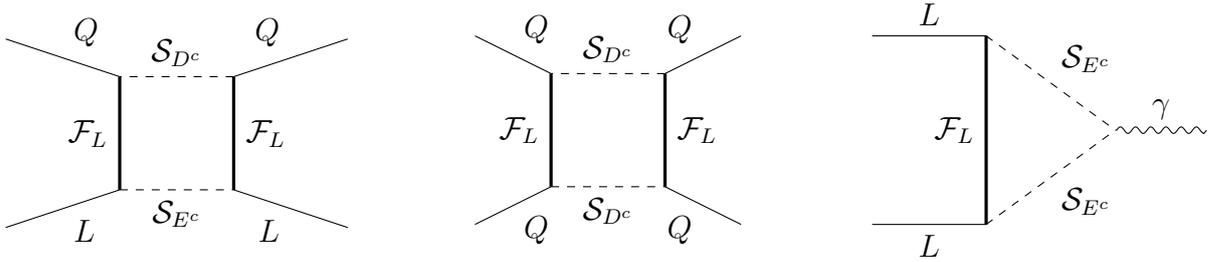

	\subsection{Naturalness: From radiative models to composite Higgs}\label{FPC} 

Naturalness has been a driving principle for the past decades when modelling and searching for new physics. The observation of the Higgs boson, with properties similar to the ones predicted by the SM, has led to partially abandon it as guidance principle.  In fact, the natural SM extensions dismissed by data are vanilla ones  such as the constrained minimal supersymmetric SM and composite Higgs models with several light top partners. 

As argued in \cite{Cacciapaglia:2021gff} the anomalies revitalize naturalness in fueling new models of fundamental interactions. By sewing together Technicolor-like models \cite{Weinberg:1975gm,Dimopoulos:1979es} with SM fermion partial compositeness \cite{Kaplan:1991dc}, one can accommodate the observed anomalies \cite{Cacciapaglia:2021gff} (for earlier attempts see \cite{Appelquist:2003hn,Appelquist:2004mn}). In these constructions the composite scale is around $2$~TeV, which is the natural techni-fermion condensation scale. As for the Higgs nature, there are several possibilities such as a composite dilatonic state  \cite{Bardeen:1985sm,Yamawaki:1985zg,Sannino:1999qe,Hong:2004td,Dietrich:2005jn,Appelquist:2010gy,Goldberger:2007zk} coming from walking Technicolor dynamics \cite{Holdom:1988gs,Cohen:1988sq}.   Here the associated effective action is obtained  by saturating the underlying trace anomaly of the theory as shown in famous Coleman graduate lecture book Coleman~\cite{coleman_1985}.  There has been a flurry of interest in this  framework~\cite{Hong:2004td,Dietrich:2005jn,Goldberger:2007zk,Appelquist:2010gy,Hashimoto:2010nw,Matsuzaki:2013eva,Golterman:2016lsd,Hansen:2016fri,Golterman:2018mfm} partially fuelled by lattice studies. In particular these investigations focused on $SU(3)$ gauge theories with $N_f= 8$ fundamental Dirac fermions~\cite{Appelquist:2016viq,Appelquist:2018yqe,Aoki:2014oha,Aoki:2016wnc}, as well as $N_f = 3$ symmetric 2-index Dirac fermions (sextets)~\cite{Fodor:2012ty,Fodor:2017nlp,Fodor:2019vmw}. The latter are known as  Minimal Walking Technicolor~\cite{Hong:2004td,Sannino:2004qp,Dietrich:2005jn,Evans:2005pu} models. The lattice collaborations corroborate, for these models,  the existence of a light singlet scalar particle to be identified with the dilaton.  Additionally, 
a light Higgs state can emerge due to top quark corrections \cite{Foadi:2012bb} or from other non QCD-like dynamics \cite{Hong:2004td,Foadi:2012bb}.  The composite Goldstone Higgs paradigm \cite{Kaplan:1983fs,Contino:2003ve} offers another alternative which is, as we shall review, disfavoured by the muon $g-2$ anomaly \cite{Cacciapaglia:2021gff}. A recent comprehensive review of composite dynamics has appeared where the reader can find in depth explanations of composite (Goldstone) Higgs dynamics and its link to new fundamental gauge dynamics \cite{Cacciapaglia:2020kgq}. We refer to \cite{Baer:2021aax} for 
a discussion of naturalness and the muon $g-2$  anomaly in supersymmetric models while for other perturbative explanations we refer to  \cite{Athron:2021iuf}.

Rather than treating separately the radiative models from the composite ones, following \cite{Cacciapaglia:2021gff} we compare elementary and composite extensions of the SM and show that Technicolor-like models \cite{Weinberg:1975gm,Dimopoulos:1979es} yield a natural interpretation of these results with  consequences for collider physics. We consider the  model of \cite{Cacciapaglia:2021gff} that, depending on the underlying dynamics, can be interpreted as either elementary, or partially elementary, or fully composite.  We therefore add to the SM Lagrangian the following interactions: 
    
\begin{equation}
\begin{split}
    -L_{NP}=&y^{ij}_L\ L^i\mathcal{F}_L\big(\mathcal{S}_E\big)^\ast+y^{ij}_E\ \big(E^i\big)^c\mathcal{F}^c_N\mathcal{S}^j_E+\\&y^{ij}_Q Q^i\mathcal{F}_L\big(S^j_D\big)^\ast+y^{ij}_U\big(U^i\big)^c\mathcal{F}^c_E\mathcal{S}^j_D+y^{ij}_D\big(D^i\big)^c\mathcal{F}^c_N \mathcal{S}^j_D+\\&\sqrt{2}k\big(\mathcal{F}_L\mathcal{F}^c_N+\mathcal{F}_E\mathcal{F}^c_L\big)\Phi_H+\qq{h.c.}  \ .
\end{split}\label{eq:LYuk2}
\end{equation}
Here $\cal F$ and $\cal S$ are (Weyl) fermions and scalars, respectively, with multiplicity $N_{\rm TC}$.  This multiplicity can either be seen as a global number or as the dimension of the fundamental representation of a new strongly coupled gauge group $\SU_{N_{\rm TC}}$. The particle's  quantum numbers are listed in Table~\ref{tab:5frags2}. In the Lagrangian \eqref{eq:LYuk2}, $L$, $E$ and $Q$ are the chiral SM fields, while $\phi_H$ carries the Higgs doublet quantum numbers. 
\begin{table}
\begin{center}
\begin{tabular}{l|c|c|c|c}
  & $\mathcal{G}_{\rm TC}$ & $SU(3)_c$ & $SU(2)_L$ & $U(1)_Y$ \\\hline
$\mathcal{F}_{L}$ & $\bf F$ & $1$ & $2$ & $Y$ \\ 
$\mathcal{F}_{N}^c$ &  $\bf \bar F$ & $1$ & $1$ & $-Y-1/2$ \\ 
$\mathcal{F}_{E}^c$ &  $\bf \bar F$ & $1$ & $1$ & $-Y+1/2$ \\  \hline
$\mathcal{S}_{E}^j$ & $\bf F$ & $1$ & $1$ & $Y-1/2$ \\
$\mathcal{S}_{D}^j$ & $\bf F$ & $3$ & $1$ & $Y+1/6$  \\\hline
\end{tabular}
\end{center}
\caption{\em\label{tab:5frags2} Quantum numbers of the new fermions $\cal F$ and scalars $\cal S$ in the model. $\mathcal{G}_{\rm TC}$ can be considered either gauged, as in composite scenarios, or global in a renormalisable model of flavour.} 
\end{table}
The model can be interpreted in several ways: 
\begin{itemize}
\item[i)]{\underline{Radiative model}: The introduction of new elementary fermions and scalars with renormalizable interactions that contribute radiatively to the SM fermion masses $\cal F$ are vector-like fermions (thus they have both chiralities) and consequently have a tree-level Dirac mass $M_{\cal F}$.}
\item[ii)]{\underline{Fundamental partial compositeness}: The Higgs of the SM is replaced by a Technicolor composite state of (non) Goldstone nature, while the new scalars are still elementary, as put forward in fundamental partial compositeness \cite{Sannino:2016sfx} (see also \cite{Kagan:1994qg,Dobrescu:1995gz,Altmannshofer:2015esa}). 
This implies that $\kappa$ models the coupling between the  effective Higgs field and its constituents. The fermions $\cal F$ can be chiral or vector-like, depending on the details of the model.}
\item[iii)]{ \underline{Gauge-fermion compositeness}: As in ii) but with the new scalars interpreted as effective operators  made of new extended Technicolor fermions \cite{Cacciapaglia:2020kgq}. This limit covers also models of partial compositeness presented in \cite{Barnard:2013zea,Ferretti:2013kya}.}
\end{itemize}
In the first case, due to the perturbative nature of the loops, the masses of the heavy fermions (top, and possibly also bottom, tau and charm) must still originate from direct Higgs Yukawa interactions, while lighter masses and flavour mixing may even be radiatively generated.
For the other two cases, $\SU({N_{\rm TC}})$ confines at low energies and generates the Higgs as a composite (Goldstone) techni-meson. In this case, besides their possible bare masses, the new fermions $\cal F$ and scalars $\cal S$ inherited a composite dynamical mass of the order of $\Lambda_{\rm TC}$.  

The relevant data used in \cite{Cacciapaglia:2021gff} for the flavour and muon $g-2$ anomalies are summarised in  Table~\ref{tab:cCH}.  Here $c_{b_L \mu_L}$ is responsible for the leptonic $B$ decays entering the $R_K$ and $R_{K^\ast}$ measurements, while $C_{B\bar B}$ encodes the strongest constraint from the $B_s$ mixing \cite{DAmico:2017mtc}. The latter yields an upper bound on the quark Yukawa combination $(y_Q y_Q^\dagger)_{bs}$.
We also consider the corrections to the $Z$ coupling to muons since the Yukawa combination $(y_L y_L^\dagger)_{\mu \mu}$ also generates corrections to the muon coupling to the Higgs. The latter is however not yet significantly constrained by the measurements.   $K$-$\bar K$ mixing and/or lepton flavour violating processes further impose very strong bounds, but on different combinations of the Yukawas (for instance, $(y_Q y_Q^\dagger)_{sd}$ for Kaon mixing and $(y_L y_L^\dagger)_{\mu e}$ for $\mu \to e \gamma$ and the likes). Because their relation to muon anomalies is model-dependent, we neglect them here.
Second column of the Table summarizes the radiative estimates for the muon $g-2$ stemming from loops of the heavy fermions and scalars.  The associated Naive Dimensional Analysis (NDA) estimate for composite dynamics is shown in the third column.  For $\Delta a_\mu$, the relevant charges read $q_{\cal F} = Y+1/2$ and $q_{{\cal S}_E} = Y-1/2$ with $Y$ the hypercharge of the fermion doublet ${\cal F}_L$. For the elementary case the hypercharge is chosen to yield integer charges for the non QCD-coloured states, in particular for the plots we assume $Y=1/2$. For the composite case the hypercharge depends on the details of the model, in particular on $N_{\rm TC}$. 
For example, for $N_{\rm TC}=2$ we have $\SU(2_{\rm TC})$ with $Y=0$ that is the model studied in \cite{Ryttov:2008xe,Galloway:2010bp}. Here with two Dirac techni-fermions, one can simultaneously accommodate composite Goldstone Higgs and Technicolor models \cite{Cacciapaglia:2014uja}. Additionally, lattice investigations have confirmed the pattern of chiral symmetry breaking $SU(4)$ to $Sp(4)$ \cite{Lewis:2011zb,Hietanen:2013fya,Arthur:2016dir}.  For the time-honoured Technicolor limit of $N_{\rm TC}=3$ one can take the value  $Y=1/2$. Additionally, in the composite theory, $g_{\rm TC} \sim \frac{4\pi}{\sqrt{N_{\rm TC}}}$ takes care of varying number of Technicolors and the Yukawa couplings in Eq.~\eqref{eq:LYuk2} appear in the combinations $\frac{y y'}{g_{\rm TC}}$, which is assumed to be a perturbative coupling in the effective field theory description of the composite models. 
In the NDA estimates, it is not possible to know, a priori, the sign of the coefficients. Hence, we take the sign from the loop effects with equal masses: for $\Delta a_\mu$ the contribution is positive, as required by the measured muon $g-2$, for $Y = 0$ or positive.
 
\begin{table*}[t]
$$ \begin{array}{c|c|c}
\hbox{Coefficient} & \hbox{Perturbative one-loop result} & \hbox{Non-perturbative NDA}\\ \hline
c_{b_L \mu_L}
&  \displaystyle N_{\rm TC}  \frac{(y_{L} y_L^\dagger)_{\mu\mu} (y_{Q} y_{Q}^\dagger)_{bs}}{(4\pi)^2 M_{{\cal F}}^2} \frac{1}{4} F(x,y)
&  \displaystyle \frac{(y_{L} y_L^\dagger)_{\mu\mu} (y_{Q} y_{Q}^\dagger)_{bs}}{g_{\rm TC}^2 \Lambda_{\rm TC}^2}\\[5mm]
C_{B\bar B}
&  \displaystyle N_{\rm TC}  \frac{ (y_{Q} y_{Q}^\dagger)_{bs}^2}{(4\pi)^2 M_{{\cal F}}^2} \frac{1}{8} F(x,x)
& \displaystyle  \frac{ (y_{Q} y_{Q}^\dagger)_{bs}^2}{g_{\rm TC}^2 \Lambda_{\rm TC}^2}\\[3mm]
\delta g_{Z \mu_L}
&  \displaystyle N_{\rm TC} g_2 \frac{M_Z^2 (y_{L} y_L^\dagger)_{\mu\mu}}{(4\pi)^2 2 (1-2s_W^2) M_{{\cal F}}^2} F_9(Y,y)
&  \displaystyle g_2 \frac{M_Z^2 (y_{L} y_L^\dagger)_{\mu\mu}}{g_{\rm TC}^2 \Lambda_{\rm TC}^2}\\[5mm]
\multirow{2}{*}{$\Delta a_\mu$}
& \displaystyle  N_{\rm TC} \frac{m_\mu (y_{L} y_E^\dagger)_{\mu\mu} \kappa v_{\rm SM}}{(4\pi)^2 M_{{\cal F}}^2 }\left[2 q_{{\cal S}_E} F_{LR} (y) + 2 q_{{\cal F}}  G_{LR}(y)\right] +
& \multirow{2}{*}{$\displaystyle   \frac{m_\mu^2}{ \Lambda_{\rm TC}^2 } \left( 1 + \frac{(y_{L} y_L^\dagger)_{\mu\mu}}{{g_{\rm TC}^2}}\right)$}\\[4mm]
& \displaystyle    N_{\rm TC} \frac{m_\mu^2 (y_{L} y_L^\dagger)_{\mu\mu}}{(4\pi)^2 M_{{\cal F}}^2 }\left[2 q_{{\cal S}_E}F_7(y) + 2 q_{\cal F} \tilde{F}_7(y)\right]  &  \\
\end{array}
$$
\caption{\em\label{tab:cCH}  Coefficients of the low-energy operators impacting  leptonic $B$ decays ($c_{b_L \mu_L}$), $B_s$ mixing ($C_{B\bar B}$), correction to the $Z$ coupling to muons and the muon $g-2$, estimated via perturbative loops of the heavy fermions and scalars (second column) \cite{DAmico:2017mtc,Calibbi:2018rzv,Arnan:2019uhr} or via their strongly-interacting NDA (third column).  For $\Delta a_\mu$, the relevant charges read $q_{\cal F} = Y+1/2$ and $q_{{\cal S}_E} = Y-1/2$. In the composite models, $m_\mu \sim N_{\rm TC} \frac{(y_{L} y_E^\dagger)_{\mu\mu} \kappa v_{\rm SM}}{(4 \pi)^2}$. The loop functions, where $x=M_{{\cal S}_D}^2/M_{\cal F}^2$ and $y=M_{{\cal S}_E}^2/M_{\cal F}^2$, are given in the appendix of \cite{Cacciapaglia:2021gff}.}
\end{table*}

For the the composite nature of the  Higgs via  (fundamental) partial compositeness  we refer to Fig.~1 of \cite{Cacciapaglia:2021gff} obtained by assuming $N_{\rm TC} = 2$ and $Y=0$. The main result is that the muon $g-2$ anomaly prefers a  low composite scale of the order of $2$~TeV. Such a scale, as observed in \cite{Cacciapaglia:2021gff},  disfavours composite Goldstone Higgs (CGH) dynamics. The latter, in fact, typically requires a new composite scale around and above five TeVs (see \cite{Cacciapaglia:2021gff,Antipin:2014mda,Frigerio:2018uwx} for further details). Therefore, traditional Technicolor-like explanations are better suited to explain simultaneously LFU violations and the $g-2$ of the muon. 

If, however, the recent lattice results  from the BMW collaboration \cite{Borsanyi:2020mff} are taken as the correct contribution to the hadronic vacuum polarisation of the photon \cite{Davier:2010nc,Davier:2017zfy,Davier:2019can} the tension with the SM reduces and, in this case, it also improves the agreement with the CGH model interpretation. Furthermore, in the BMW scenario it is difficult to explain also  $R_K$ for fixed $N_{\rm TC}$ because a very large muon left-handed Yukawa coupling is needed. The BMW result is also in tension with electroweak precision data  \cite{Passera:2008jk, Keshavarzi:2020bfy, Crivellin:2020zul}.  
Overall, the large value required for the muon left-handed Yukawa coupling is an issue common to both  models ii) and iii) and can be ameliorated by increasing $N_{\rm TC}$.

Turning to the results for the radiative explanation \cite{Cacciapaglia:2021gff}, we observe that it depends on three heavy particles masses, two scalars and one fermion. Additionally, the Higgs Yukawa coupling to the new heavy fermions $\kappa$ is responsible for the dominant contribution to the muon $g-2$ and therefore one can maximize it by requiring it to saturate the muon mass \cite{Cacciapaglia:2021gff} \cite{Baker:2021yli} via 
\begin{equation} \label{eq:murad}
m_\mu = N_{\rm TC}\frac{(y_{L} y_E^\dagger)_{\mu\mu} \kappa v_{\rm SM}}{(4 \pi)^2} \ln \frac{\Lambda^2}{M_{\cal F}^2} \ ,
\end{equation}
with $\Lambda$ an UV cutoff interpreted as the scale at which the effective muon Yukawa vanishes. In the absence of Higgs corrections the model was studied in \cite{Arnan:2016cpy} and because of this it is unable to explain the observed $g-2$ of the  muon. 

The reader can find the summary of the radiative results in Fig.~2 of \cite{Cacciapaglia:2021gff}. The upshot is that fitting $R_K$ asks for either large Yukawas or large $ N_{\rm TC}$ which is then  constrained by the $B_s$ mixing. 
Additional this scenario leads to Yukawas that develop Landau poles at  low energy scales, thus hampering the high energy validity of the approach \cite{DAmico:2017mtc}.

\bigskip 
We learned that LFU violations and the $g-2$ muon anomaly can be explained via Technicolor-like models with a scale of new physics around $2$~TeV. This is a welcome news for collider experiments since it implies  that new massive states, such as the techni-rho, can very well be  within  reach of the LHC direct searches \cite{Belyaev:2018jse,Belyaev:2018qye}. The composite Goldstone Higgs scenario is, however, disfavoured by the low composite scale while radiative corrections stemming from elementary extensions of the SM require very large muon left-handed Yukawa couplings and unnaturally large multiplicity of new fields. 

Intriguingly, if the anomalies are confirmed, naturalness still plays a guiding role when constructing extensions of the SM.

	\subsection{Theoretical explanations of the W mass anomaly}
	The recent CDF collaboration measurement of the $W$ mass is finding support in a range of theoretical models. In line with the spirit of the previous phenomenological overview, we refer to  \cite{Bhaskar:2022vgk} for  
whether all the anomalies, including the W-mass result can be explained in a model featuring a weak-singlet $S_1$ and a weak-triplet $S_3$
scalar LQ, as well as the explanation via a vector LQ $V_2$ \cite{Cheung:2022zsb}  as the one addressing also  $R_D$. Models including  $Z^\prime$  were considered in \cite{Strumia:2022qkt, Cai:2022cti} with their impact summarised via the $S, T$ and $U$ oblique parameters. General electroweak precision data analysis was performed  also in \cite{Cacciapaglia:2022xih} with direct applications to non-standard Higgs models such as technicolor Higgs, techniglueballs and dilaton-like Higgs via the effective field theory of \cite{Sannino:2015yxa}.  Additionally, other explanations involving  scalar triplets \cite{Ghoshal:2022vzo, Popov:2022ldh, Borah:2022obi, Heeck:2022fvl,Kanemura:2022ahw} were invoked to also provide neutrino masses via an electroweak scale Higgs triplet. In \cite{Song:2022xts, Heo:2022dey, Han:2022juu, Arcadi:2022dmt, Chowdhury:2022moc, Ghorbani:2022vtv, Lee:2022gyf}  two-Higgs-doublet-models explanations were considered while in \cite{Nagao:2022oin}  vectorlike leptons were used. For  general classifications of the impact of new physics via electroweak oblique parameters we refer to \cite{Strumia:2022qkt, DiLuzio:2022xns} and \cite{Cacciapaglia:2022xih}, while the use of Standard Model Effective Theories we refer to \cite{deBlas:2022hdk, Bagnaschi:2022whn}. 
	
	{ {It is instructive to perform  effective analyses in terms of higher dimensional operators similar to the case of flavour observables.  In  \cite{Fan:2022yly} a set of six-dimensional operators in the SM effective field theory was considered. These operators are relevant to the electroweak precision tests. It was shown that an upward shift in W-boson mass is driven by the  $2{\cal{O}}_T = (H^\dagger \overset{\leftrightarrow}{D_\mu}
H)^2$  with a coefficient $c_T ({\rm TeV}/\Lambda)^2 \geq 0.01$. The result further suggests that the new physics scale favored by the CDF result is within the reach of current and near-future LHC searches.  }}

	\subsection{Dark side of the anomalies}
	
	It is exciting to explore the possibility that the same NP sector able to  account for the anomalies in flavour and/or in $(g-2)_{e,\mu}$ may also explain the dark matter puzzle. It has been shown that this can occur via either freeze-in \cite{Hall:2009bx} or freeze-out dark matter production mechanisms.  For the latter mechanism, where the dark matter is a thermal relic that went out of the primordial thermal plasma equilibrium at later stages of the universe evolution, we have models in which an extra $Z^\prime$ explains the flavor anomalies \cite{1511.07447, Darme:2018hqg, Belanger:2015nma, Biswas:2019twf, AristizabalSierra:2015vqb, Kawamura:2017ecz, Falkowski:2018dsl, Altmannshofer:2016jzy} featuring also extra vector-like fermions, other models correspond to theories with elementary extra fermions and scalars, where the anomalies are explained by radiative corrections  \cite{Cline:2017qqu, Cerdeno:2019vpd, Calibbi:2019bay, Huang:2020ris, Arcadi:2021cwg}, and to theories with scalar LQs \cite{Choi:2018stw, DEramo:2020sqv}.  In these models, the dark matter candidate (typically in the GeV mass range region) can either be a real scalar or a neutral fermion.  For the freeze-in mechanism, in which the coupling with the SM plasma is feeble and therefore the dark matter is never in thermal equilibrium,  the dark matter density, starting from zero or very small values, slowly grows up to the point it gets frozen-in at its cosmological abundance. Models of this type have a dark matter candidate in the BSM sector featuring also a vector LQ addressing the anomalies \cite{Belfatto:2021ats}. Here the dark matter state is a light (order 10 KeVs) right-handed neutrino, produced via freeze-in through its interactions with the massive (order TeVs) LQ.
	
For the W-mass anomaly,  the authors of reference \cite{Wang:2022dte}, consider extra electroweak triplet scalar dark matter multiplets in which the anomaly is explained via their 
 the loop corrections. Here, the lightest neutral particle in the multiplets serves as a candidate for cold dark matter (DM).

	\section{Impact of future LHCb and Belle II measurements }\label{Belle}

We now discuss the future impact of the LHCb and Belle II experiments improved statistics according to the expected time line presented in \cite{Belle-II:2018jsg},\cite{LHCb:2018roe}.
During the LHC Run1+Run2 periods, LHCb experiment has  collected data corresponding
to an integrated luminosity of $9~fb^{-1}$. It is expected to reach a total of $23~fb^{-1}$ by the end of the Run3. Roughly during the same time line  Belle II experiment is expected to collect an order of magnitude of about $5~ab^{-1}$, corresponding to five times the statistics achieved by the past B-factories Belle and BaBar.  
In order to acquire a rough understanding of the impact on the Wilson coefficients due to combined  accumulated statistics of both LHCb and Belle II we maintained the central values for the HI and HS observables while we estimated the projected uncertainties using the values provided in \cite{Belle-II:2018jsg}.

The impact of LHCb measurements of exclusive $B \rightarrow {K^{(*)}} \mu^+ \mu^-$ decays is expected to dominate, but the contribution from Belle II will be important in exclusive 
$B \rightarrow {K^{(*)}}  e^+ e^-$, since the Belle II apparatus has similar performances in both the di-electrons and di-muons final states. Moreover, Belle II will add precise 
measurements of $B \rightarrow X_s e^+ e^-$ and $B \rightarrow X_s \mu^+ \mu^-$ inclusive decays.

\begin{figure}[htbp]
    \centering
    \begin{subfigure}{0.4\textwidth} 
\centering
\includegraphics[height=5.2cm,keepaspectratio]{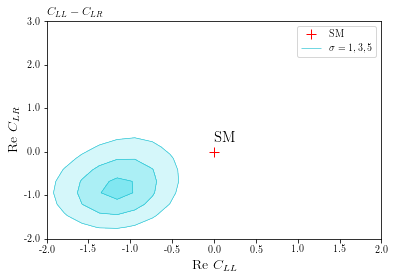}
\end{subfigure}
\hspace{2.2 cm} 
\begin{subfigure}{0.4\textwidth} 
\centering
\includegraphics[height=5.2cm,keepaspectratio]{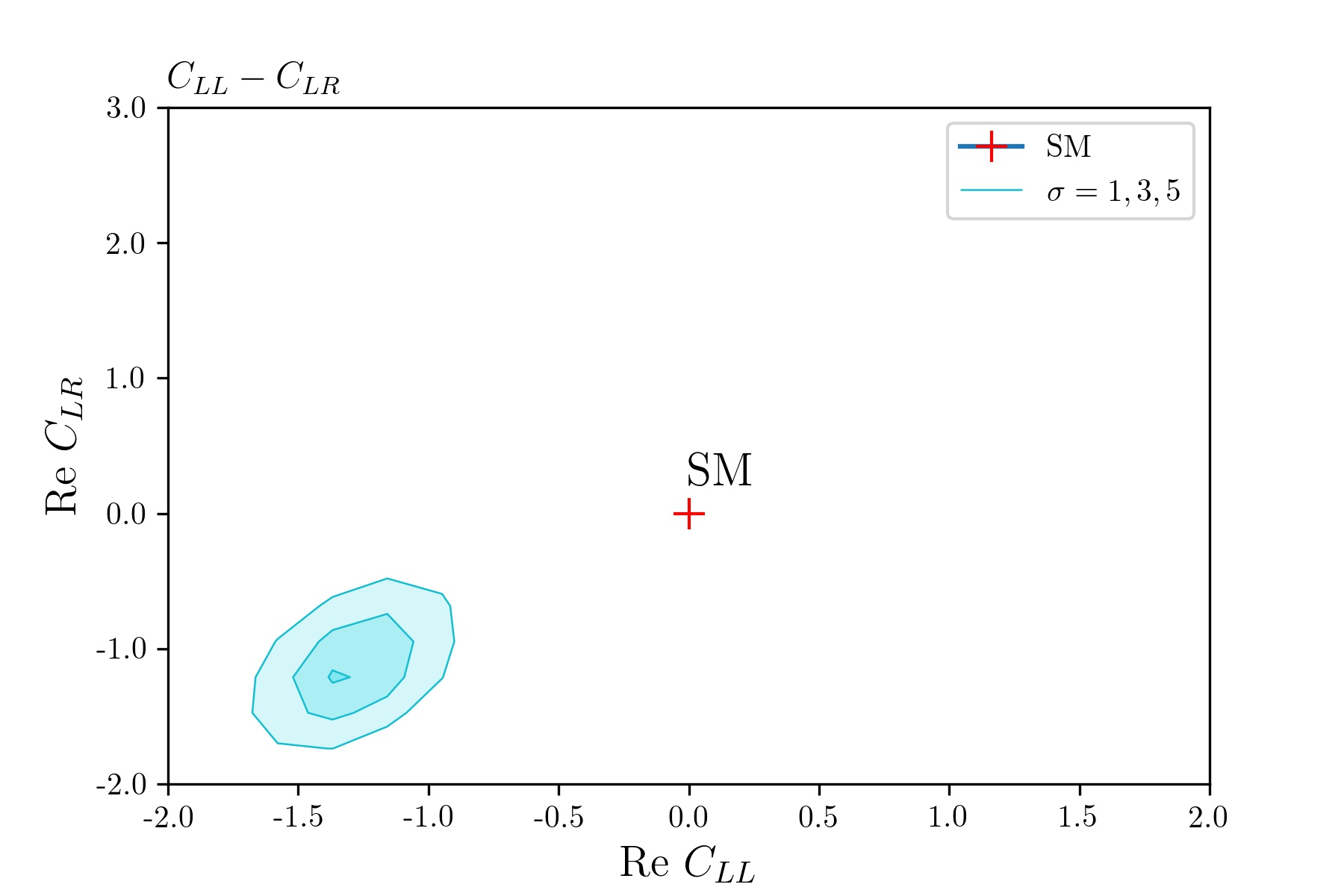}
\end{subfigure}
\\
\begin{subfigure}{0.4\textwidth} 
\centering
\includegraphics[height=5.2cm,keepaspectratio]{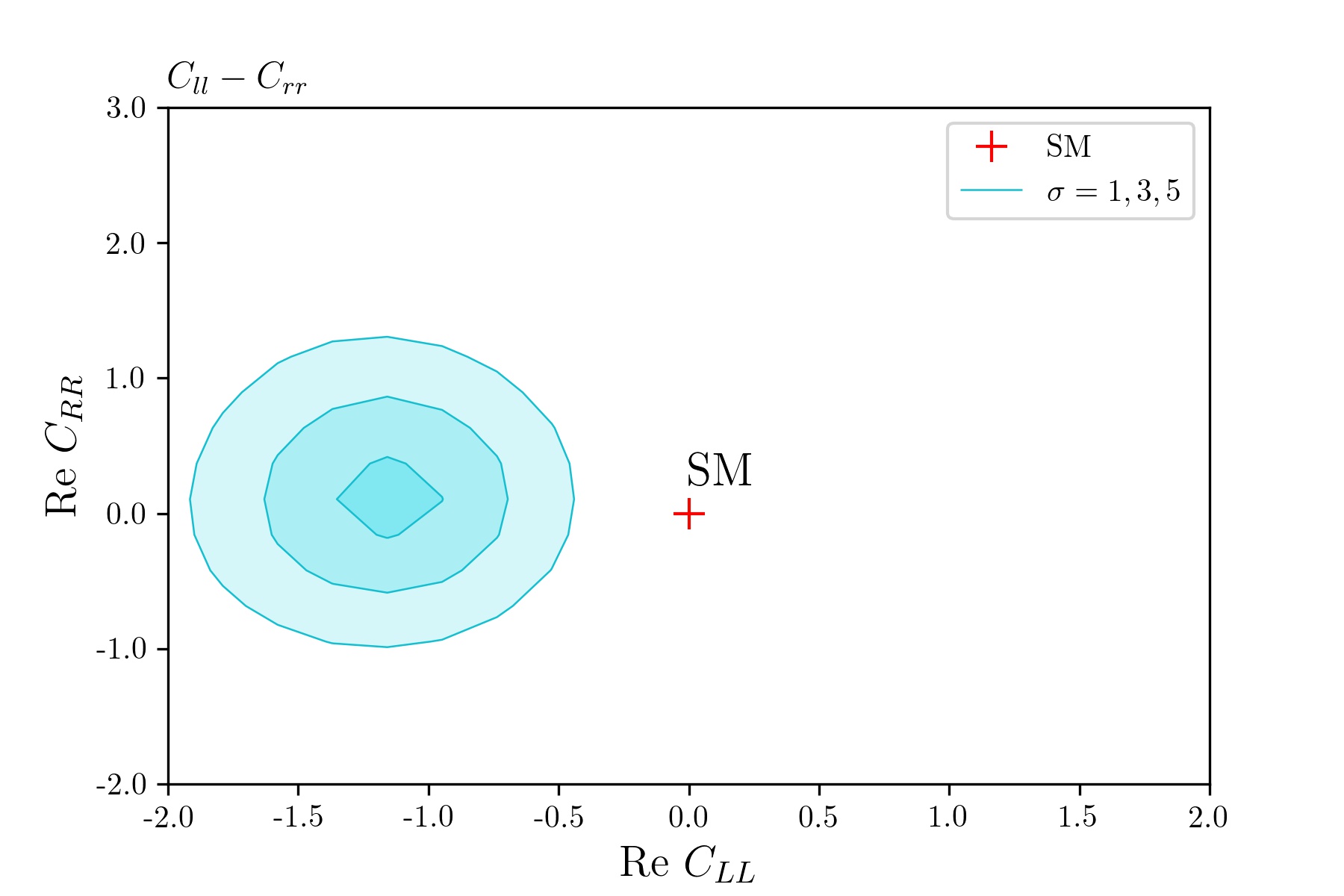}
\end{subfigure}
\hspace{2.2 cm} 
\begin{subfigure}{0.4\textwidth} 
\centering
\includegraphics[height=5.2cm,keepaspectratio]{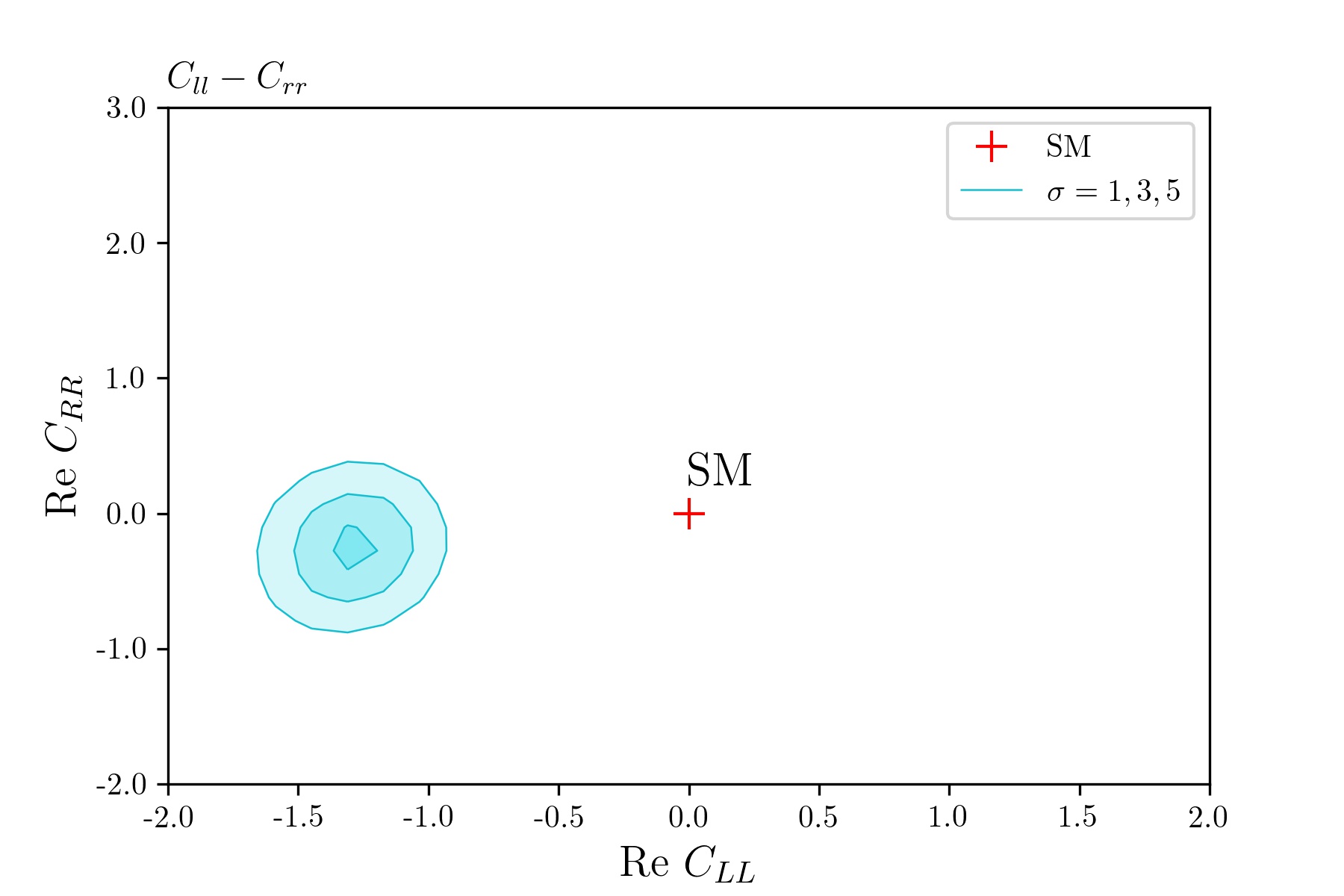}
\end{subfigure}\\ 
\begin{subfigure}{0.4\textwidth} 
\centering
\includegraphics[height=5.2cm,keepaspectratio]{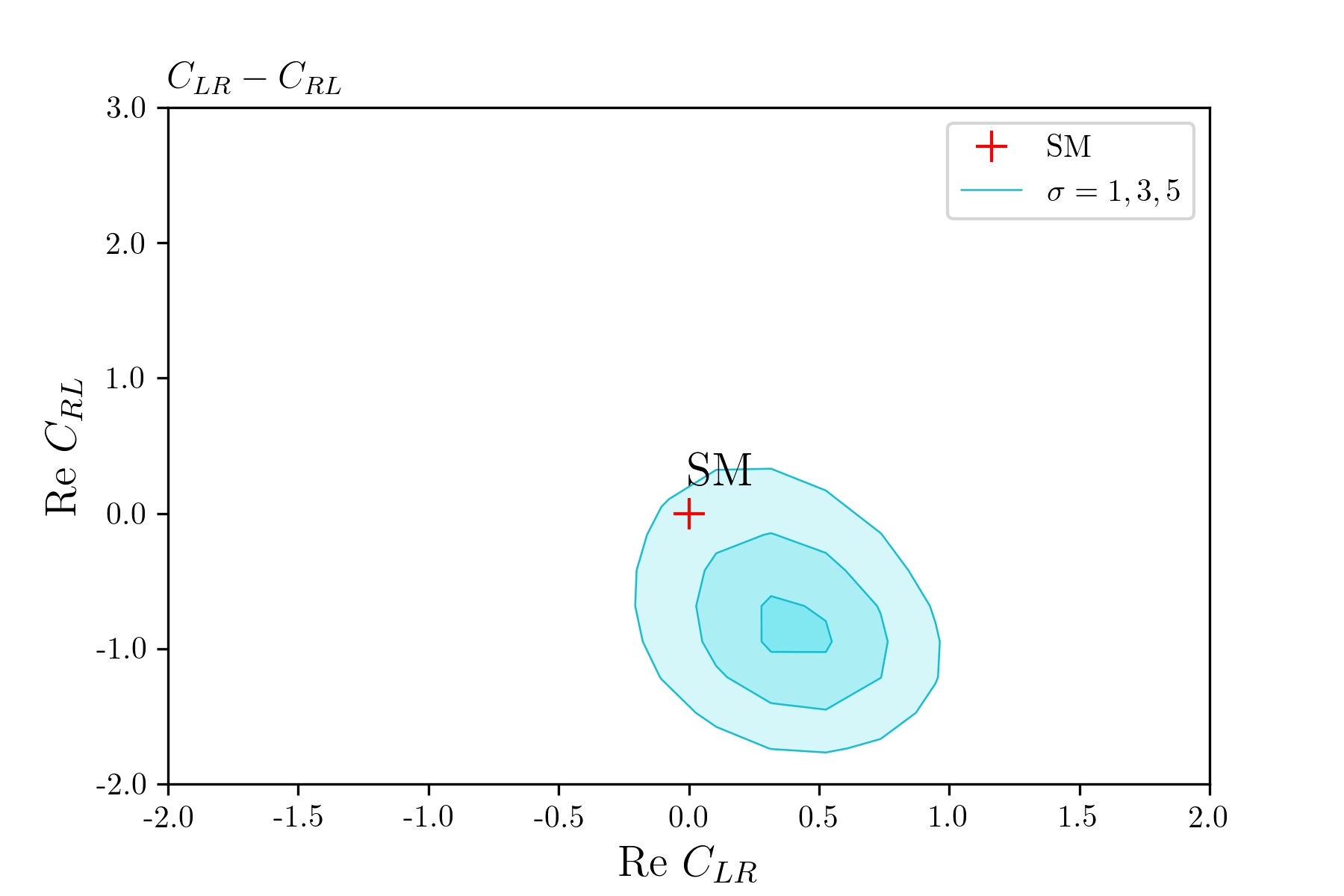}
\end{subfigure}
\hspace{2.2 cm} 
\begin{subfigure}{0.4\textwidth} 
\centering
\includegraphics[height=5.2cm,keepaspectratio]{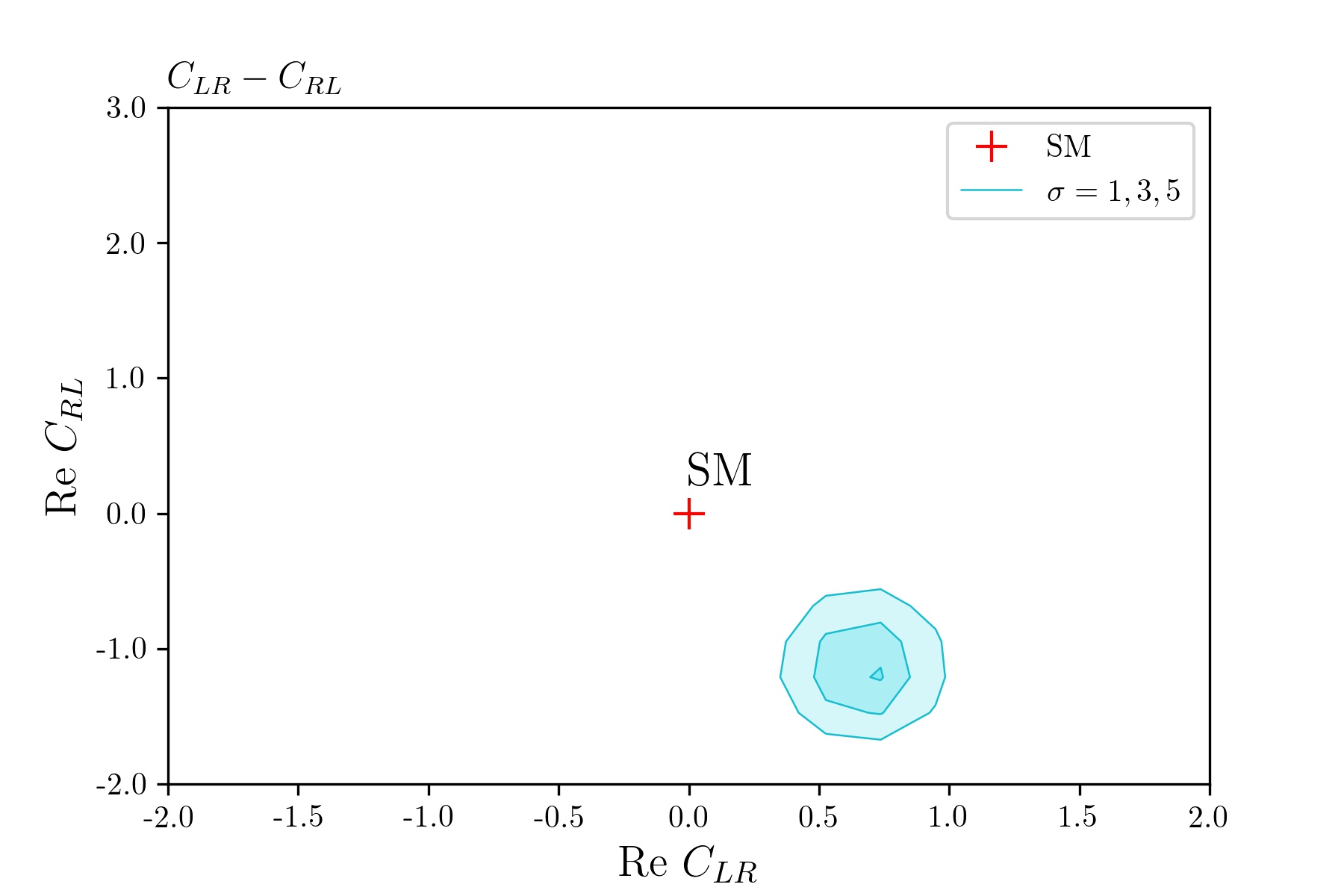}
\end{subfigure}
    
    \caption{Contour plots of 3 Wilson coefficient combinations with actual measurements (left) and future LHCb and Belle II measurements (right) with hadronic sensitive and insensitive observables. The central values of the future measures are kept to the today's values while the uncertainties adjusted for the previewed accumulated statistics.}
    \label{fig:Future3}
    \end{figure}
    \begin{figure}[htbp]
    \centering
    \begin{subfigure}{0.4\textwidth} 
\centering
\includegraphics[height=5.2cm,keepaspectratio]{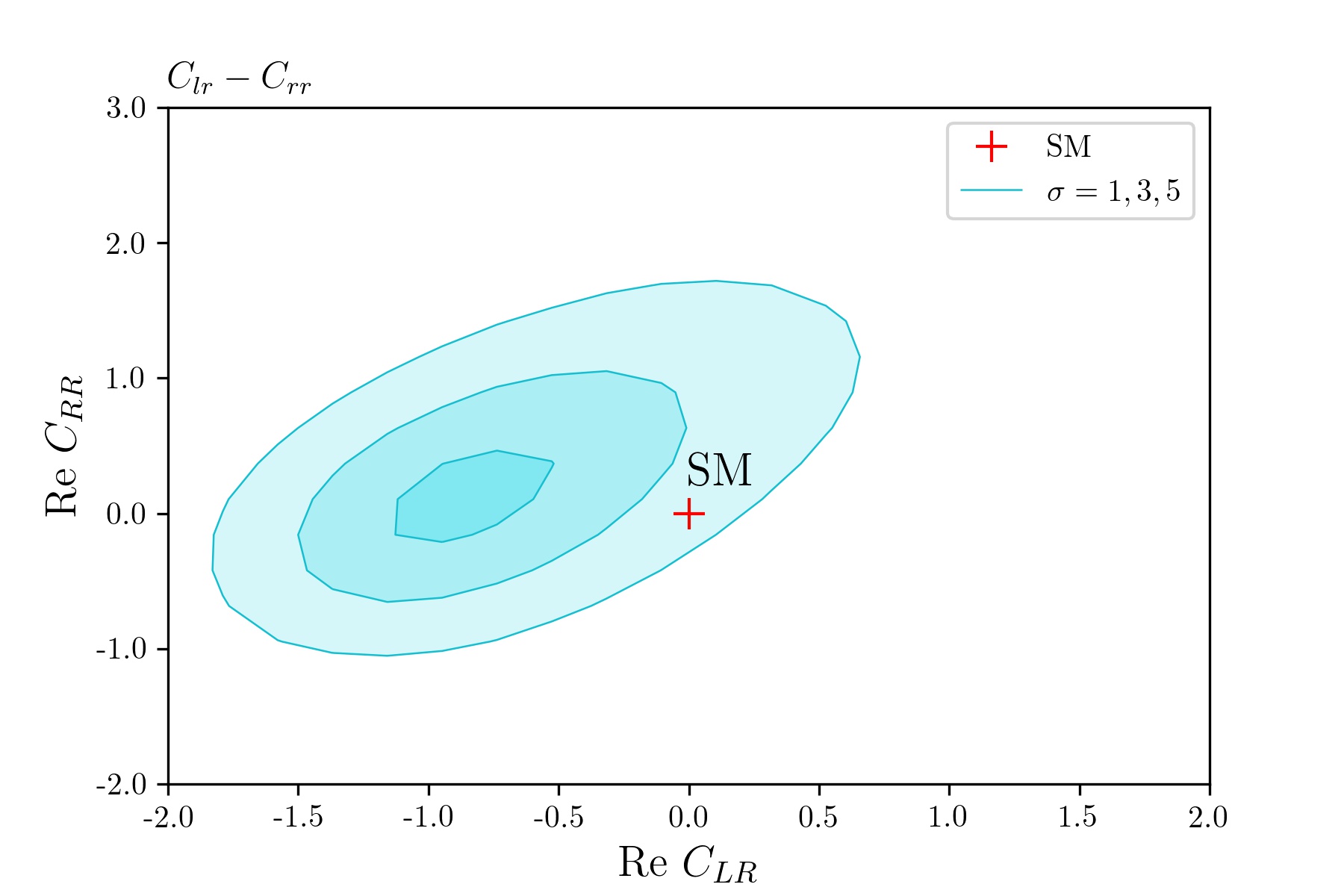}
\end{subfigure}
\hspace{2.2 cm} 
\begin{subfigure}{0.4\textwidth} 
\centering
\includegraphics[height=5.2cm,keepaspectratio]{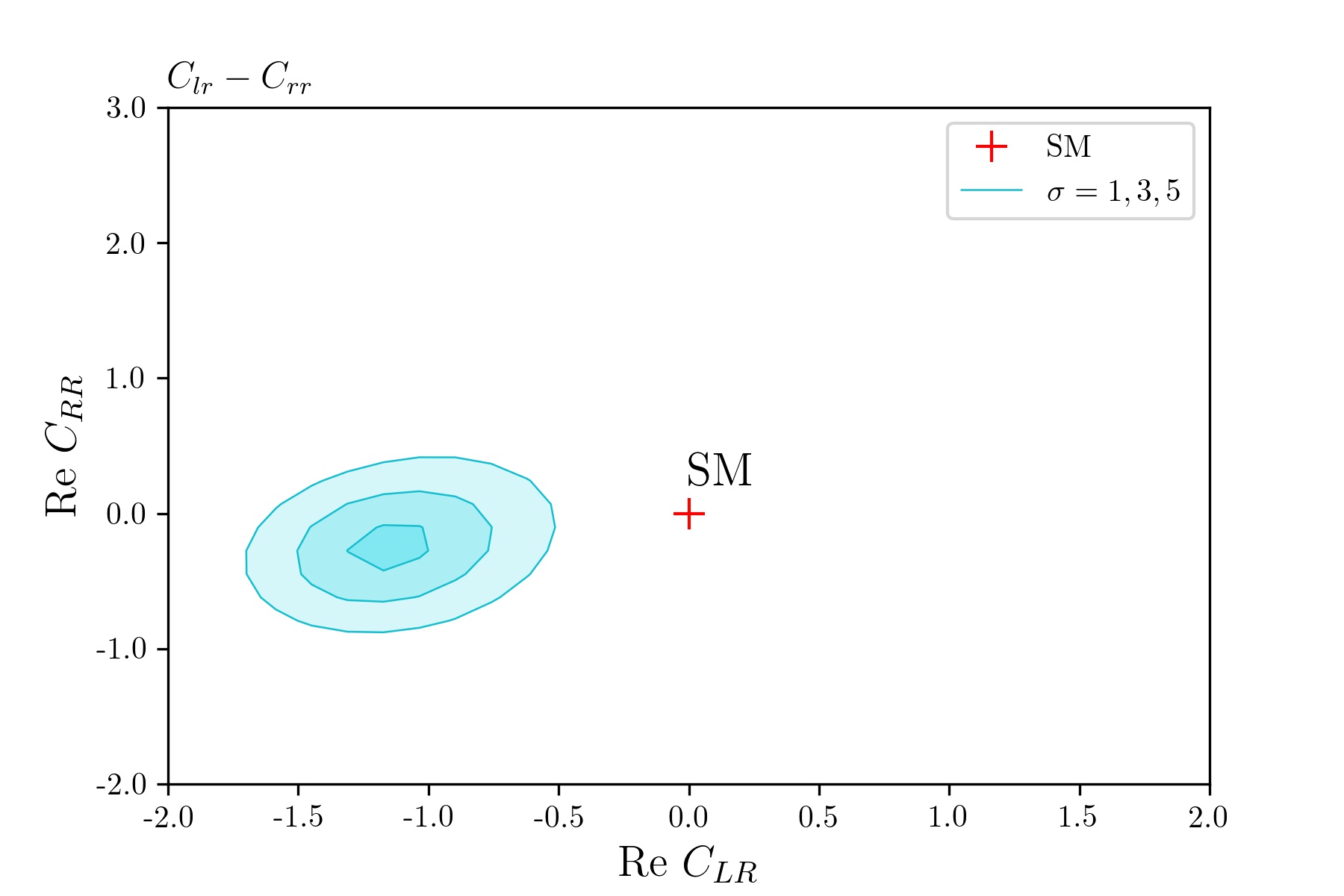}
\end{subfigure}
\\
\begin{subfigure}{0.4\textwidth} 
\centering
\includegraphics[height=5.2cm,keepaspectratio]{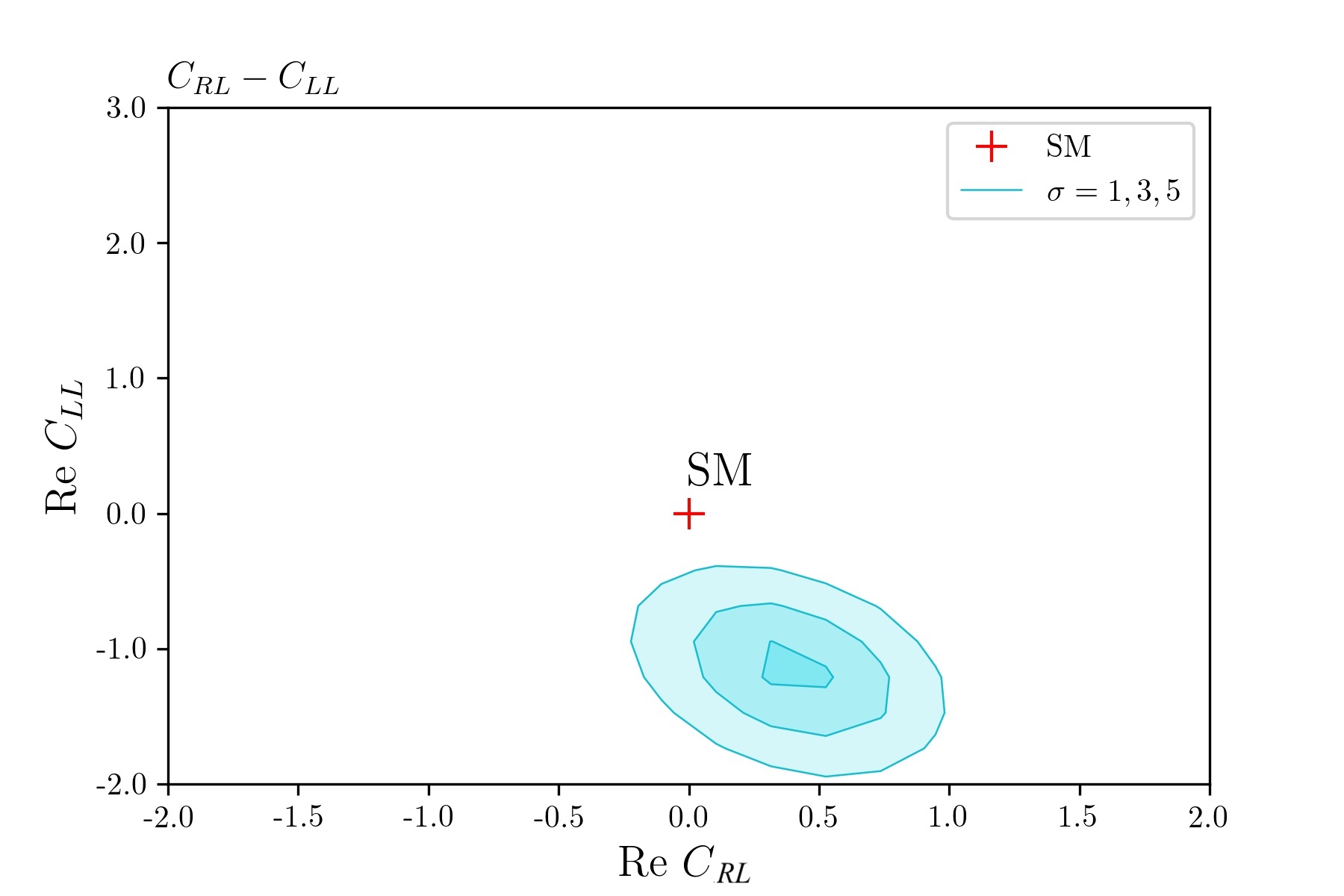}
\end{subfigure}
\hspace{2.2 cm} 
\begin{subfigure}{0.4\textwidth} 
\centering
\includegraphics[height=5.2cm,keepaspectratio]{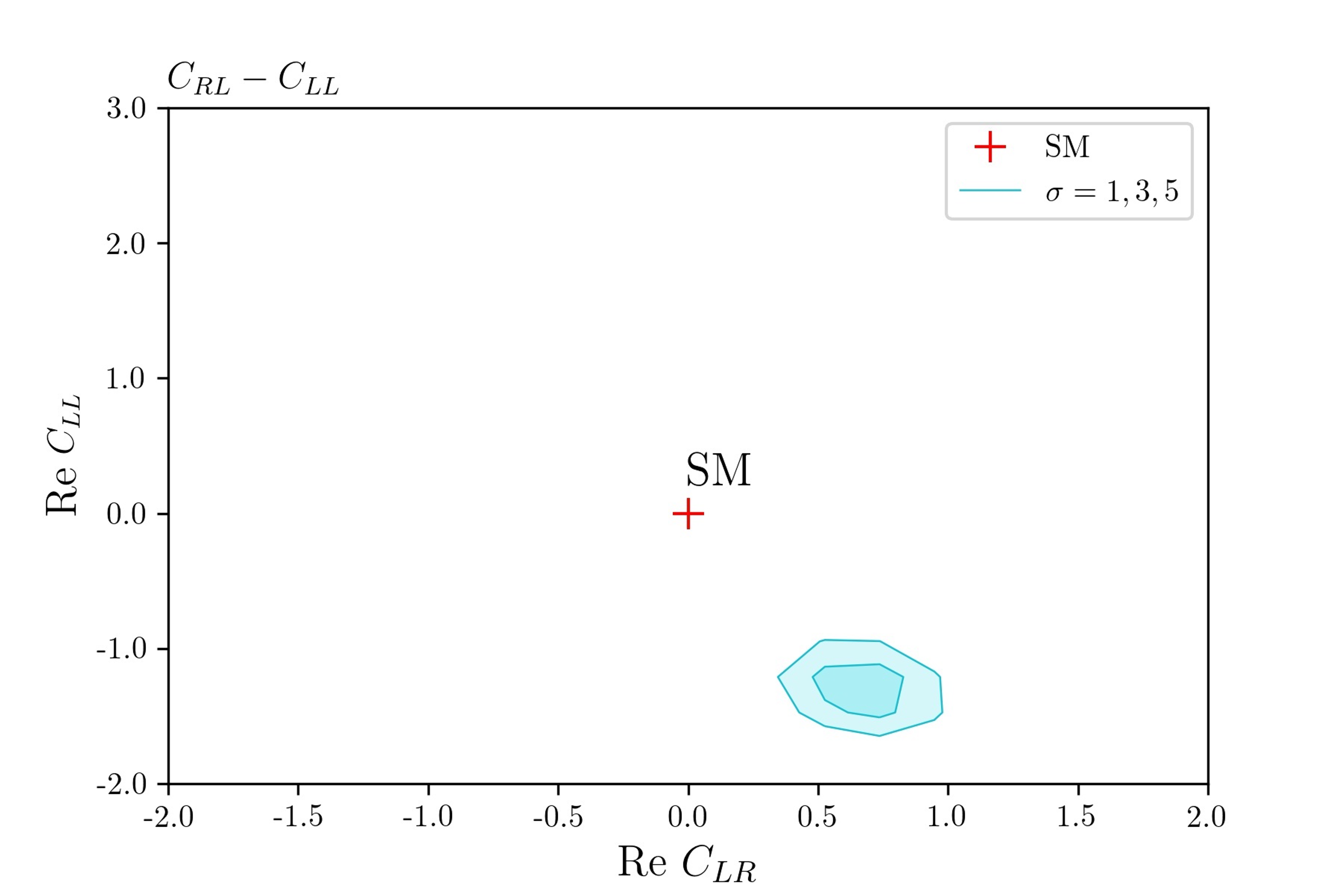}
\end{subfigure}\\ 
\begin{subfigure}{0.4\textwidth} 
\centering
\includegraphics[height=5.2cm,keepaspectratio]{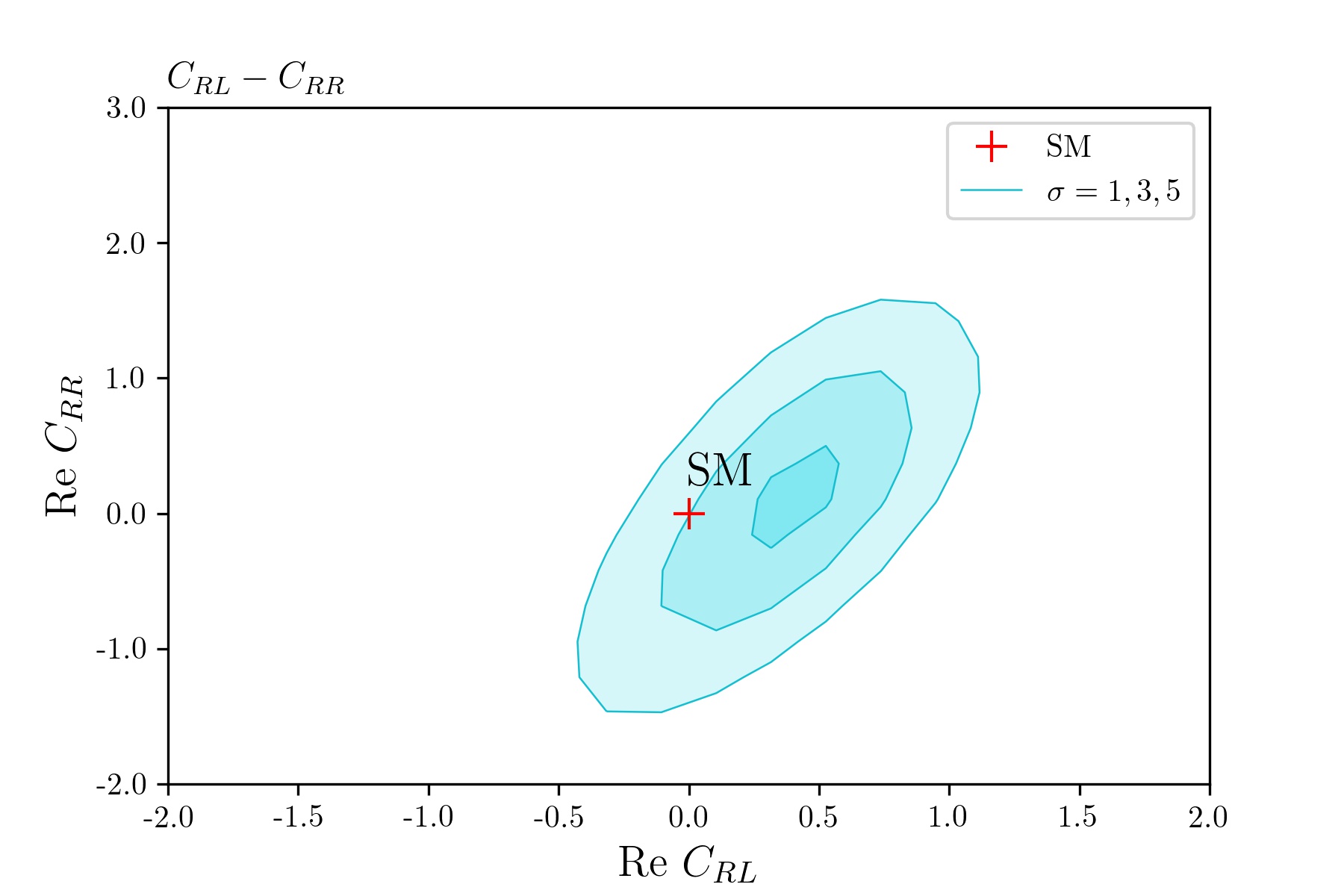}
\end{subfigure}
\hspace{2.2 cm} 
\begin{subfigure}{0.4\textwidth} 
\centering
\includegraphics[height=5.2cm,keepaspectratio]{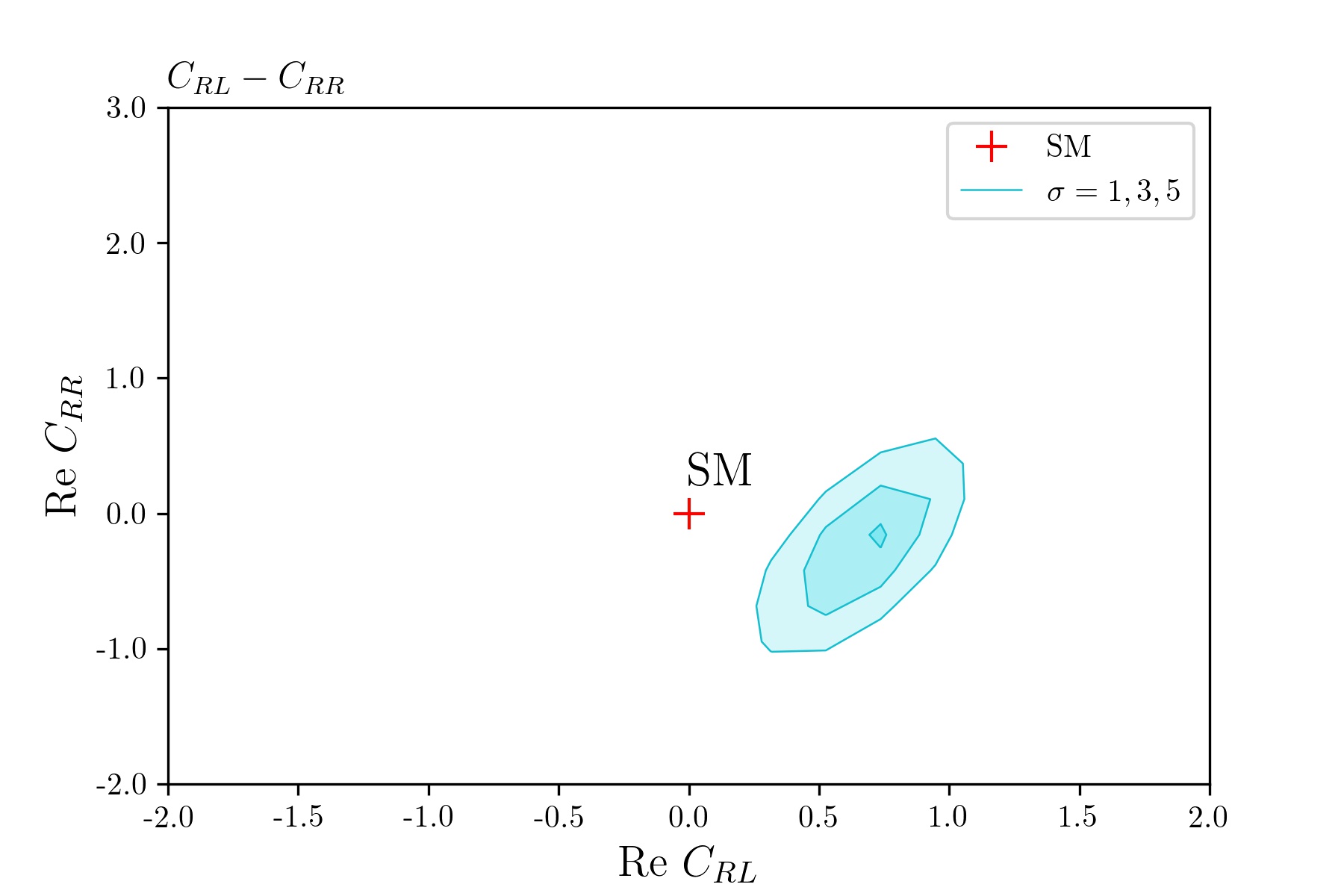}
\end{subfigure}
    
    \caption{Contour plots of the remaining 3 Wilson coefficient combinations not present in the previous figure with actual measurements (left) and future LHCb and Belle II measurements (right) with hadronic sensitive and insensitive observables. The central values of the future measures are kept to the today's values while the uncertainties adjusted for the previewed accumulated statistics.}
    \label{fig:Future4}
    \end{figure}

Performing a global fit, in the spirit of the previous sections, to the four relevant Wilson coefficients we obtain the contour plots presented in Fig.~\ref{fig:Future3} and  Fig.~\ref{fig:Future4}.\\
Each subfigure corresponds to the two-dimensional contour plot for each pair of Wilson coefficients with the other two pairs fixed at their central values.  The projections show that potential deviations w.r.t. the SM can be as large as $20\sigma$ when considering the results of the global fit summarized below:
\beq
	\begin{array}{rcl}
		C_ {b_L\mu _L}^{\text{BSM}} &=& -1.283\pm 0.007\\ 
		C_ {b_L\mu _R}^{\text{BSM}} &=& -1.15 \pm 0.14 \\ 
		C_ {b_R\mu _L}^{\text{BSM}} &=& 0.67 \pm 0.02 \\ 
		C_ {b_R\mu _R}^{\text{BSM}} &=& -0.25 \pm 0.06 \\
		\label{FutureGlobalmuon}
	\end{array}\quad \begin{aligned}
     &\chi^{2}_{SM}=1000.55,\\
     &\tilde{\chi}^{2}=619.19.
     \end{aligned} \quad
	\rho = 
	\left(
\begin{array}{cccc}
1 	 &0.001 	 &0.000 	 &-0.002 	
 \\  
0.001 	 &1 	 &-0.013 	 &-0.123 
 \\  
-0.000 	 &-0.013 	 &1 	 &0.003 
 \\  
-0.002 	 &-0.123 	 &0.003 	 &1 
 \\  
\end{array}
\right)
	.\eeq
The reduced chi-squared values are
\beq
\frac{\chi^{2}_{SM}}{\text{\# d.o.f.}}=\frac{1000.55}{275}=3.64, \qquad \frac{\tilde{\chi}^{2}}{\text{\# d.o.f.}}=\frac{629.19}{271}=2.28
\eeq
These results can be compared with the results in \eqref{Globalmuon} and  it shows  the overall improvement of the fitted values. Of course, these projections should be taken {\it cum grano salis} since we have kept fixed the central values of the different measures to today's values. Nevertheless, the overall picture shows that future measures will be able to settle the issue on whether NP is present in these flavour observables.

	\section{Conclusions}\label{conclusions}
	  
	  In this work we have analyzed  the LHCb experimental results for the $\mu/e$ ratios $R_{K}$ and  $R_{K^*}$,  the $g-2$ lepton magnetic moments results including the most recent data as well as the CDF result for the $W$ mass. These measures substantially deviate from the expected values predicted by the SM. We have, however, also   included in the analysis all the other observables related to this physics. 
	  
	  We started with a deep theoretical investigation of the $\mu/e$ ratios $R_{K}$ and  $R_{K^*}$  as well as the process $B_s \rightarrow \mu^+ \mu^-$. We went beyond the state-of-the-art by studying the impact of complex Wilson coefficients and  derived constraints on both their imaginary and real parts. This analysis has been then followed by a comprehensive comparison with experimental results. We find that:
	  
\begin{enumerate}
    \item The  deviations from the Standard Model are at the $4.7\sigma$ level when including only the hadronic insensitive observables. 
    \item Including the hadronic sensitive observables the deviation from the SM increases to a $6.1\sigma$ level.

    \item When considering simultaneously all relevant Wilson coefficients and combining both hadronic sensitive and insensitive data into the fit, the deviation from the SM peaks at 7.2$\sigma$ and decreases at the $4.9\sigma$ level if we assume that the central values of $R_K$ and $R_K^{\ast}$ are taken to be unity.
    
    \item By an independent method we estimated the SM   contributions unaccounted for in {\sc Flavio}~\cite{Straub:2018kue} which can stem, for an example, from hadronic physics and found that the result still supports the presence of NP. 
\end{enumerate}    
 { In the future we plan to perform a similar analysis in terms of SM effective field theories also for the $g-2$ and $W$-mass anomalies.}
 
 \vskip .3cm
We then moved to review different theoretical models apt at explaining the deviations from the Standard Model with the following conclusions: 
\begin{enumerate}
    \item  A minimal model featuring a $Z^\prime$ is unable to explain all the anomalies simultaneously. Additionally, even when  trying to address only the LFU violation, one finds that this is challenged by the $B_s$ mass mixing and LHC direct bounds. More involved models can ameliorate the situation. 
    
    \item Scalar leptoquarks, either a singlet or a doublet of weak interactions, can explain all the anomalies. In the first case one has to forbid di-quark couplings and in the latter the contribution to flavour anomalies is radiatively generated.  
    
    \item  Technicolor-like models, with a scale of new physics around $2$~TeV, can explain LFU violations, the $g-2$ of the muon as well as the weighted average of the $W$-mass result. 
    
    \item Composite Goldstone Higgs scenarios are challenged by the low energy composite scale required by the $g-2$ anomaly, and therefore  are  disfavoured.  
    
    \item Radiative models typically require very large muon left-handed Yukawa couplings and unnaturally large new fields multiplicity, thus are also challenged by the data. 

\item It is possible to accommodate dark matter and anomalies by either freeze-in or freeze-out type of constructions. The first mechanism can be realized in leptoquark type of models.
 
\end{enumerate}
 
Last but not least we have estimated the impact of the forthcoming data from LHCb Belle II in few years from now, when LHC Run3 will be completed and Belle II will have accumulated $5 ab^{-1}$.

	\subsubsection*{Acknowledgments}
	We thank Marzia Bordone for enlightening  discussions, Massimo Passera and Chien-Peng Yuan for comments on the manuscript, and
	Peter Stangl for clarifications about the {\sc Flavio} program.
	
	\appendix
	\section{Observables}\label{appendixA}
	In Table~\ref{tab:ang_obs} and \ref{tab:br_obs} we summarize the observables used in addition to the `hadronic insensitive' observables. 
	All bins are treated in the experimental analyses as independent, even if overlapping. It is clear that a correlation should 
	exists between measurements in overlapping bins, however this is not estimated by the experimental collaborations. 
	For this reason we include in our fit the measurements in all relevant bins, even if overlapping, without including any correlation 
	beyond the ones given in the experimental papers. Notice that, for instance in the case of the LHCb analysis \cite{1512.04442}, 
	the result in the bin $[1.1,6]$ GeV$^2$ has a smaller error than the measurements in the bins $[1.1,2.5],[2.5,4],[4,6]$ GeV$^2$, 
	even when the information from these three bins is combined. 
	
	\begin{table}[pt]
		\begin{center}
			\begin{tabular}{c|c}
				\toprule
				\multicolumn{2}{c}{\textbf{Angular observables}}   \\
				\midrule
				Observable  & {{$[q^{2}_{\text{min}},q^{2}_{\text{max}}]$ [GeV$^{2}$]}} \\ 
				\midrule
				\multicolumn{2}{c}{{\color{black}{LHCb $B^{+}\to K^{*+}\mu\mu$ 2020 \cite{LHCb:2020gog}, $B^0\to K^{*0}\mu\mu$ 2020 S \cite{LHCb:2020gog}}}} \\
				\midrule
				\text{$\langle F_{L} \rangle $} & [1.1, 6], [15, 19], [0.1, 0.98], [1.1, 2.5], [2.5, 4], [4, 6], [15, 17], [17, 19]  
				 \\  
				\text{$\langle S_{3} \rangle $} & [1.1, 6], [15, 19], [0.1, 0.98], [1.1, 2.5], [2.5, 4], [4, 6], [15, 17], [17, 19]  
				 \\  
				\text{$\langle S_{4} \rangle $} & [1.1, 6], [15, 19], [0.1, 0.98], [1.1, 2.5], [2.5, 4], [4, 6], [15, 17], [17, 19]  
				 \\  
				\text{$\langle S_{5} \rangle $} & \textcolor{blue}{[1.1, 6]}, [15, 19], [0.1, 0.98], [1.1, 2.5], [2.5, 4], \textcolor{blue}{[4, 6]}, [15, 17], [17, 19]  
				 \\  
				\text{$\langle S_{7} \rangle$} & [1.1, 6], [15, 19], [0.1, 0.98], [1.1, 2.5], [2.5, 4], [4, 6], [15, 17], [17, 19]  
				 \\  
				\text{$\langle S_{8} \rangle$} & [1.1, 6], [15, 19], [0.1, 0.98], \textcolor{blue}{[1.1, 2.5]}, [2.5, 4], [4, 6], [15, 17], [17, 19]  
				 \\  
				\text{$\langle S_{9} \rangle $} & [1.1, 6], [15, 19], [0.1, 0.98], [1.1, 2.5], [2.5, 4], [4, 6], [15, 17], [17, 19] \\  
				\text{$\langle  A_{FB} \rangle $} & [1.1, 6], [15, 19], [0.1, 0.98], [1.1, 2.5], [2.5, 4], [4, 6], [15, 17], [17, 19] \\  
				 \text{$\langle  P_1 \rangle $} & [1.1, 6], [15, 19], [0.1, 0.98], [1.1, 2.5], [2.5, 4], [4, 6], [15, 17], [17, 19] \\  
				\text{$\langle  P_2 \rangle $} & [1.1, 6], [15, 19], [0.1, 0.98], [1.1, 2.5], [2.5, 4], [4, 6], [15, 17], [17, 19] \\  
				\text{$\langle  P_3 \rangle $} & [1.1, 6], [15, 19], [0.1, 0.98], [1.1, 2.5], [2.5, 4], [4, 6], [15, 17], [17, 19] \\  
				\text{$\langle  P'_4 \rangle$} & [1.1, 6], [15, 19], [0.1, 0.98], [1.1, 2.5], [2.5, 4], [4, 6], [15, 17], [17, 19] \\  
				\text{$\langle  P'_5 \rangle $} & \textcolor{blue}{[1.1, 6]}, [15, 19], [0.1, 0.98], [1.1, 2.5], [2.5, 4], \textcolor{blue}{[4, 6] }, [15, 17], [17, 19] \\  
				\text{$\langle  P'_6 \rangle $} & [1.1, 6], [15, 19], [0.1, 0.98], [1.1, 2.5], [2.5, 4], [4, 6], [15, 17], [17, 19] \\  
				\text{$\langle  P'_8 \rangle$} & [1.1, 6], [15, 19], [0.1, 0.98], [1.1, 2.5], [2.5, 4], [4, 6], [15, 17], [17, 19] \\  
				\bottomrule
				\multicolumn{2}{c}{{\color{black}{CMS $B\to K^{*}\mu\mu$ 2017 \cite{CMS:2017ivg}}}} \\
				\midrule
				\text{$\langle P_{1} \rangle (B^{0}\to K^{*}\mu\mu)$} & [1, 2], [2, 4.3], [4.3, 6], [16, 19] \\  
				\text{$\langle P_{5}' \rangle (B^{0}\to K^{*}\mu\mu)$} & [1, 2], [2, 4.3], [4.3, 6], [16, 19] \\  
				\bottomrule
				\multicolumn{2}{c}{{\color{black}{ATLAS $B\to K^{*}\mu\mu$ 2017 \cite{ATLAS:2017dlm}}}} \\
				\midrule
				\text{$\langle F_{L} \rangle (B^{0}\to K^{*}\mu\mu)$} & [0.04, 2], [2, 4], [4, 6], [0.04, 4], \textcolor{blue}{[1.1, 6]}, [0.04, 6] \\  
				\text{$\langle S_{3} \rangle (B^{0}\to K^{*}\mu\mu)$} & [0.04, 2], [2, 4], [4, 6], [0.04, 4], [1.1, 6], [0.04, 6] \\  
				\text{$\langle S_{4} \rangle (B^{0}\to K^{*}\mu\mu)$} & [0.04, 2], [2, 4], \textcolor{blue}{[4, 6]}, [0.04, 4], [1.1, 6], [0.04, 6] \\  
				\text{$\langle S_{5} \rangle (B^{0}\to K^{*}\mu\mu)$} & [0.04, 2], [2, 4], \textcolor{blue}{[4, 6]}, [0.04, 4], [1.1, 6], [0.04, 6] \\  
				\text{$\langle S_{7} \rangle (B^{0}\to K^{*}\mu\mu)$} & [0.04, 2], [2, 4], [4, 6], [0.04, 4], [1.1, 6], [0.04, 6] \\  
				\text{$\langle S_{8} \rangle (B^{0}\to K^{*}\mu\mu)$} & [0.04, 2], \textcolor{blue}{[2, 4]}, [4, 6], [0.04, 4], [1.1, 6], [0.04, 6] \\  
				\text{$\langle P_{1} \rangle (B^{0}\to K^{*}\mu\mu)$} & [0.04, 2], [2, 4], [4, 6], [0.04, 4], [1.1, 6], [0.04, 6] \\  
				\text{$\langle P_{4}' \rangle (B^{0}\to K^{*}\mu\mu)$} & [0.04, 2], [2, 4], \textcolor{blue}{[4, 6]}, [0.04, 4], [1.1, 6], [0.04, 6] \\  
				\text{$\langle P_{5}' \rangle (B^{0}\to K^{*}\mu\mu)$} & [0.04, 2], [2, 4], \textcolor{blue}{[4, 6]}, [0.04, 4], [1.1, 6], [0.04, 6] \\  
				\text{$\langle P_{6}' \rangle (B^{0}\to K^{*}\mu\mu)$} & [0.04, 2], [2, 4], [4, 6], [0.04, 4], [1.1, 6], [0.04, 6] \\  
				\text{$\langle P_{8}' \rangle (B^{0}\to K^{*}\mu\mu)$} & [0.04, 2], \textcolor{blue}{[2, 4]}, [4, 6], [0.04, 4], [1.1, 6], [0.04, 6] \\  
				\bottomrule

			\end{tabular}
		\end{center}
		\caption{\em List of angular observables used in the global fit in addition to the `hadronic insensitive' observables. We explained in Section~\ref{fit} how the theoretical and experimental observations are treated.
		\label{tab:ang_obs}}	
	\end{table}%

	\normalsize
	
	\begin{table}[!htb!]
		\begin{center}
			\small\begin{tabular}{c|c}
				\toprule
				\multicolumn{2}{c}{\textbf{Branching ratios}}   \\
				\midrule
				Observable  & {{$[q^{2}_{\text{min}},q^{2}_{\text{max}}]$ [GeV$^{2}$]}} \\ \midrule
				
				\multicolumn{2}{c}{{\color{black}{LHCb $B^{\pm}\to K\mu\mu$ 2014 \cite{1403.8044}}}} \\
				\midrule
				\multirow{2}{*}{$\frac{d}{dq^{2}}\text{BR}(B^{\pm}\to K\mu\mu)$} & [0.1, 0.98], [1.1, 2], [2, 3], [3, 4], 
				[4, 5], [5, 6], \textcolor{blue}{[15, 16]}, [16, 17],\\
				& [17, 18], [18, 19], \textcolor{red}{[19, 20]}, \textcolor{blue}{[20, 21]}, [21, 22], [1.1, 6], [15, 22]  \\   
				\bottomrule
				
				\multicolumn{2}{c}{{\color{black}{LHCb $B^{0}\to K\mu\mu$ 2014 \cite{1403.8044}}}} \\
				\midrule
				$\frac{d}{dq^{2}}\text{BR}(B^{0}\to K\mu\mu)$ & \textcolor{blue}{[0.1, 2]}, [2, 4], \textcolor{blue}{[4, 6]}, [15, 17], [17, 22], [1.1, 6], \textcolor{blue}{[15, 22]}\\  \bottomrule
				\multicolumn{2}{c}{{\color{black}{LHCb $B^{\pm}\to K^{*}\mu\mu$ 2014 \cite{1403.8044}}}} \\
				
				\midrule
				$\frac{d}{dq^{2}}\text{BR}(B^{\pm}\to K^{*}\mu\mu)$ & [0.1, 2], \textcolor{blue}{[2, 4]}, [4, 6], [15, 17], \textcolor{red}{[17, 19]}, [1.1, 6], 
				[15, 19] \\  \bottomrule
				
				\multicolumn{2}{c}{{\color{black}{LHCb $B^{0}\to K^{*}\mu\mu$ 2016 \cite{1606.04731}}}} \\
				\midrule
				$\frac{d}{dq^{2}}\text{BR}(B^{0}\to K^{*}\mu\mu)$ & [0.1, 0.98], [1.1, 2.5], [2.5, 4], [4, 6], [15, 17], 
				[17, 19], [1.1, 6], [15, 19] \\  \bottomrule
				
				\multicolumn{2}{c}{{\color{black}{LHCb $B_{s}\to \phi\mu\mu$ 2021 \cite{2105.14007}}}} \\
				\midrule
			    $\frac{d}{dq^{2}}\text{BR}(B_{s}\to \phi \mu\mu)$ & \textcolor{blue}{[0.1, 0.98]}, \textcolor{red}{[1.1, 2.5]}, \textcolor{red}{[2.5, 4]}, 
				\textcolor{blue}{[4, 6]}, [15, 17], [17, 19], \textcolor{red}{[1.1, 6]}, [15, 19] \\  \bottomrule
				
				\multicolumn{2}{c}{{\color{black}{Babar $B\to X_{s} ll$ 2013 \cite{1312.5364}}}} \\
				\midrule
				\text{$\frac{d}{dq^{2}}\text{BR} (B\to X_{s} ll)$} &\textcolor{blue}{[0.1, 2]}, [2.0, 4.3], [4.3, 6.8],  [1, 6], [14,2, 25] \\  
				\text{$\frac{d}{dq^{2}}\text{BR}(B\to X_{s}\mu\mu)$} &[0.1, 2], [2.0, 4.3], [4.3, 6.8],  [1, 6], [14,2, 25]\\ 
				\text{$\frac{d}{dq^{2}}\text{BR}(B\to X_{s} ee)$}&\textcolor{blue}{[0.1, 2]}, [2.0, 4.3], [4.3, 6.8],  [1, 6], [14,2, 25]\\ 
				\bottomrule
				\multicolumn{2}{c}{{\color{black}{Belle $B \to X_{s}ll$ 2005 \cite{hep-ex/0503044}}}} \\
				\midrule
				$\frac{d}{dq^{2}}\text{BR}(B\to X_{s} ll)$ & [0.04, 1], [1, 6], [14.4, 25]\\  \bottomrule
				
			\end{tabular}
		\end{center}
		\caption{\em List of differential branching ratios used in the global fit in addition to the `hadronic insensitive' observables. The bins highlighted in red refer to measurements which deviate more than $3.5$ $\sigma$ from the theoretical prediction.
			\label{tab:br_obs}}
	\end{table}%

\end{document}